\documentclass[aps,prd,preprintnumbers,amsmath,amssymb]{revtex4}

\usepackage{epsfig}
\usepackage[colorlinks=true,linkcolor=blue]{hyperref} 
\usepackage{color}
\usepackage{ulem}
\usepackage{dcolumn}
\usepackage{graphicx}
\usepackage{subfig}

\usepackage{amssymb}
\usepackage{pifont}
\usepackage{multirow}

\DeclareGraphicsExtensions{.pdf}

\newcommand{\comma}{,}

\begin{document}

\title{Global analysis of fragmentation functions to charged hadrons with high-precision data from the LHC}

\author{Jun~Gao$^{1,2}$,~ChongYang~Liu$^{1,2}$,~XiaoMin~Shen$^{1,2,3}$,~Hongxi Xing$^{4,5,6}$,~Yuxiang Zhao$^{6,7,8,9}$}

\affiliation{
    $^1$ INPAC, Shanghai Key Laboratory for Particle Physics and Cosmology \\ \& School of Physics and Astronomy, Shanghai Jiao Tong University, Shanghai 200240, China \\
    $^2$Key Laboratory for Particle Astrophysics and Cosmology (MOE), Shanghai 200240, China\\
    $^3$Deutsches Elektronen-Synchrotron DESY, Notkestr. 85, 22607 Hamburg, Germany\\
    $^4${Key Laboratory of Atomic and Subatomic Structure and Quantum Control (MOE), Guangdong Basic Research Center of Excellence for Structure and Fundamental Interactions of Matter, Institute of Quantum Matter, South China Normal University, Guangzhou 510006, China}\\
    $^5${Guangdong-Hong Kong Joint Laboratory of Quantum Matter, Guangdong Provincial Key Laboratory of Nuclear Science, Southern Nuclear Science Computing Center, South China Normal University, Guangzhou 510006, China}\\
    $^6${Southern Center for Nuclear-Science Theory (SCNT), Institute of Modern Physics, Chinese Academy of Sciences, Huizhou 516000, China}\\
    $^7${Institute of Modern Physics, Chinese Academy of Sciences, Lanzhou, Gansu 730000, China}\\
    $^8${University of Chinese Academy of Sciences, Beijing 100049, China}\\
    $^9${Key Laboratory of Quark and Lepton Physics (MOE) and Institute of Particle Physics, Central China Normal University, Wuhan 430079, China}
 }

\email{\\ {jung49@sjtu.edu.cn} \\ {liucy1999@sjtu.edu.cn}\\
{xmshen137@sjtu.edu.cn}\\
{hxing@m.scnu.edu.cn}\\
{yxzhao@impcas.ac.cn}\\}

\begin{abstract}

Fragmentation functions (FFs) are essential non-perturbative QCD inputs for predicting hadron production cross sections in high energy scatterings. 
In this study, we present a joint determination of FFs for light charged hadrons through a global analysis at next-to-leading order (NLO) in QCD.
Our analysis incorporates a wide range of precision measurements from the LHC, as well as data from electron-positron collisions and semi-inclusive deep inelastic scatterings.
By including measurements of jet fragmentation at the LHC in our global analysis, we are able to impose strong constraints on the gluon FFs. 
A careful selection of hadron kinematics is applied to ensure the validity of factorization and perturbative calculations of QCD.
In addition, we introduce several methodological advances in fitting, resulting in a flexible parametrization form and the inclusion of theoretical uncertainties from perturbative calculations.
Our best-fit predictions show very good agreement with the global data, with $\chi^2/N_{pt}\sim 0.90$. We also generate a large number of Hessian error sets to estimate uncertainties and correlations of the extracted FFs.
FFs to charged pions (kaons and protons) are well constrained for momentum fractions down to 0.01 (0.1).
Total momentum of partons carried by light charged hadrons are determined precisely.
Their values for $u$, $d$ quarks and gluon saturate at about 50\% for a lower cut of the momentum fraction of 0.01.
Comparing our determinations to results from other groups, we find significant discrepancies, particularly for the proton fragmentation functions, as well as for gluon and unfavored quarks to pions and kaons. 
Pulls from individual datasets and impact of various choices of the analysis are also studied in details.
Our analysis raises questions concerning existing and future experimental measurements, including the need for clarifications on the definitions of various hadron final states and the necessity of further measurements on fragmentation from heavy quarks.
Additionally, we present an update of the FMNLO program used for calculating hadron production cross sections.
We demonstrate the broad applications of the FMNLO program combined with our new FFs, including NLO predictions on jet charges in proton collisions and reference cross sections for heavy-ion collisions, which show good agreement with LHC data.
Our FFs, including the error sets (denoted as NPC23), are publicly available in the form of LHAPDF6 grids.  

\end{abstract}
\maketitle

\pagebreak
\tableofcontents\newpage

\section{Introduction}

Hadronization of quarks and gluons into hadrons is essential to understanding
color confinement of the quantum chromodynamics (QCD). 
As non-perturbative quantities, various Fragmentation Functions (FFs) are proposed
to describe the phenomenon of hadronization, namely, the probability density that an outgoing
parton transforms into a colorless hadron in the simplest picture \cite{METZ2016136}. 
The properties of FFs are also reflected in the associated sum rules, which are derived from the conservation laws in QCD \cite{Collins:1981uw}.
For instance, the conservation of momentum in the collinear direction, 
combined with the probability interpretation of FFs, 
leads to the momentum sum rule.
Such a sum rule can help to constrain FFs for phenomenological studies 
that encompass all relevant hadron flavors and cover a wide range of momentum fraction.
Recent investigations have sparked new interest in assessing the validity of these properties for both single-hadron \cite{Collins:2023cuo} and dihadron FFs \cite{Pitonyak:2023gjx}, based on their operator definitions. \\

Apart from being crucial for comprehending color confinement in QCD, FFs have a wide application on many aspects of QCD.  
On one hand, we can use FFs to probe the transport properties of partons inside quark-gluon plasma created in heavy ion collision~\cite{JET:2013cls}.
On the other hand, they play a critical role in accurately interpreting experimental data
and for the study of nucleon structure, by disentangling different quark flavors from the initial
state.
Their importance becomes especially significant in the era of electron-ion colliders \cite{Accardi:2012qut,Anderle:2021wcy}, because
of the unprecedented demand for the precise determination of various parton distribution functions (PDFs).
FFs are also key inputs to precision programs at future electron-positron colliders such as CEPC~\cite{Liang:2023yyi,Zhu:2023xpk}.
Due to their non-perturbative and time dependent nature, as well as the intricacy of defining all out-states \cite{Collins:2023cuo}, direct calculation of FFs in lattice QCD using traditional Monte Carlo method has never been realized. Recently, new ideas of using quantum computing has been proposed \cite{Li:2024nod,Grieninger:2024axp}. Considering the limitations of quantum hardware, global analysis remains the most robust methodology for precise determination of FFs.
Therefore, it is of paramount importance to precisely determine FFs through global analyses of worldwide data.
\\

Thanks to the QCD factorization theorem, in the presence of hard scales, cross sections can be approximated as a convolution of perturbatively calculable short-distance scattering involving partons with universal long-distance functions~\cite{Collins:1989gx}, including PDFs and FFs. 
Therefore, different processes involving FFs can be related, such as single inclusive hadron production in $e^+e^-$ annihilation (SIA), semi-inclusive
deep-inelastic scattering (SIDIS), and proton-proton collisions. 
Tremendous efforts have been devoted to perturbatively calculable parts of these cross sections.
The hard coefficient functions for SIA have been calculated 
at next-to-next-to-leading order (NNLO) in QCD in Refs.~\cite{Rijken:1996vr,Rijken:1996ns,Mitov:2006wy,Blumlein:2006rr,Gehrmann_2022,Bonino_2024}.

For SIDIS, the next-to-leading order (NLO) corrections to the coefficients are given in Refs.~\cite{Altarelli:1979kv,Nason:1993xx,Furmanski:1981cw,Graudenz:1994dq,deFlorian:1997zj,deFlorian:2012wk}.
The NNLO corrections have been calculated recently in Refs.~\cite{Goyal:2023xfi, Bonino:2024qbh}.
Approximate NNLO and N$^3$LO corrections have also been obtained from expansion of the threshold resummation expressions \cite{Abele:2021nyo, Abele:2022wuy}.
The coefficient functions for single-inclusive hadron production at hadron-hadron collisions are known at NLO~\cite{Aversa:1988vb,deFlorian:2002az,Jager:2002xm}.
In addition to single-inclusive hadron production, jet fragmentation at hadron colliders provides a direct means to probe the FFs as functions of momentum fraction.
Numerical study of jet fragmentation at NLO has been carried out in Ref.~\cite{Arleo:2013tya} and recently in Refs.~\cite{Liu:2023fsq,Zidi:2024lid,Caletti:2024xaw} due to development on QCD subtraction method.

Analytic results for inclusive jet fragmentation have been calculated in both perturbative QCD with narrow jet approximation~\cite{Kaufmann:2015hma}, and in soft-collinear effective theory (SCET)~\cite{Chien:2015ctp}.
There also exists experimental measurements of hadron production in association with an isolated photon or $Z$ boson. 
The factorization formula for $Z+h$ production within SCET has been derived in Ref.~\cite{Kang:2019ahe}.
Despite of non-perturbative nature of FFs, their scale dependence 
follows the Dokshitzer-Gribov-Lipatov-Altarelli-Parisi (DGLAP) evolution equation with time-like splitting kernels,
which can be perturbatively calculated.
These time-like splitting kernels have been calculated at  ${\cal O}(\alpha_s^3)$ in the strong coupling constant in Refs.~\cite{Mitov:2006ic,Moch:2007tx,Almasy:2011eq, Chen:2020uvt,Ebert:2020qef,Luo:2020epw}.
\\

Equipped with perturbative QCD calculations and various experimental measurements, a comprehensive fit involving various data samples to extract FFs is feasible.
The representative efforts can be found in the works of
DSS \cite{deFlorian:2007ekg,deFlorian:2014xna,deFlorian:2017lwf,Borsa:2021ran}, HKNS \cite{Hirai:2007cx}, AKK \cite{Albino:2008fy}, NNFF \cite{Bertone:2018ecm}, MAPFF~\cite{Khalek:2021gxf}, and JAM \cite{Moffat:2021dji}.

They are carried out at NLO in QCD with different data samples and different theoretical prescriptions.
The HKNS analysis only includes SIA data, while AKK and NNFF analyses incorporate both SIA and proton-proton collision data. 
The MAPFF study uses SIA and SIDIS data. 
The DSS study uses data from SIA, SIDIS, and proton-proton collisions.
For JAM collaboration, they performed
a simultaneous fit of PDFs and FFs by including DIS, SIDIS, SIA, and Drell-Yan data. 
It is interesting to note that the AKK group has
extended their studies to include hadrons of $K_S^0$, $\Lambda(\bar{\Lambda})$, in addition to pion, kaon and (anti-)proton.
There also exists determinations of FFs at NNLO with SIA data only~\cite{Bertone:2017tyb,Soleymaninia:2018uiv}, and at approximate NNLO with SIA and SIDIS data~\cite{Borsa:2022vvp,AbdulKhalek:2022laj}.
Nevertheless, one of the least explored directions is the simultaneous fit by including different identified final states from different scattering processes~\cite{Soleymaninia:2020bsq,Soleymaninia:2019sjo}.\\
In this paper, we present a global analysis at the next-to-leading order on FFs of light charged hadrons, 
specifically $\pi^{\pm}$, $K^{\pm}$ and $p/\bar{p}$.
Note that a summary of the analysis together with an application on test of momentum sum rule have been reported in our earlier publication~\cite{Gao:2024nkz}.
To the best of our knowledge, our work represents the first joint determination of FFs of identified charged hadrons by simultaneously including relevant datasets from SIA, SIDIS, and $pp$ collisions.
There are several innovations of our analysis. 
Firstly, a stringent selection criterion has been implemented on the kinematics of the fragmentation processes to ensure the validity of leading twist collinear factorization and the associated perturbative calculations of QCD.
Secondly, residual theory uncertainties have been incorporated into the analysis. 
Thirdly, a fast interpolation technique for
the calculations of the cross sections has been utilized to significantly increase the efficiency of the global fit, instead of
the traditional method using Mellin transforms. 
Additionally, for the dataset selections, the production cross section ratios of various charged hadrons as well as measurements on charged hadron 
production in jets from the $pp$ collisions at the LHC have been included.  
The latter datasets provide strong constraints on the gluon FFs.
It is for the first time that the jet measurements have been included in a global analysis for light charged hadrons.
\\

The remainder of the paper is structured as follows: Sec. \ref{sec:overview} provides an overview of our global analysis, while Sec. \ref{sec:theory}
describes the theoretical framework. 
The results of the extracted FFs are presented in Sec. \ref{sec:results}, 
and the fit quality and alternative fits are discussed in Sec. \ref{sec:fit} and Sec. \ref{sec:alter_fit}, respectively.
Finally we summarize our main results in Sec. \ref{sec:summary}.
Our study has resulted in a new collection of FFs, which are publicly available on LHAPDF6. 
This work marks the inaugural effort of a newly established ``Non-Perturbative Physics Collaboration (NPC)'' for global analyses of FFs and PDFs.
The FF sets will be labeled as ``NPC23''.
\\

\section{Overview of the NPC23 analysis of FFs}
\label{sec:overview}
\subsection{Executive Summary}
In this section, we provide a summary of the primary results and characteristics of our global analysis on the fragmentation functions to light charged hadrons (NPC23).
We begin by presenting the best-fit FFs and their associated uncertainties at 68\% confidence level, followed by a concise overview of several advancements in our analysis, including the selection of experimental data, theoretical uncertainties, and fitting methodologies.

\subsubsection{Delivered FFs}

\begin{figure}[htbp]
\centering
  \includegraphics[width=0.8\textwidth]{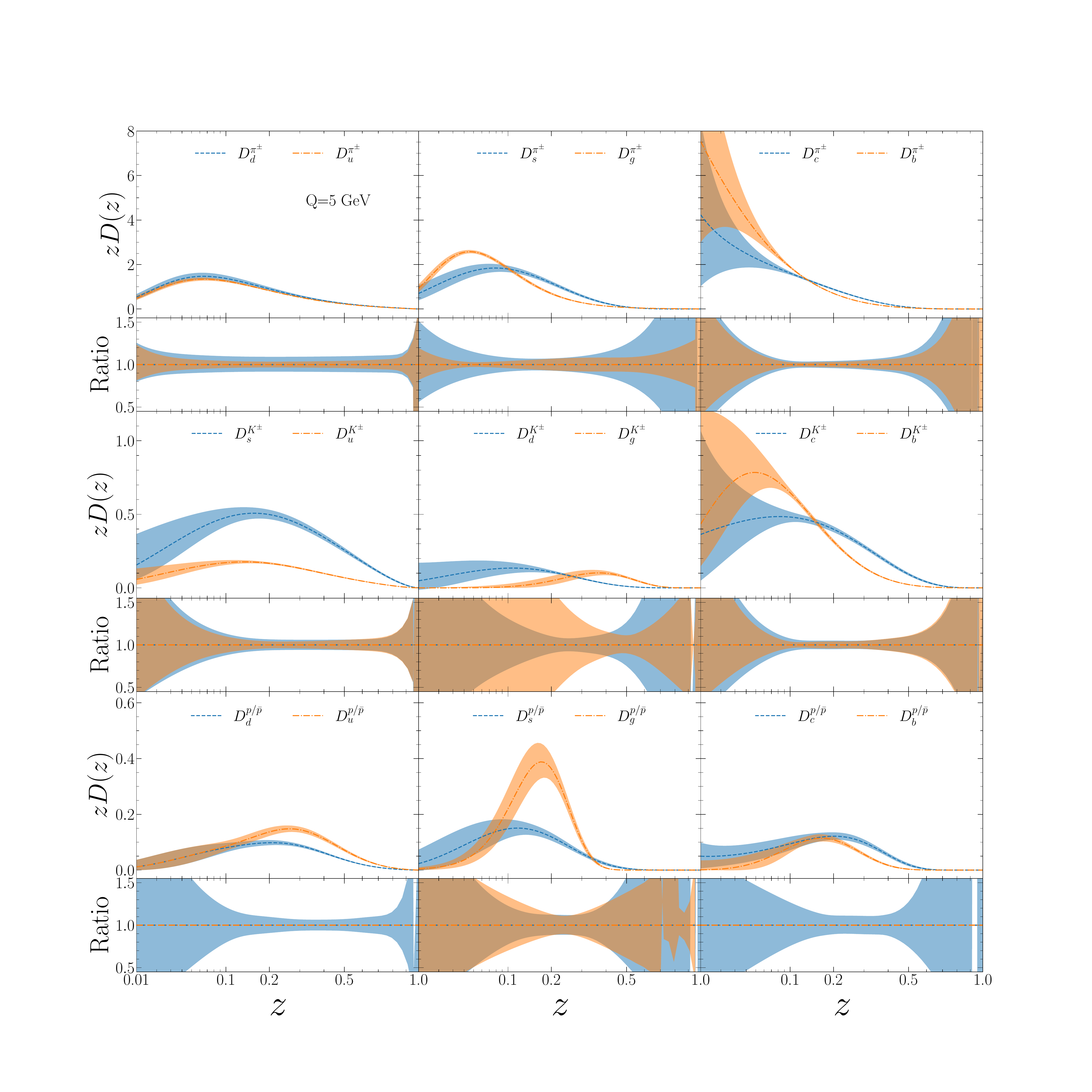}
	\caption{
Fragmentation functions from this study illustrating the diverse parton fragmentation outcomes, including $\pi^\pm$, $K^\pm$, and $p/\bar{p}$, at an energy of 5 GeV. The colored band represents the uncertainties estimated with the Hessian method at 68\% C.L. , with corresponding relative uncertainties displayed in the lower panel of each subplot.
	}
  \label{Fig:ff42_5GeV}
\end{figure}

The extracted fragmentation functions for various partons to pion ($\pi^{\pm}$), kaon ($K^{\pm}$), proton and anti-proton ($p/\bar p$), are shown at an initial scale of $Q=5$~GeV in Fig.~\ref{Fig:ff42_5GeV}. For simplicity, the results for positively and negatively charged particles are summed over. 
In each row, the sub-figures in the left column show FFs of the constituent quarks to the charged hadrons, namely $u$ and $d$ quarks to pion and proton, and $u$ and $s$ quarks to kaon.
The rest sub-figures correspond to FFs of un-favored quarks and gluon, as well as $c$ and $b$ quarks to the hadrons respectively.
Both the absolute values of momentum fraction times the FFs and their ratios normalized to the best-fit results are shown in the top and bottom panel of each sub-figure, respectively.
In general, the FFs to charged hadrons are well constrained for momentum fractions $z\sim 0.1-0.7$.
The FFs to charged pions have been precisely determined due to the large statistics from various measurements as well as the dominance in production rates over other charged hadrons.
Our results show an uncertainty of 3\%, 4\% and 8\% for FFs from gluon to pion at $z=0.05, 0.1$ and 0.3, respectively.
This high precision is primarily attributed to the data of jet fragmentation from the LHC, as will be explained later. 
Meanwhile, our determination shows an uncertainty of 4\%, 4\% and 7\% for FFs from $u$ quark to pion, kaon and proton at $z=0.3$, respectively.
We allow for breaking of isospin symmetry in FFs to pions.
The data indicate a slight preference for larger FFs from $d$ quark compared to the $u$ quark, although the differences remain within uncertainties.
The uncertainties for the latter are smaller due to the higher production rate of the $u$ quark in SIDIS and $pp$ collisions.
We observe a significant flavor asymmetry for the FFs to kaons from the $s$ and $u$ quarks, with the former being generally larger by a factor of $2 \sim 4$. 
This is expected given the larger mass of the strange quark, indicating that it is much harder for a $u$ quark to pick up a $\bar s$ quark from the vacuum.
For the FFs to protons as shown in the lower-left corner of Fig.~\ref{Fig:ff42_5GeV}, we simply assume that the FFs from the $u$ quark are twice those from the $d$ quark, considering the valence quarks in the proton. 
The current data are unable to distinguish between them.
The FFs to kaons and (anti-)protons from gluons are less well constrained due to their smallness in comparison to the FFs to pions from gluons.
The FFs of gluons to (anti-)protons exhibit a narrow peak around $z\sim 0.2$, albeit with large uncertainties.
Notably, we find that the FFs from the strange quark to pions can be even larger than those from the $d$ quark.
That is due to the pull from SIA measurements on the charged hadron cross sections.
It is possible that the SIA measurements also include contributions from production of short-lived strange hadrons, especially $K^0_S$ and $\Lambda$. 
Further discussion on this will be provided in later sections. 
The FFs from heavy quarks ($c$ and $b$) are well constrained for $z$ between 0.1 $\sim$ 0.5, thanks to the measurements from heavy-flavor tagged hadronic events in $e^+e^-$ collisions at the Z-pole.
These measurements significantly reduce the degeneracy of fragmentation from different quark flavors when fitting to the inclusive SIA data.
However, the uncertainties on FFs from heavy quarks increase significantly for $z$ below 0.1 due to our kinematic selection for the SIA data at the Z-pole.
Additionally, FFs from heavy quarks are softer than those from light quarks due to cascade decays of the heavy-flavor mesons, especially for the bottom quark. 

\subsubsection{Selection of experiments\label{sec:data-selection}}

A wide range of measurements exist for the single inclusive production of charged hadrons from SIA, SIDIS and $pp$ collisions at various center-of-mass energies.
These measurements can be categorized into several groups based on the identified light charged hadrons, including $\pi^{\pm}$, $K^{\pm}$, $p/\bar p$, as well as unidentified charged hadrons or charged tracks. 
For the latter two categories, we simply assume that they correspond to a sum of the light charged hadrons due to their dominance in production, while acknowledging that their exact definitions may vary between experiments.
It is important to note that even for the identified charged hadrons, there are ambiguities regarding the inclusion or exclusion of certain secondary contributions from short-lived hadrons, which differ, for instance, between experiments at LEP~\cite{OPAL:1994zan,ALEPH:1994cbg,DELPHI:1998cgx,OPAL:2002isf,DELPHI:2000ahn} and ALICE~\cite{ALICE:2014juv, ALICE:2019hno,ALICE:2020jsh}.
Uncertainties due to these possible mismatches between different measurements should be clarified in future experiments.  
We will briefly outline a few key differences in our choice of datasets and points compared to previous determinations. 
Our primary goal is to ensure the validity of leading-twist factorization and the convergence of perturbative calculations.
To achieve this, we have implemented stringent selection criteria on hadron kinematics, requiring both the hadron momentum fraction $z>0.01$ and the hadron energies or transverse momenta $E_h(p_{T,h})>4$ GeV.
The hadron energies are measured in the center-of-mass frame and the Breit frame for SIA and SIDIS, respectively.
For $pp$ collisions, the transverse momenta of the hadrons are used instead of the energies.
This approach allows us to safely neglect hadron mass corrections in our analysis.
For SIA measurements on single hadron production, we only incorporate those based on inclusive event samples except for the ones from heavy-flavor tagged events. 
Notably, we have not included the OPAL measurement on fragmentation of individual light quarks~\cite{OPAL:1999zfe} since those extractions are based on a leading-order analysis.
Our kinematic cuts exclude most of the SIDIS measurements except for those from H1 and ZEUS at high $Q^2$ values.
However, we choose to retain a subset of the COMPASS data, which have the highest inelasticity and the largest Bjorken-$x$, even though they could have been excluded by our kinematic selection. 
For $pp$ collisions, we are able to include, for the first time, a variety of jet fragmentation measurements from the LHC.
Additionally, we have also incorporated traditional measurements on differential cross sections of single inclusive hadron production as functions of hadron transverse momenta.
However, we have only incorporated measured ratios of different hadron species instead of the absolute cross sections, in order to
avoid additional complications and uncertainties arising from the inputs of various normalizations.

\subsubsection{Advances in fitting methodology}

Our analysis represents a joint determination of all three light charged hadrons, rather than separate fits for individual hadron species.  
Thus both correlations between measurements on different hadron species, and correlations between their theory predictions can be properly taken into account.  
To estimate uncertainties and correlations of different FFs, a large number of Hessian FFs have been published.
In the determination of the Hessian uncertainties, we have applied a dynamic tolerance criterion, which involves examining potential tensions among different data or possible underestimation of experimental uncertainties. 
Accesses to our FFs are available via LHAPDF6 grids, as explained in appendix~\ref{sec:lha6}.
NLO QCD predictions of all processes included in our analysis are calculated with the FMNLO program~\cite{Liu:2023fsq}, which enables fast
convolution with FFs using stored grids on hard coefficient functions.
This ensures efficient scans over the high-dimensional parameter space, allowing us to provide a large number of alternative fits for investigations into various systematic effects or choices of parameters. 
Moreover, the convolution with FFs is performed directly in $z$-space rather than in the space of Mellin moments. 
This allows for much more flexible parametrization forms of the FFs at the initial scale and the application of a positivity constraint.
Uncertainties on the theoretical predictions are also calculated by taking the envelope of an ensemble of predictions with different QCD scale choices, and included as part of the full covariance matrix together with experimental uncertainties.   
\subsection{Experimental datasets fitted}
Our new determination of FFs has been obtained via a global analysis of 
single-inclusive hadron production and jet fragmentation measurements on hadron-hadron colliders,
SIDIS on lepton-ion colliders, and SIA data on lepton colliders.
All these datasets can be divided into subsets according to the range of jet $p_T$ for jet fragmentation,
$Q^2$ for SIDIS, 
and collision energy for single-inclusive hadron production at $pp$ and SIA.
In total, this results in 138 subsets.
We provide detailed explanations of our choices of the experimental datasets used in the following.
\subsubsection{Hadron collisions}
For hadron-hadron collision, we include in our fit  $\pi^{\pm}$, $K^{\pm}$, $p/\bar p$, and unidentified hadron production data from
both jet fragmentation and single-inclusive hadron production.
We make the assumption that the measured cross sections of unidentified charged hadrons $h^{\pm}$ or charged tracks are the sum of charged pions, kaons and protons, as explained earlier.

\begin{table}[h]
  \begin{tabular}{|l|l|l|l|l|l|l|l|l|}
    \hline
    exp. & $\sqrt{s}$(TeV) & luminosity & hadrons & final states & $R_j$ &
    cuts for jets/hadron & observable & $N_{\rm{pt}}$\\
    \hline
    ATLAS{\cite{ATLAS:2019dsv}} & 5.02 & 25 ${\rm pb}^{- 1}$ & $h^{\pm}$ &
    $\gamma + j$ & 0.4 & $\Delta \phi_{j, \gamma} > \frac{7 \pi}{8}$ &
    $\frac{1}{N_{{\rm jet}}} \frac{d N_{{\rm ch}}}{d p_{T, h}}$ & 6\\
    \hline
    CMS{\cite{CMS:2018mqn}} & 5.02 & 27.4 ${\rm pb}^{- 1}$ & $h^{\pm}$ &
    $\gamma + j$ & 0.3 & $\Delta \phi_{j, \gamma} > \frac{7 \pi}{8}, \Delta
    R_{h, j} < R_j$ & $\frac{1}{N_{{\rm jet}}} \frac{d N_{{\rm ch}}}{d \xi}$
    & 4\\
    \hline
    ATLAS{\cite{ATLAS:2020wmg}} & 5.02 & 260 ${\rm pb}^{- 1}$ & $h^{\pm}$ &
    $Z + h$ & no jet & $\Delta \phi_{h, Z} > \frac{3}{4} \pi$ & $\frac{1}{n_Z}
    \frac{d N_{{\rm ch}}}{d p_{T, h}}$ & 9\\
    \hline
    CMS{\cite{CMS:2021otx}} & 5.02 & 320 ${\rm pb}^{- 1}$ & $h^{\pm}$ & $Z +
    h$ & no jet & $\Delta \phi_{h, Z} > \frac{7}{8} \pi$ & $\frac{1}{n_Z}
    \frac{d N_{{\rm ch}}}{d p_{T, h}}$ & 11\\
    \hline
    LHCb{\cite{LHCb:2022rky}} & 13 & 1.64 ${\rm fb}^{- 1}$ & $\pi^{\pm}, K^{\pm}, p/\bar{p}$& $Z
    + j$ & 0.5 & $\Delta \phi_{j, \gamma} > \frac{7 \pi}{8}, \Delta R_{h, j} <
    R_j$ & $\frac{1}{n_Z} \frac{d N_{{\rm ch}}}{d \zeta}$ & 20\\
    \hline
    ATLAS{\cite{ATLAS:2018bvp}} & 5.02 & 25 ${\rm pb}^{- 1}$ & $h^{\pm}$ &
    inc. jet & 0.4 & - & $\frac{1}{N_{{\rm jet}}} \frac{d N_{{\rm ch}}}{d
    \zeta}$ & 63\\
    \hline
    ATLAS{\cite{ATLAS:2011myc}} & 7 & 36 ${\rm pb}^{- 1}$ & $h^{\pm}$ & inc.
    jet & 0.6 & $\Delta R_{h, j} < R_j$ & $\frac{1}{N_{{\rm jet}}} \frac{d
    N_{{\rm ch}}}{d \zeta}$ & 103\\
    \hline
    ATLAS{\cite{ATLAS:2019rqw}} & 13 & $33$ ${\rm fb}^{- 1}$ & $h^{\pm}$ &
    dijet & 0.4 & $p_T^{{\rm lead}} / p_T^{{\rm sublead}} < 1.5$ &
    $\frac{1}{N_{{\rm jet}}} \frac{d N_{{\rm ch}}}{d \zeta}$ & 280\\
    \hline
  \end{tabular}
  \caption{Jet fragmentation datasets used in the fit, together with the
  c.m. energy, luminosity, identified hadrons, final states,
  anti-$k_T$ jet radius $R_j$, cuts on jets or/and hadrons, the observable,
  and the number of data points after data selection. 
}
  \label{tab:data-jet-frag}
\end{table}

In the context of jet fragmentation, the tagged hadron is required to be produced either 
within a reconstructed jet or inside a cone that is in an opposing azimuthal direction to a $Z$ boson.
The jet fragmentation data used in our fit are summarized in Tab.~\ref{tab:data-jet-frag}, including the center-of-mass energy, luminosity, identified hadrons, final states, anti-$k_T$ jet radius $R_j$, cuts on jets and/or hadrons, the observable, and the number of data points after data selection described in Sec.\ref{sec:data-selection}.
ATLAS and CMS ~\cite{ATLAS:2019dsv, CMS:2018mqn} measured the hadron multiplicity in transverse momentum $p_T$ and 
$\xi\equiv \ln [-p_{T,\gamma}^2 / (\vec{p}_{T,\gamma}\cdot\vec{p}_{T,h})]$ of charged hadrons inside reconstructed jets, produced 
in association with an isolated photon, separated by $\Delta\phi_{j,\gamma} > 7\pi/8$.
For jet fragmentation with a tagged $Z$ boson, we include measurements of charged track multiplicity from the ATLAS and CMS collaborations
 ~\cite{ATLAS:2020wmg, CMS:2021otx}.
The charged tracks in these events
reside primarily in the leading jet azimuthally opposite to the $Z$ boson {\cite{ATLAS:2020wmg}}.
Though no jet reconstructions are explicitly performed in these two measurements,
the tagged charged tracks are required to be azimuthally well separated to the $Z$ boson 
by $\Delta\phi_{\rm{h},\gamma} > 3\pi/4$ (ATLAS) or $\Delta\phi_{\rm{h},\gamma} > 7\pi/8$ (CMS). 
We also include $Z+j$  dataset at 13 TeV from the LHCb \cite{LHCb:2022rky}, measured separately for $\pi^{\pm}$, $K^{\pm}$ and $p/\bar p$ production.
Note that for the LHCb measurements several selection cuts on kinematics of muons from decays of $Z$ boson are applied which have been realized in the FMNLO program.
Another category of jet fragmentation included in our fit is parton fragmentation inside inclusive jets or dijet events, 
which have been measured by the ATLAS at 5.02~TeV, 7~TeV, and 13~TeV~\cite{ATLAS:2018bvp, ATLAS:2011myc, ATLAS:2019rqw}.
The results are presented as charged track multiplicity in $\zeta\equiv p_{T,h}/p_{T,j}$, with $p_{T,j}$ being the transverse momentum of the probed jet.
For the 13-TeV di-jet measurements, the two leading jets are required to 
satisfy the balance condition $p_{T}^{\rm lead}/p_{T}^{\rm sublead} < 1.5$.

\begin{figure}[htbp]
\centering
  \includegraphics[width=0.6\textwidth]{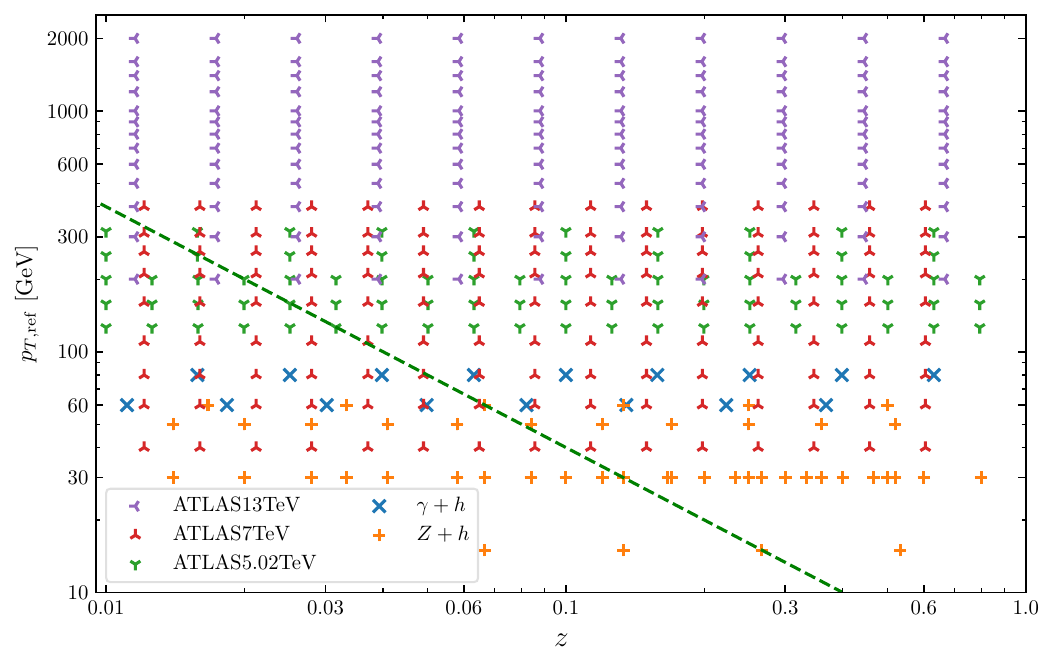}
	\caption{
     $z$-$p_T$ coverage of the jet fragmentation in  $\gamma/Z$-tagged hadron production, and ATLAS inclusive jets or dijet events. 
     $p_{T,{\rm ref}}$  is the reference momentum of the initiating parton at LO in QCD.
     The green dashed lines correspond to the cut  $p_{T,h}>4$ GeV.
	}
  \label{Fig:z-Q-jetfrag}
\end{figure}

These jet fragmentation datasets cover a wide kinematic region, as shown in Fig.~\ref{Fig:z-Q-jetfrag}, where only data points with LO hadron momentum fraction $z>0.01$ are displayed.
The reference momentum $p_{T,{\rm ref}}$ of the initiating parton at LO in QCD is chosen as the 
transverse momentum of the photon/$Z$ boson for $\gamma/Z$-tagged hadron production.
For ATLAS inclusive jets or dijet events, $p_{T,\rm{ref}}$ is the transverse momentum of the jet.
The green dashed line corresponds to the cut $p_{T,h}>4$ GeV at LO in QCD, as mentioned in Sec.~\ref{sec:data-selection}, to ensure the validity of the leading twist factorization formalism and convergence of perturbative calculations.

We also include inclusive hadron production measurements from ALICE~\cite{ALICE:2014juv, ALICE:2019hno,ALICE:2020jsh} and STAR~\cite{STAR:2011iap}, 
which are summarized in Tab.~\ref{tab:data-pp-inc}.
The table includes the center-of-mass energy, number of events,
kinematic cuts, the identified hadrons, the observable used in the fit, and
the number of data points after data selection. 
Here, $\pi, K, p$ denote $\pi^{\pm}, K^{\pm}, p +
  \bar{p}$, respectively. 
As mentioned in Sec.~\ref{sec:data-selection}, only bins with $p_{T,h}>4~{\rm GeV}$ are included in our fit.
We exclusively examine ratios involving production cross sections of different charged hadrons or of different collision energies, aiming to sidestep additional complexities and uncertainties arising from inputs of various normalizations.
For example, Ref.~\cite{ALICE:2020jsh} includes ATLAS measurements at both 7~TeV and 13~TeV. 
We consider the cross section ratio of different charged hadrons at 13~TeV, 
and the cross section ratio of the same charged hadron between 13~TeV and 7~TeV.
\begin{table}[h]
  \begin{tabular}{|l|l|l|l|l|l|l|}
    \hline
    exp. & $\sqrt{s_{N N}} ({\rm TeV})$ & \# events (million) & $p_{T, h}$ &
    hadrons & observable & $N_{{\rm pt}}$\\
    \hline
    ALICE{\cite{ALICE:2020jsh}} & 13 & 40-60($p p$) & $[2, 20]$ GeV & $\pi, K,
    p$, $K_S^0$ & $K / \pi, p / \pi, K^0_S / \pi$ & 49\\
    \hline
    ALICE{\cite{ALICE:2020jsh}} & 7 & 150($p p$) & $[3, 20]$ GeV & $\pi, K, p$
    & 13TeV/7TeV for $\pi, K, p$ &
    37\\
    \hline
    ALICE{\cite{ALICE:2019hno}} & 5.02 & 120($p p$) & $[2, 20]$ GeV & $\pi, K,
    p$ & $K / \pi, p / \pi$ & 34\\
    \hline
    ALICE{\cite{ALICE:2014juv}} & 2.76 & 40($p p$) & $[2, 20]$ GeV
    & $\pi, K, p$ & $K / \pi, p / \pi$ & 27\\
    \hline
    STAR{\cite{STAR:2011iap}} & 0.2 & 14($p p$) & $[3, 15]$ GeV & $\pi, K, p,
    K_S^0$ & $K / \pi$, $p / \pi^+$, $\bar{p} / \pi^-$, $K_S^0 / \pi$, $\pi^-
    / \pi^+$, $K^- / K^+$ & 60\\
    \hline
  \end{tabular}
  \caption{Inclusive hadron production datasets on hadron colliders used in the fit, 
  together with the center-of-mass energy, number of events,
  kinematic cuts, the identified hadrons, the observable used in the fit, and
  the number of data points after data selection. 
  Here, $\pi, K, p$ denote $\pi^{\pm}, K^{\pm}, p +
  \bar{p}$, respectively. 
  }
  \label{tab:data-pp-inc}
\end{table}

\subsubsection{SIDIS}

For SIDIS, we incorporate 
measurements of scaled momentum distributions, $x_p$, of unidentified charged hadrons from H1 and ZEUS at high $Q^2$~\cite{H1:2007ghd,H1:2009lef, ZEUS:2010mrq}.
Both the H1 and ZEUS experiments measured the normalized scaled momentum distributions $D \equiv \frac{1}{N} \frac{d n_{h^{\pm}}}{dx_p}$,
where the scaled momentum $x_p$ is defined as $2 p_h / Q$, $p_h$ is the momentum of the identified charged track in the current region of the Breit frame, 
$n_{h^{\pm}}$ is the number of charged tracks, and $N$ is the number of events.
The H1 experiment also measured the charge asymmetry of the scaled momentum distributions, defined as 
$(D^+ - D^-)/(D^+ + D^-)$,  with $D^{+(-)}$ being the normalized scaled momentum distributions of the positively (negatively) charged tracks.

There are also measurements on the differential multiplicity of identified charged hadrons from COMPASS at relatively low $Q^2$
with isoscalar (06I) \cite{COMPASS:2016xvm,COMPASS:2016crr}, or proton (16p) \cite{Pierre:2019nry} targets.
The multiplicity for a specific hadron of type h is defined as
\begin{equation}\label{Eq:hmul}
    \frac{\mathrm{d} M^{\mathrm{h}} (x, z, Q^2)}{\mathrm{d} z} =
\frac{\mathrm{d}^3 \sigma^{\mathrm{h}} (x, z, Q^2) / \mathrm{d} x \mathrm{d}
Q^2 \mathrm{d} z}{\mathrm{d}^2 \sigma^{\mathrm{DIS}} (x, Q^2) / \mathrm{d} x \mathrm{d} Q^2}
\end{equation}
where $x=Q^2/(2P\cdot q)$ is the Bjorken variable,
$z=(P\cdot P_h)/(P\cdot q)$, 
and $Q^2=-q^2$ is the negative square of the lepton momentum transfer,
$y=s/(x Q^2)$,
$P,~p_h,~q_\gamma$ are the 4-momenta of the incoming hadron/ion, the tagged hadron, and the virtual photon, respectively.
We only include the COMPASS data with the highest inelasticity and the large Bjorken-$x$. 
In the measurement with a proton target, we include data on the cross section ratio of $p / \bar{p}$ instead of the absolute cross sections considering the non-negligible hadron mass corrections.
All these SIDIS datasets are summarized in Tab.~\ref{tab:data-SIDIS}.
\begin{table}[h]
  \begin{tabular}{|l|l|l|l|l|l|l|}
    \hline
    exp. & $\sqrt{s}$(GeV) & luminosity & kinematic cuts & hadrons & obs &
    $N_{{\rm pt}}$\\
    \hline
    H1{\cite{H1:2007ghd}} & 318 & 44 ${\rm pb}^{- 1}$ & $Q^2 \in$[175,20000]
    ${\rm GeV}^2$ & $h^{\pm}$ & $D \equiv \frac{1}{N} \frac{d n_{h^{\pm}}}{d
    x_p}$ & 16\\
    \hline
    H1{\cite{H1:2009lef}} & 318 & 44 ${\rm pb}^{- 1}$ & $Q^2 \in$[175,8000]
    ${\rm GeV}^2$ & $h^{\pm}$ & $A \equiv \frac{D^+ - D^-}{D^+ + D^-}$ & 14\\
    \hline
    ZEUS{\cite{ZEUS:2010mrq}} & 300,318 & 440 ${\rm pb}^{- 1}$ & $Q^2
    \in$[160,40960] ${\rm GeV}^2$ & $h^{\pm}$ & $D$ & 32\\
    \hline
    COMPASS06{\cite{COMPASS:2016xvm,COMPASS:2016crr}}& 17.3 & 540 ${\rm pb}^{-
    1}$ & $x \in [0.14, 0.4], y \in [0.3, 0, 5]$ & $\pi, K, h^\pm$ & $\frac{d
    M^h}{d z}$ & 124\\
    \hline
    COMPASS16{\cite{Pierre:2019nry}}& 17.3 & - & $x \in [0.14, 0.4], y \in
    [0.3, 0, 5]$ & $\pi, K, p$ & $\frac{d M^h}{d z}$ & 97\\
    \hline
  \end{tabular}
  \caption{Selected SIDIS datasets used in the fit, together with the center-of-mass energy, 
  luminosity, kinematic cuts, the identified hadrons, the observable,
  and the number of data points after data selection. 
  }
      \label{tab:data-SIDIS}
\end{table}

\begin{figure}[htbp]
\centering
  \includegraphics[width=0.6\textwidth]{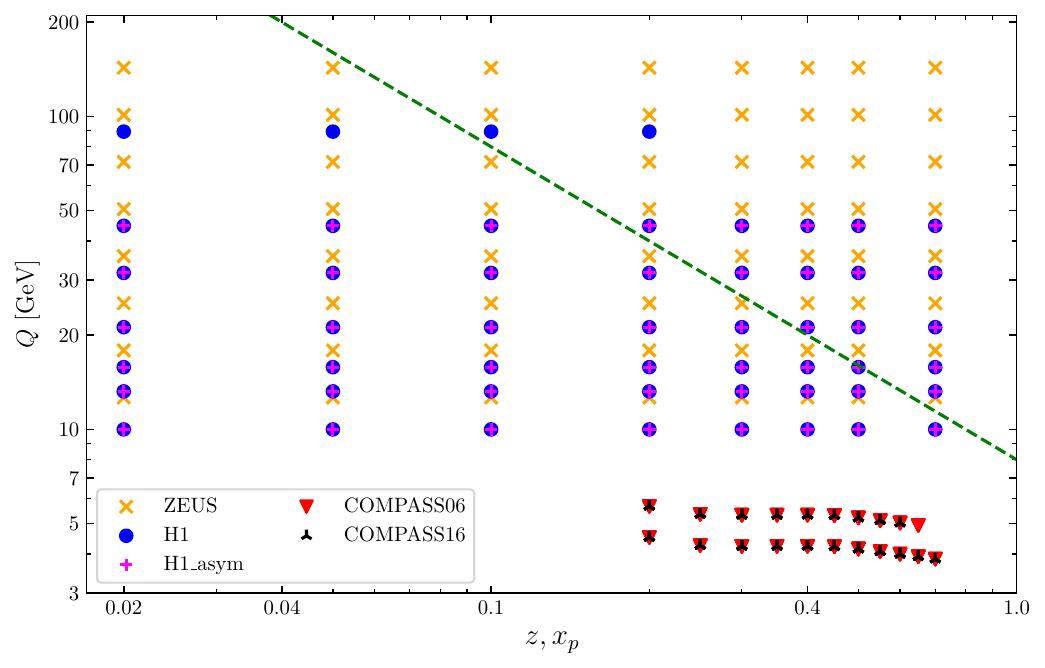}
	\caption{
$x_p$-$Q$ ($z$-$Q$ ) coverage of the HERA (COMPASS) datasets. 
The green dashed line corresponds to the cut $x_p Q/2 > 4$ GeV, which only applies to H1 (H1 asymmetry) and ZEUS datasets.
	}
  \label{Fig:z-Q-SIDIS}
\end{figure}

The $x_p(z)$-$Q$ coverage of all the SIDIS datasets is summarized in Fig.~\ref{Fig:z-Q-SIDIS}, 
where the $x$ axis corresponds to the scaled momentum variable $x_p$ and $z$ for HERA datasets (ZEUS, H1) and COMPASS datasets, respectively. 
The green dashed line corresponds to our kinematic cut $x_p Q/2 > 4$ GeV, mentioned in Sec.~\ref{sec:data-selection}, 
which only applies to ZEUS and H1 data points, but not to the COMPASS data points.

\subsubsection{SIA}
The SIA datasets included in our fit are summarized in Tab.~\ref{tab:data-SIA}, along with the center-of-mass
energy, luminosity, partonic final states, the identified hadrons, and the number of data points after data selection. 
For $e^+ e^-$ collisions at the $Z$ mass pole, we show the number of hadronic $Z$ events instead of the luminosity. 
We have included a large collection of data from TAS, TPC below the $Z$-pole~\cite{TASSO:1988jma,TPCTwoGamma:1988yjh}, 
OPAL, ALEPH, DELPHI, and SLD at $Z$-pole~\cite{OPAL:1994zan,ALEPH:1994cbg,DELPHI:1998cgx,SLD:2003ogn}, 
and OPAL and DELPHI above the $Z$-pole~\cite{OPAL:2002isf,DELPHI:2000ahn}.
They measured the production of $\pi^{\pm}$, $K^{\pm}$ and $p/\bar p$ separately, except for OPAL at a collision energy of 202 GeV.
All measurements are presented as hadron multiplicity in the scaled momentum defined as $2 p_h / \sqrt s$.
We have applied kinematic cuts on the hadron momentum fraction $z>0.01$ and hadron energies $E_h>4$ GeV.
Apart from measurements on the inclusive hadron production, which receives contributions from all active flavor of quarks, SLD also measured the hadron production from events of $Z$ bosons decaying into heavy quarks ($c$ or $b$ quark).
DELPHI conducted similar measurements but for $b$ quark only.
These heavy flavor tagged measurements provide key inputs to the determination of FFs from heavy quarks and play important roles in the separation of quark flavors in FFs, 
as will be explained in later sections.   
\begin{table}[h]
\begin{tabular}{|l|l|l|l|l|l|}
  \hline
  exp. & $\sqrt{s}$ & lum.($n_Z$) & final states & hadrons & $N_{{\rm pt}}$\\
  \hline
  OPAL{\cite{OPAL:1994zan}} & $m_Z$ & 780 000 & $Z \rightarrow q \bar{q}$ &
  $\pi^{\pm} \comma K^{\pm}$ & 20\\
  \hline
  ALEPH{\cite{ALEPH:1994cbg}} & $m_Z$ & 520 000 & $Z \rightarrow q \bar{q}$ &
  $\pi^{\pm}, K^{\pm}, p (\bar{p})$ & 42\\
  \hline
  DELPHI{\cite{DELPHI:1998cgx}} & $m_Z$ & 1 400 000 & $Z \rightarrow q
  \bar{q}$ & $\pi^{\pm} \comma K^{\pm}, p (\bar{p})$ & 39\\
  \cline{4-6}
   &  &  & $Z \rightarrow b \bar{b}$ & $\pi^{\pm} \comma K^{\pm}, p
  (\bar{p})$ & 39\\
    \hline
  \multirow{3}{*}{SLD{\cite{SLD:2003ogn}}} & \multirow{3}{*}{$m_Z$} & \multirow{3}{*}{400 000} & $Z \rightarrow q \bar{q}$ &
  $\pi^{\pm} \comma K^{\pm}, p (\bar{p})$ & 66\\
  \cline{4-6}
   &  &  & $Z \rightarrow b \bar{b}$ & $\pi^{\pm} \comma K^{\pm}, p
  (\bar{p})$ & 66\\
  \cline{4-6}
   &  &  & $Z \rightarrow c \bar{c}$ & $\pi^{\pm} \comma K^{\pm}, p
  (\bar{p})$ & 66\\
  \hline
  TASSO{\cite{TASSO:1988jma}} & 34GeV & 77 ${\rm pb}^{- 1}$ & inc. had. &
  $\pi^{\pm} \comma K^{\pm}, p (\bar{p})$ & 3\\
  \hline
  TASSO{\cite{TASSO:1988jma}} & 44GeV & 34 ${\rm pb}^{- 1}$ & inc. had. &
  $\pi^{\pm}, \pi^0$ & 5\\
  \hline
  TPC{\cite{TPCTwoGamma:1988yjh}} & 29GeV & 70 ${\rm pb}^{- 1}$ &
  inc. had. & $\pi^{\pm} \comma K^{\pm}$ & 12\\
  \hline
  OPAL{\cite{OPAL:2002isf}} & 201.7GeV & 433 ${\rm pb}^{- 1}$ & inc. had. &
  $h^{\pm}$ & 17\\
  \hline
  DELPHI{\cite{DELPHI:2000ahn}} & 189GeV & 157.7 ${\rm pb}^{- 1}$ & inc. had.
  & $\pi^{\pm} \comma K^{\pm}, p (\bar{p})$ & 9\\
  \hline
\end{tabular}
  \caption{Selected SIA datasets used in the fit, together with the center-of-mass
  energy, luminosity, partonic final states, the identified hadrons, and the
  number of data points after data selection. 
  For $e^+ e^-$ collisions at the  $Z$ mass pole, we show the number of hadronic $Z$ events instead of the
  luminosity. 
  }
    \label{tab:data-SIA}
\end{table}

\begin{figure}[htbp]
\centering
  \includegraphics[width=0.6\textwidth]{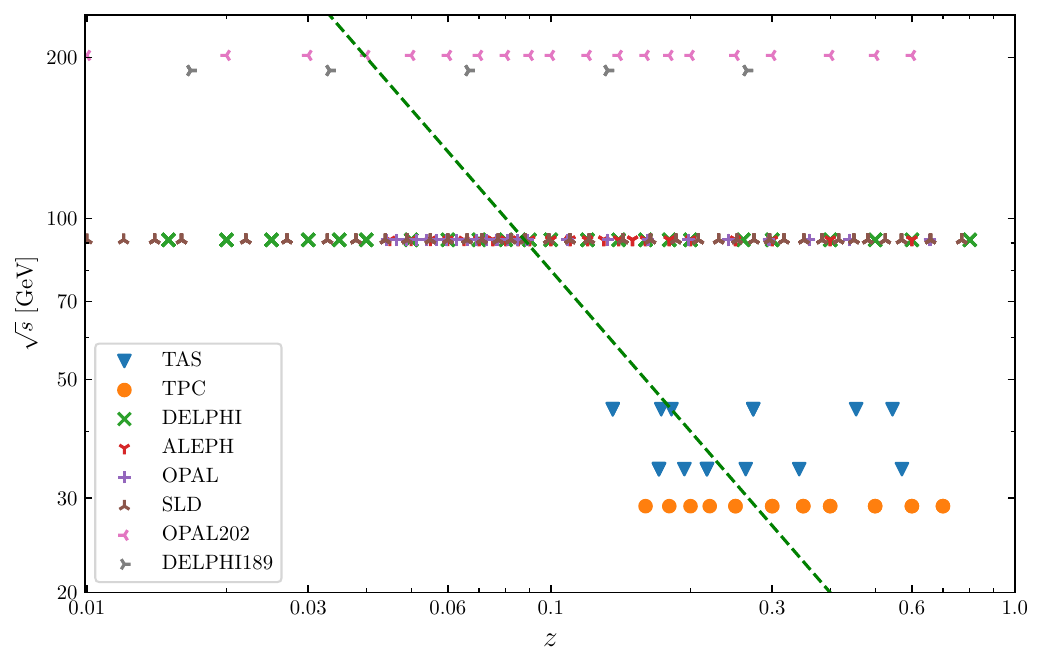}
	\caption{
     $z$-$\sqrt{s}$ coverage of the SIA datasets. 
     The green dashed line corresponds to the kinematic cut $E_h>4$ GeV.
	}
  \label{Fig:z-Q-SIA}
\end{figure}

We have not included SIA measurements with energies lower than 34~GeV for which the theory predictions may receive large power corrections, see e.g. \cite{Li:2024etc}.  
The $z$-$\sqrt{s}$ coverage of all the SIA datasets is plot in Fig.~\ref{Fig:z-Q-SIA}. 
The green dashed line corresponds to the kinematic cut $E_h>4$ GeV.
A large number of data points from SIA have been excluded in our current analysis.
A posterior comparison shows good agreement between theory predictions and the data down to $z\sim 0.05$, suggesting the possibility of relaxing our kinematic cuts for SIA measurements in future analyses.

\section{Theoretical inputs to NPC23}
\label{sec:theory}

\subsection{Parametrization form}

The parameterization form of fragmentation functions of parton $i$ to hadron $h$ used at the initial scale $Q_0$ is given by 
\begin{equation}\label{eq:para}
z D^{h}_{i}\left(z, Q_{0}\right)=z^{\alpha^h_i}(1-z)^{\beta^h_i} \exp \left(\sum_{n=0}^m a^h_{i,n}(\sqrt {z})^{n}\right),
\end{equation}
where $\{\alpha,~\beta,~a_{n}\}$ are free parameters in the fit.
We choose $Q_0=5$ GeV and use a zero-mass scheme for heavy quarks with $n_f=5$.
One advantage of the above parametrization form is that the fragmentation functions are positively defined, thus no additional positivity constraints need to be applied.
The total number of free parameters is 63 for $\pi^+$, $K^+$, and $p$ combined.
The number of independent parameters for all the parton-to-hadron FFs is 
summarized in Tabs. \ref{tab:param-pion}, \ref{tab:param-kaon} and  \ref{tab:param-proton}.
We increase the degree of polynomials $m$ in Eq. (\ref{eq:para}) until no significant improvements of fit are observed, and the final values vary from 0 to 2 depending on the flavors of parton and hadron.
By charge conjugation, the FFs of negatively charged hadrons can be related to those of positively charged hadrons. 
To reduce the number of free parameters, we also assume partial flavor symmetries among FFs from favored (unfavored) light (anti-)quarks, indicated by $=$ and $\simeq$ in the tables.
Taking the parton-to-$\pi^+$ FFs (Tabs. \ref{tab:param-pion} ) as examples, we have assumed approximate (indicated by $\simeq$) flavor symmetries among favored (anti)quarks, i.e., the $u, \bar{d}$ quarks. 
They share the same $\alpha,\beta,a_1,a_2$ parameters (indicated by  ``$-$'' ) at the initial scale $Q_0=5~{\rm GeV}$, but are allowed to have different overall factors $a_0$. 
This proportion relation will be violated by QCD evolution.
We have also assumed exact flavor symmetries (indicated by $=$) between two of the unfavored (anti-)quarks $\bar{u}, d$, which is protected under QCD evolution.
$D_s^{\pi^+} (z, Q_0)$ and $D_{\bar{u}}^{\pi^+} (z, Q_0)$, however, only share the same $\alpha$ parameter at the starting scale. 
The $\beta$ parameter for the gluon-to-charged hadron FFs is fixed to optimal values as determined from the fit in estimation of uncertainties of FFs, due to their strong correlation with other parameters.
The number of independent fit parameters for each parton-to-$\pi^+$ FF is summarized in the last column. 
There are 25 d.o.f. for the parton-to-$\pi^+$ FFs in total in our final fit.

\begin{table}[h]  \begin{tabular}{|c|c|c|c|c|c|c|c|}
  \hline
  parton-to-$\pi^+$ & favored & $\alpha$ & $\beta$ & $a_0$ & $a_1$ & $a_2$ &
  d.o.f.\\
  \hline
  $u$ & Y &  &  &  &  &  & 5\\
  \hline
  $\bar{d} \simeq u$ & Y & - & - &  & - & - & 1\\
  \hline
  $\bar{u} = d$ & N &  &  &  &  & x & 4\\
  \hline
  $s = \bar{s} \simeq \bar{u}$ & N & - &  &  &  & x & 3\\
  \hline
  $c = \bar{c}$ & N &  &  &  &  & x & 4\\
  \hline
  $b = \bar{b}$ & N &  &  &  &  & x & 4\\
  \hline
  $g$ & N &  & F &  &  &  & 4\\
  \hline
\end{tabular}
  \caption{Non-zero parameters for the parton-to-$\pi^+$ FFs. 
  Approximate (indicated by $\simeq$) or exact (indicated by $=$) flavor symmetries among
  favored (anti)quarks ($u, \bar{d}$) or unfavored light (anti)quarks
  ($\bar{u}, d, s, \bar{s}$) are assumed. 
 ``$-$'' indicates parameters fixed by the approximate flavor symmetry.  
  ``x'' corresponds to vanishing parameters, whose presence does not significantly improve the fit quality.
  The $\beta$ parameter for the gluon-to-$\pi^+$ FF is non-zero but is fixed
  during the fit. 
 All the other parameters are free and independent of each other.
  The number of independent fit parameters for each  parton-to-$\pi^+$ FF is summarized in the last column. 
  }
  \label{tab:param-pion}
\end{table}

For the parton-to-$K^+$ FFs,  we have assumed approximate flavor symmetry among favored (anti-)quarks $\bar{s}$ and $u$, and exact flavor symmetries among unfavored light (anti-)quark, and unfavored heavy quark, respectively.
Adding more parameters in the sector of unfavored quarks brings little improvement on fit quality due to both their small contributions and the less sensitivity of data to flavors of unfavored quarks.  
In our final fit, there are a total of 20 independent parameters for parton-to-$K^+$ FFs.

\begin{table}[h]  
  \begin{tabular}{|c|c|c|c|c|c|c|c|}
    \hline
    parton-to-$K^+$ & favored & $\alpha$ & $\beta$ & $a_0$ & $a_1$ & $a_2$ & d.o.f.\\
    \hline
    $u$ & Y &  &  &  &  & x & 4\\
    \hline
    $\bar{s} \simeq u$ & Y & - & - &  & - & x & 1\\
    \hline
    $\bar{u} \!=\! d \!=\! \bar{d} \!=\! s$ & N &  &  &  &  & x & 4\\
    \hline
    $c = \bar{c}$ & N &  &  &  &  & x & 4\\
    \hline
    $b = \bar{b}$ & N &  &  &  &  & x & 4\\
    \hline
    $g$ & N &  & F & &  & x & 3\\
    \hline
  \end{tabular}
  \caption{Similar to Tab.~\ref{tab:param-pion}, but for the parton-to-$K^+$ FFs. 
  There are 20 d.o.f. in total.}
  \label{tab:param-kaon}
\end{table}
For the parton-to-proton FFs,  we have assumed exact flavor symmetry among the two valence quarks 
$D_u^p (z, Q_0) = 2 D_d^p (z, Q_0)$ at the starting scale $Q_0$.
We have also assumed exact flavor symmetries among unfavored light (anti-)quark, and unfavored heavy quark, respectively. 
Relaxing flavor symmetry between valence quarks brings in little improvement on fit quality, but makes the fit unstable.
That indicates a poor constraint on the separation of quark flavors in proton FFs.  
In our final fit, there are a total of 18 independent parameters for parton-to-$K^+$ FFs.

\begin{table}[h]  
\begin{tabular}{|c|c|c|c|c|c|c|c|}
  \hline
  parton-to-$p$ & favored & $\alpha$ & $\beta$ & $a_0$ & $a_1$ & $a_2$ & d.o.f.\\
  \hline
  $u = 2 d$ & Y &  &  &  &  & x & 4\\
  \hline
  $\bar{u} = \bar{d} = s = \bar{s}$ & N &  &  &  & x & x & 3\\
  \hline
  $c = \bar{c}$ & N &  &  &  &  & x & 4\\
  \hline
  $b = \bar{b}$ & N &  &  &  &  & x & 4\\
  \hline
  $g$ & N &  & F &  &  & x & 3\\
  \hline
\end{tabular}
\caption{Similar to Tab.~\ref{tab:param-pion}, but for the parton-to-$p$ FFs, 
from which the parton-to-$\bar{p}$ FFs can be determined by charge symmetry. 
We have assumed $D_u^p (z, Q_0) = 2 D_d^p (z, Q_0)$ at the starting scale $Q_0$.
There are 18 d.o.f. in total.
}
\label{tab:param-proton}
\end{table}

\subsection{Theoretical computations}

The fragmentation functions are evolved to higher scales using two-loop
time-like splitting kernels to be consistent with the NLO analysis.
The splitting functions were calculated in Refs.~\cite{Stratmann:1996hn} and are
implemented in HOPPET~\cite{Salam:2008qg, Salam:2008sz:v3}.
The numerical results of QCD evolution are also compared with APFEL~\cite{Bertone:2013vaa} and found agreement.
The hard coefficient functions for SIA and hadron collisions are calculated with the FMNLO program as detailed in Ref.~\cite{Liu:2023fsq}.
The FMNLO computation is based on a hybrid scheme of phase-space slicing method and local subtraction method, 
and accurate to NLO in QCD.
It has been interfaced to MG5\_aMC@NLO~\cite{Alwall:2014hca,Frederix:2018nkq} and made publicly available~\cite{Liu:2023fsq}. 
The hard coefficient functions for SIDIS are calculated with an updated version of the FMNLO program, which is explained in appendix \ref{sec:fmnlo} and also publicly available. 
We adopt a zero-mass scheme for heavy quarks consistently with $n_f=5$ and $\alpha_S(M_Z)=0.118$ through all calculations.
For theoretical predictions of hadron production at SIA with heavy flavor tagged events, we only include contributions from Feynman diagrams with the specified heavy quark coupled directly to the $Z$ boson or photon, which is well justified at NLO.
There are ambiguities on matching theoretical predictions to the experimental measurements when going beyond NLO, e.g., on treatment of contributions from gluon splitting into heavy quarks.
Furthermore, the FMNLO program can generate and store interpolation tables of the coefficient functions, thus ensuring fast convolution with arbitrary FFs without repeating the calculations.
This approach facilitates efficient exploration of the parameter space of high-dimensionality across numerous iterations. 
We use the CT14 NLO parton distribution functions~\cite{Dulat:2015mca} with $\alpha_S(M_Z)=0.118$ for calculations involving initial hadrons.
The central values for the renormalization and fragmentation scales ($\mu_{R,0}$ and $\mu_{D,0}$) are set to the momentum transfer $Q$ for both SIA and SIDIS.
The factorization scale ($\mu_{F,0}$) of initial hadrons for SIDIS is also set to $Q$.
In the case of hadron collisions, the central values for the factorization scale and renormalization scales are set to half the sum of the transverse mass of all final state particles.
The central value for the fragmentation scale is determined as the maximum of the transverse momentum of all final state particles for inclusive hadron production and as the transverse momentum of the jet multiplied by the jet cone size for hadron fragmentation inside the jet~\cite{Kaufmann:2015hma}.
The covariance matrix of $\chi^2$ calculations incorporates theoretical uncertainties, assumed to be fully correlated among points within each subset of the data.
These theoretical uncertainties are estimated by the half width of the envelope of theoretical predictions based on 9 scale combinations: $\mu_F/\mu_{F,0}=\mu_R/\mu_{R,0}=\{1/2,~1,~2\}$ and $\mu_D/\mu_{D,0}=\{1/2,~1,~2\}$.
It is noteworthy that the impact of different choices of the nominal scales is minimal, given the inclusion of theoretical uncertainties.

\subsection{Goodness of fit function and the covariance matrix}
The agreement between the data points $D_k$ and the corresponding theoretical predictions $T_k$ 
 is quantified by the log-likelihood function ~\cite{Pumplin:2002vw}:
\begin{equation}
\chi^2 (\{a\}, \{\lambda\}) = 
\sum_{k = 1}^{N_{\rm{pt}}} \frac{1}{s_k^2}  
\left( D_k - T_k - \sum_{\alpha = 1}^{N_{\lambda}} \beta_{k, \alpha} \lambda_{\alpha} \right)^2 + \sum_{\alpha = 1}^{N_{\lambda}} \lambda_{\alpha}^2,
\end{equation}
where $\{a\}$ are FF parameters, 
the nuisance parameters $\{\lambda\}$ describe sources of correlated errors, which are assumed to follow standard normal distributions,
$s_k$ represents the total uncorrelated systematic and statistical errors, 
and $\beta_{k,\alpha}$ quantifies the sensitivity of the $k$-th measurement to the $\alpha$-th correlated error source.
In our case, the correlated errors can be either the normalization uncertainties from the measurements or the theoretical uncertainties estimated with scale variations.
Minimizing the log-likelihood function $\chi^2 (\{a\}, \{\lambda\})$ with respect to the nuisance parameters leads to the profiled $\chi^2$ function:
\begin{equation}
\chi^2 (\{a\}) \equiv \chi^2 (\{ a \}, \{ \lambda =\hat{\lambda} \}) 
= \sum_{i, j = 1}^{N_{\rm{pt}}}
(T_i - D_i) (\mathrm{cov}^{- 1})_{ij}  (T_j - D_j),
\end{equation}
where $\{\hat{\lambda}\}$ are the best-fit nuisance parameters, 
$\rm{cov}^{-1}$ is the inverse of the covariance matrix: 
\begin{equation}
(\mathrm{cov})_{ij} \equiv s_i^2 \delta_{ij}
+ \sum_{\alpha = 1}^{N_{\lambda}} \beta_{i, \alpha} \beta_{j, \alpha}, 
\quad 
(\mathrm{cov}^{-1})_{ij} = \frac{\delta_{ij}}{s_i^2} 
- \sum_{\alpha, \beta = 1}^{N_{\lambda}}
\frac{\beta_{i, \alpha}}{s_i^2} A_{\alpha \beta}^{- 1}  \frac{\beta_{j,\beta}}{s_j^2}~.
\end{equation}
Here the $A_{\alpha\beta}$ is defined as
\begin{eqnarray}
  A_{\alpha \beta} & = & \delta_{\alpha \beta} + \sum_{k = 1}^{N_{\rm{pt}}}
  \frac{\beta_{k, \alpha} \beta_{k, \beta}}{s_k^2}.
\end{eqnarray}
To avoid the  D'Agostini bias \cite{Ball:2009qv}, which can arise from including multiplicative systematic errors $\beta_{i,\alpha}\equiv\sigma_{i, \alpha} D_i$
in the covariance matrix, we adopt the `$t$' definition of the covariance matrix~\cite{Ball:2012wy}.
This means that the correlated systematic errors are calculated as $\sigma_{i, \alpha} T_i$ instead of $\sigma_{i, \alpha} D_i$.
The best-fit fragmentation parameters are determined by minimizing the $\chi^2$ and then further validated through a series of profile scans on each of those parameters.
These parameter space scans are conducted using the MINUIT~\cite{minuit} program.
We apply a tolerance criterion of $\Delta\chi^2\sim 2$ to determine parameter uncertainties, as will be explained later.
Additionally, we employ the iterative Hessian approach~\cite{hessian} to generate error sets of
fragmentation functions, which can be used to propagate parameter uncertainties to physical observables.

\section{The NPC23 output: FFs, moments}
\label{sec:results}
\subsection{NPC23 FFs as functions of $z$ and $Q$}

\begin{figure}[htbp]
\centering
  \includegraphics[width=0.8\textwidth]{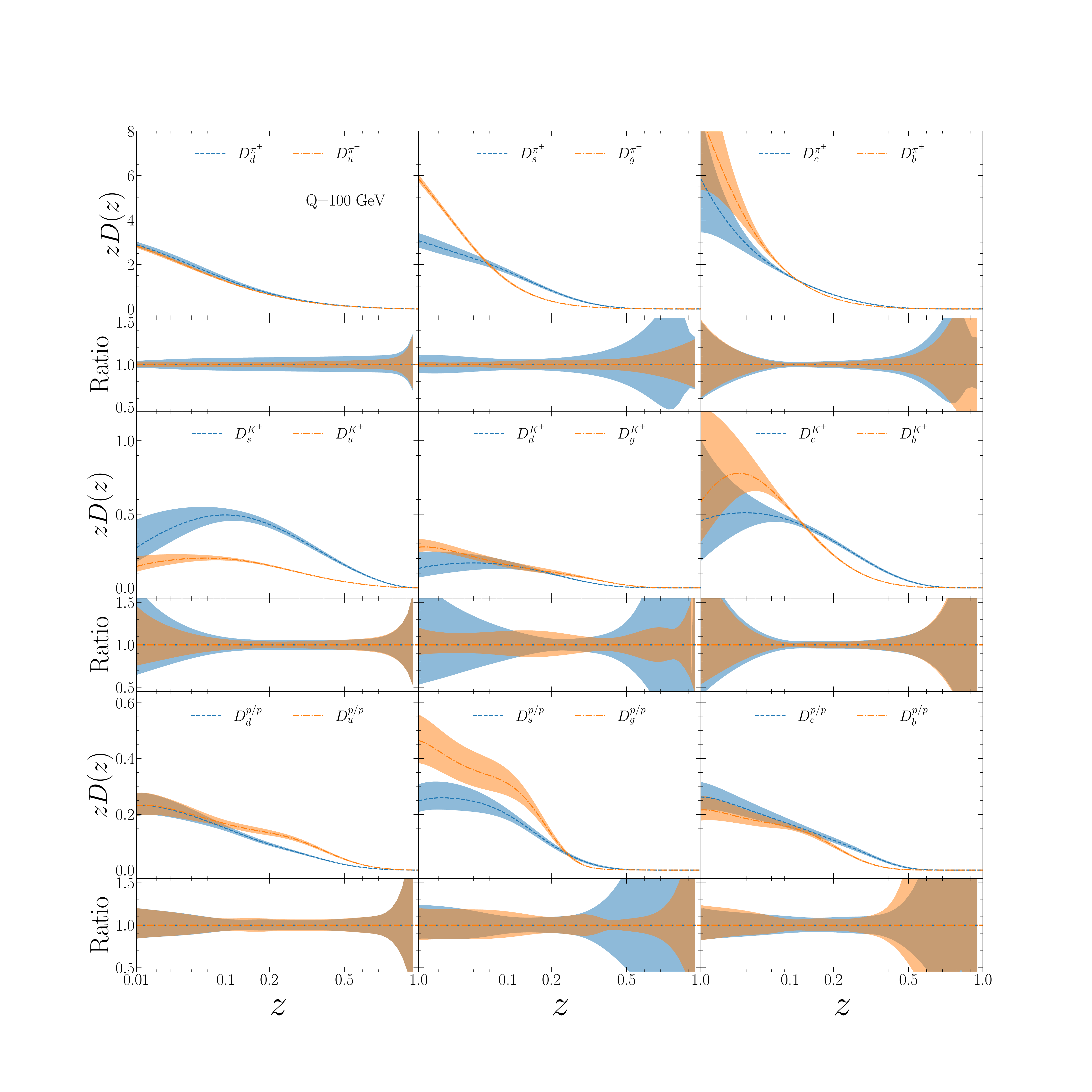}
	\caption{
 Similar to Fig. \ref{Fig:ff42_5GeV} but for FFs at 100 GeV.
	}
  \label{Fig:ff42_100GeV}
\end{figure}
\begin{figure}[htbp]
\centering
  \includegraphics[width=0.8\textwidth]{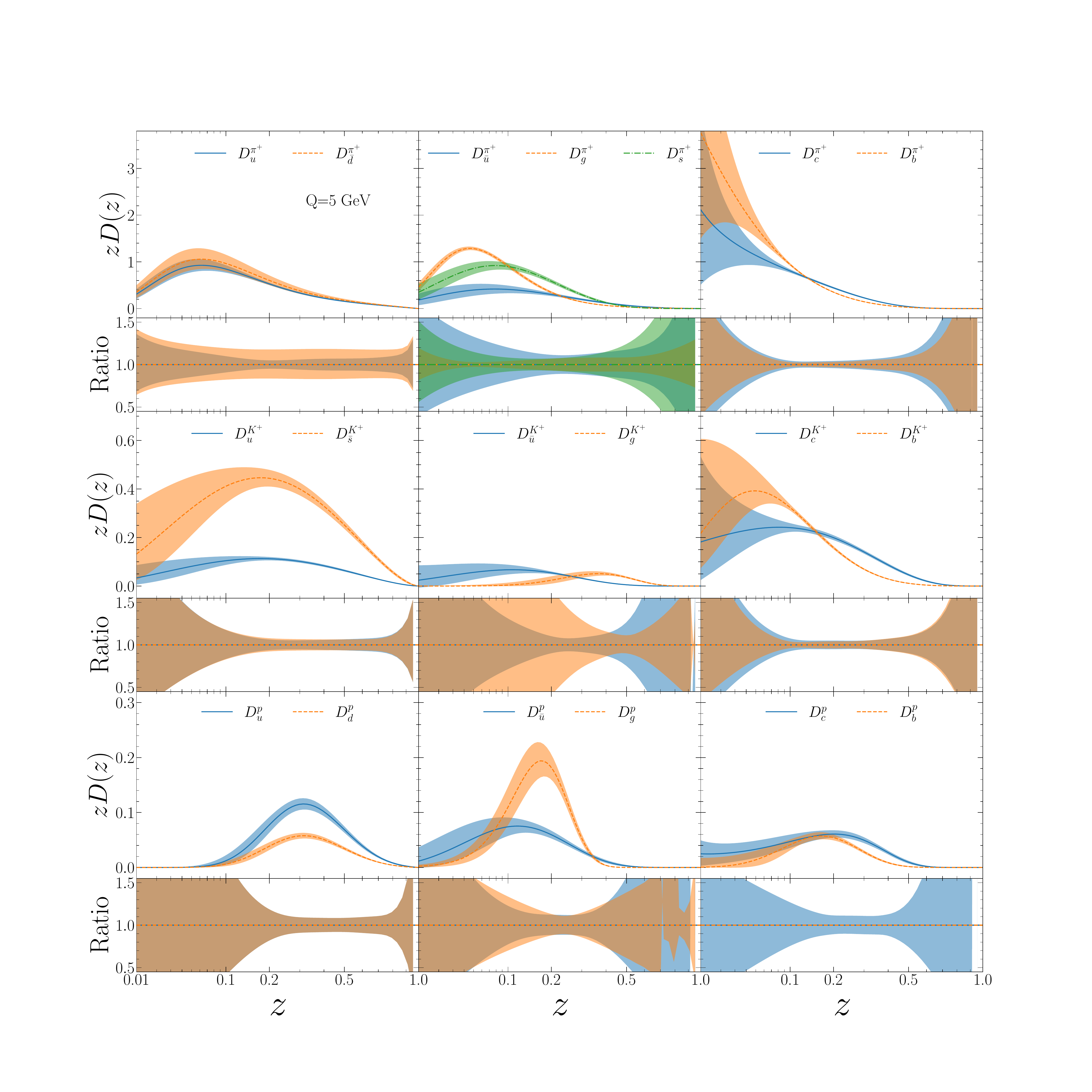}
	\caption{
Similar to Fig. \ref{Fig:ff42_5GeV} but for FFs to $\pi^+,~K^+,~p$ at 5 GeV.
	}
  \label{Fig:ff42_plus_5GeV}
\end{figure}

In this section, we provide further details of our FFs to light charged hadrons.
Fig.~\ref{Fig:ff42_100GeV} presents the FFs in a similar way as Fig.~\ref{Fig:ff42_5GeV} but at a scale of $Q=100$ GeV.
The differences thus reflect effects from DGLAP evolution on mixing of FFs of different parton flavors and redistribution of hadron momenta in different kinematic regions.
The FFs grow quickly at small-$z$ region, especially for pions and protons.
Various peaks tend into plateaus due to smearing effects from QCD evolution. 
For FFs summed over positively and negatively charged hadrons, contributions from favored and unfavored quark flavors are mixed. 
We further show FFs to positively charged hadrons only at $Q=5$ GeV in Fig.~\ref{Fig:ff42_plus_5GeV} to separate their contributions. 
FFs from gluon and heavy quarks are simply half of those shown earlier in Fig.~\ref{Fig:ff42_5GeV}, since we assume they fragment equally to positively and negatively charged hadrons. 
Relative uncertainties increase for FFs from $u$ and $\bar d$ quarks to $\pi^+$ compared to $\pi^{\pm}$, as most experimental data only constrain the sum of production rates of positively and negatively charged hadrons. 
They are only sensitive to FFs summed over quark and anti-quark. 
It is even more evident that the strange quark shows a much larger FF to $\pi^+$ compared to those from $\bar u$ and $d$ quarks by looking at the second plot in the top panel. 
We have performed alternative fits by enforcing SU(3) flavor symmetry in FFs of unfavored quarks to $\pi^+$ rather than what has been done in our nominal fits.
The extracted FFs from strange quark to pions do decrease, but with a penalty of an increase of the total $\chi^2$ by a few tens of units mostly from the SIA data.  

\subsection{Moments and sum rules of FFs}

FFs represent the number densities of hadrons and satisfy various fundamental sum rules derived from first principles, including the momentum sum rule and the charge sum rule. 
Testing the momentum sum rule based on our determination of FFs is particularly desirable due to the suppression of small-$z$ contributions.
However, the momentum sum rule also involves FFs to neutral hadrons, which are not available in this analysis.
Therefore, we focus solely on the momentum sum carried by light charged hadrons below.
Another interesting quantity is the jet charge, which equals moments of the difference of FFs to positively and negatively charged hadrons at LO in QCD. 
For instance, from the flavor dependence of FFs, we expect jets initiated by $u$ ($d$) quarks to have a positive (negative) jet charge. 

\subsubsection{Momentum sum}

The key quantity we calculate is the total momentum carried by a specific hadron or a class of hadrons for various flavors of partons using the following expression:
\begin{equation}
\langle z\rangle_i^h =\int_{z_{min}}^{1}dz zD_i^h(z,Q).
\end{equation}
The non-zero lower limit of integration $z_{min}$ is introduced since experimental data only cover a finite kinematic region.
The data are only sensitive to FFs with $z$ above $0.01$ or even larger, depending on the flavor of the parton and the hadrons.
Varying the lower limit can be used to test the convergence of the momentum sum, although the extrapolation to $z_{min}=0$ can be sensitive to the choice of parametrization forms.
The results of $\langle z\rangle_i^h$ for light quarks, gluon, and heavy quarks are shown in Table~\ref{Tab:mom}, where the central values and uncertainties are calculated from our best-fit and Hessian error FFs.
We choose the lower limit $z_{min}$ to be $0.01$ for $g$, $u$, and $d$ quarks, and $0.088$ for $s$, $c$, and $b$ quarks, based on the kinematic coverage of relevant data.
It is observed that the three charged hadrons carry approximately 50\% to 53\% of the momentum of $u$, $d$ quarks and gluon.
The uncertainties for $d$ quarks are more than twice as large as those for $u$ quarks and gluon.
The total momentum of the strange quark carried away by light charged hadrons is 57\%, even with a much higher $z_{min}$.
As mentioned earlier, one possible reason is that part of the momentum carried by short-lived neutral hadrons is also included in the SIA measurements due to the prompt decay of those hadrons, especially for $K^0_S$ decaying into $\pi^{\pm}$.
\begin{table}[]
  \centering
\begin{tabular}{|c|c|c|c|c|c|c|}
\hline
 \textbf{mom.} & $g(z>0.01)$  & $u(z>0.01)$ & $d(z>0.01)$ & $s(z>0.088)$ & $c(z>0.088)$ & $b(z>0.088)$ \\ \hline
\textbf{$\pi^+$} 
 &$0.200	^{+0.008}_{-0.008}$&$0.262^{+0.017}_{-0.016}$&$0.128^{+0.020	}_{-0.019}$&$0.161^{+0.013}_{-0.013}$&$0.130^{+0.005}_{-0.005}$&$0.111^{+0.003}_{-0.003}$\\
\textbf{$K^+$} 
 &$0.018	^{+0.004}_{-0.003}$&$0.058^{+0.005}_{-0.004}$&$0.019^{+0.004	}_{-0.004}$&$0.015^{+0.002}_{-0.002}$&$0.065^{+0.003}_{-0.003}$&$0.046^{+0.002}_{-0.002}$\\
\textbf{$p$} 
 &$0.035	^{+0.006}_{-0.005}$&$0.044^{+0.004}_{-0.004}$&$0.022^{+0.002	}_{-0.002}$&$0.015^{+0.002}_{-0.002}$&$0.018^{+0.002}_{-0.002}$&$0.012^{+0.001}_{-0.001}$\\ \hline
\textbf{$\pi^-$}  
 &$0.200	^{+0.008}_{-0.008}$&$0.128^{+0.020 }_{-0.019}$&$0.299^{+0.054	}_{-0.049	}$&$0.161^{+0.013}_{-0.013}$&$0.130^{+0.005}_{-0.005}$&$0.111^{+0.003}_{-0.003}$\\
\textbf{$K^-$} 
 &$0.018	^{+0.004}_{-0.003}$&$0.019^{+0.004}_{-0.004}$&$0.019^{+0.004	}_{-0.004}$&$0.205^{+0.014	}_{-0.013}$&$0.065^{+0.003}_{-0.003}$&$0.046^{+0.002}_{-0.002}$\\
\textbf{$\bar p$} 
 &$0.035	^{+0.006}_{-0.005}$&$0.019^{+0.003}_{-0.003}$&$0.019^{+0.003	}_{-0.003}$&$0.015^{+0.002}_{-0.002}$&$0.018^{+0.002}_{-0.002}$&$0.012^{+0.001}_{-0.001}$\\ \hline
\textbf{Sum} 
 &$0.507	^{+0.014}_{-0.013}$&$0.531^{+0.015}_{-0.013}$&$0.506^{+0.042	}_{-0.037}$&$0.572^{+0.029}_{-0.028}$&$0.425^{+0.013}_{-0.012}$&$0.338^{+0.007}_{-0.007}$\\ \hline
\end{tabular}
	\caption{
	Total momentum of the partons, including $g$, $u$, $d$, $s$, $c$ and $b$ quarks, carried by various charged hadrons ($\pi^{\pm}$, $K^{\pm}$, $p$ and $\bar p$) in the fragmentations.
	The central values and uncertainties at 68\% C.L. are calculated from our best-fit and Hessian error FFs at $Q=5$~GeV. 
	The last row is the sum over all light charged hadrons.
	}
  \label{Tab:mom}
\end{table}

\begin{figure}[htbp]
\centering
  \includegraphics[width=0.43\textwidth]{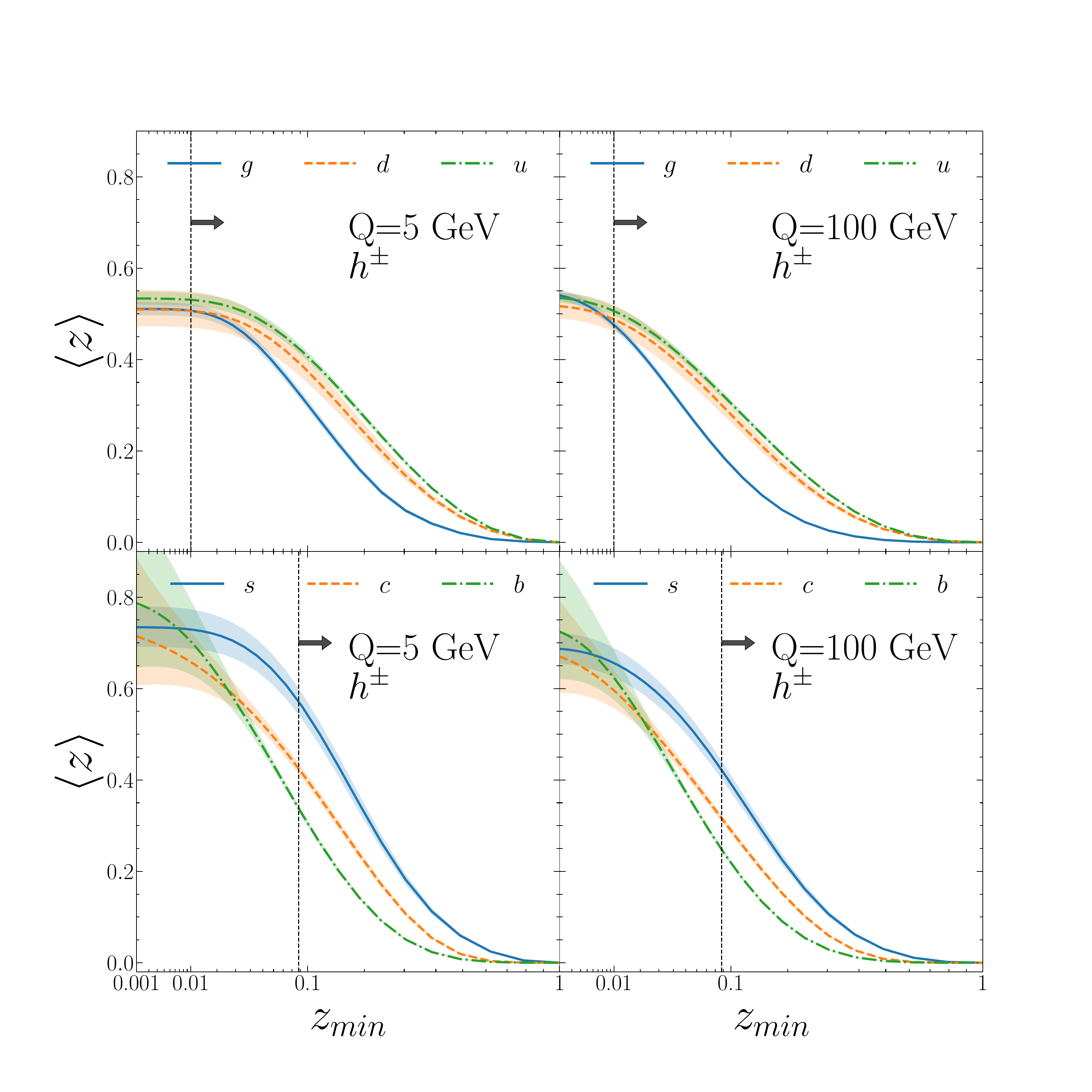}
  \hspace{0.2in}
  \includegraphics[width=0.43\textwidth]{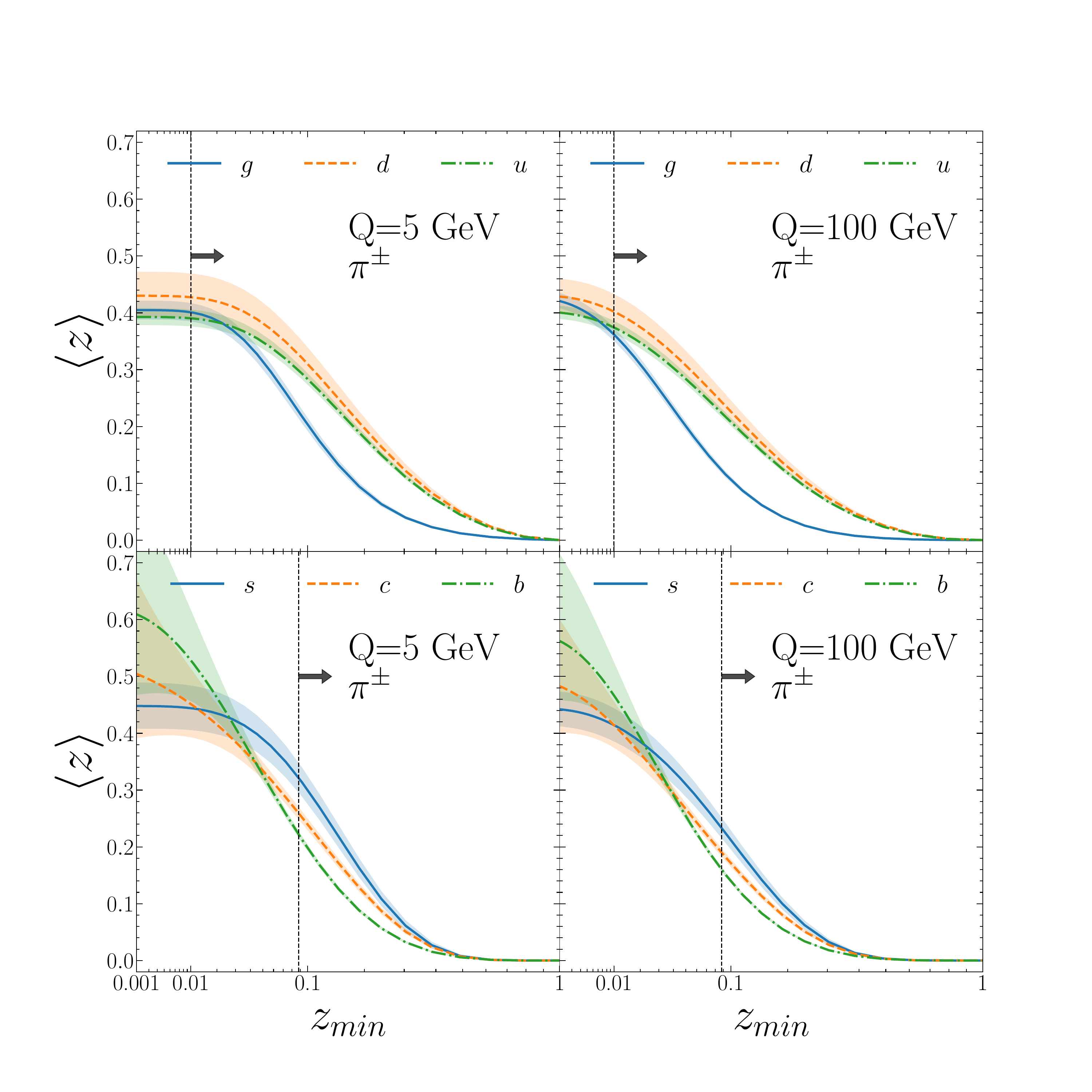}
	\caption{
 Average momentum fraction carried by charged hadrons and $\pi^\pm$ fragmented from various partons including $u, d, s, c, b, g$ as a function of $z_{min}$.
 The central values and uncertainties at 68\% C.L. are calculated from our best-fit and Hessian error FFs at $Q=5$~GeV and 100~GeV. 
 The vertical lines indicate the kinematic coverage of relevant data of constraints.
 }
  \label{Fig:ff42_moment_1}
\end{figure}

\begin{figure}[htbp]
\centering  
\includegraphics[width=0.43\textwidth]{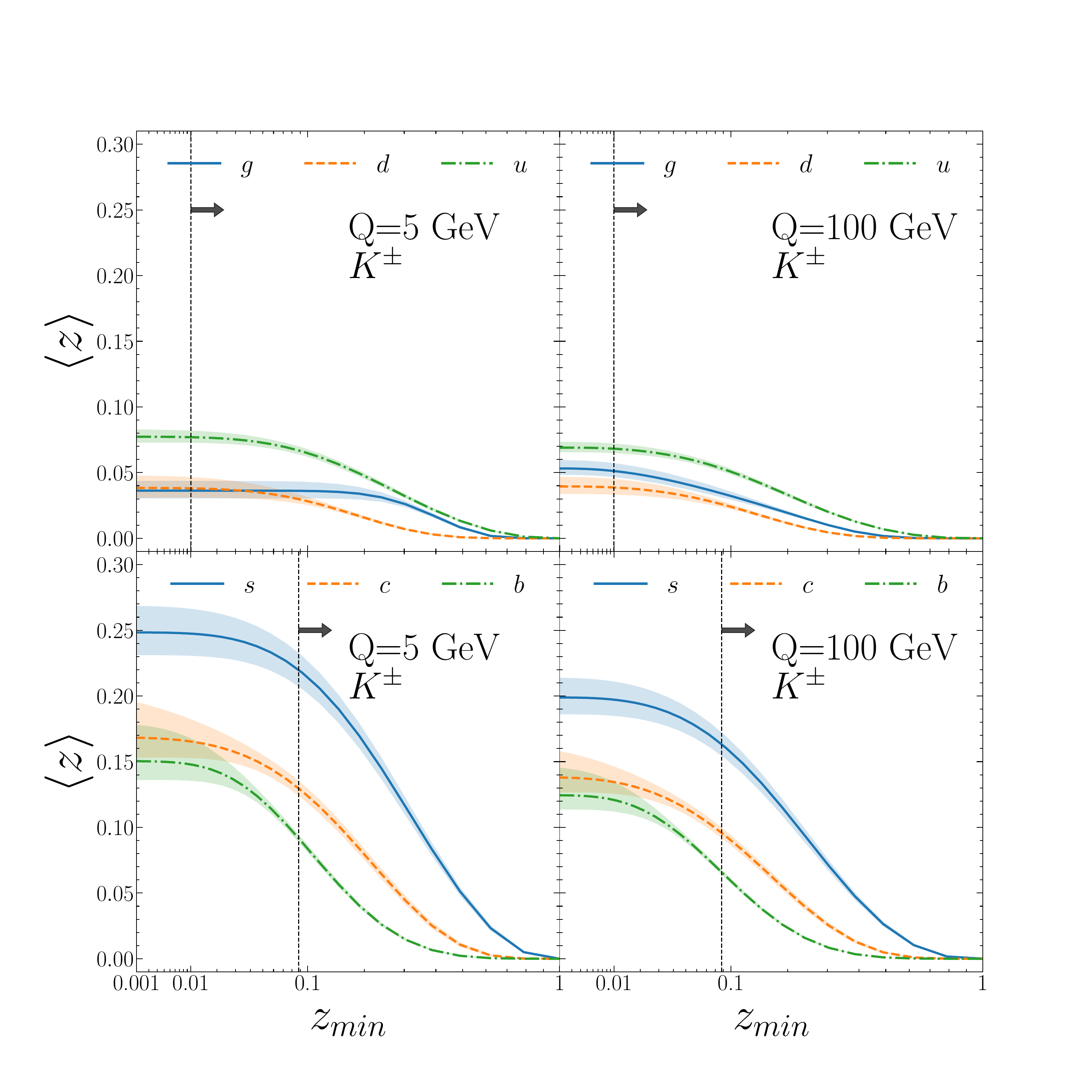}\hspace{0.2in}
  \includegraphics[width=0.43\textwidth]{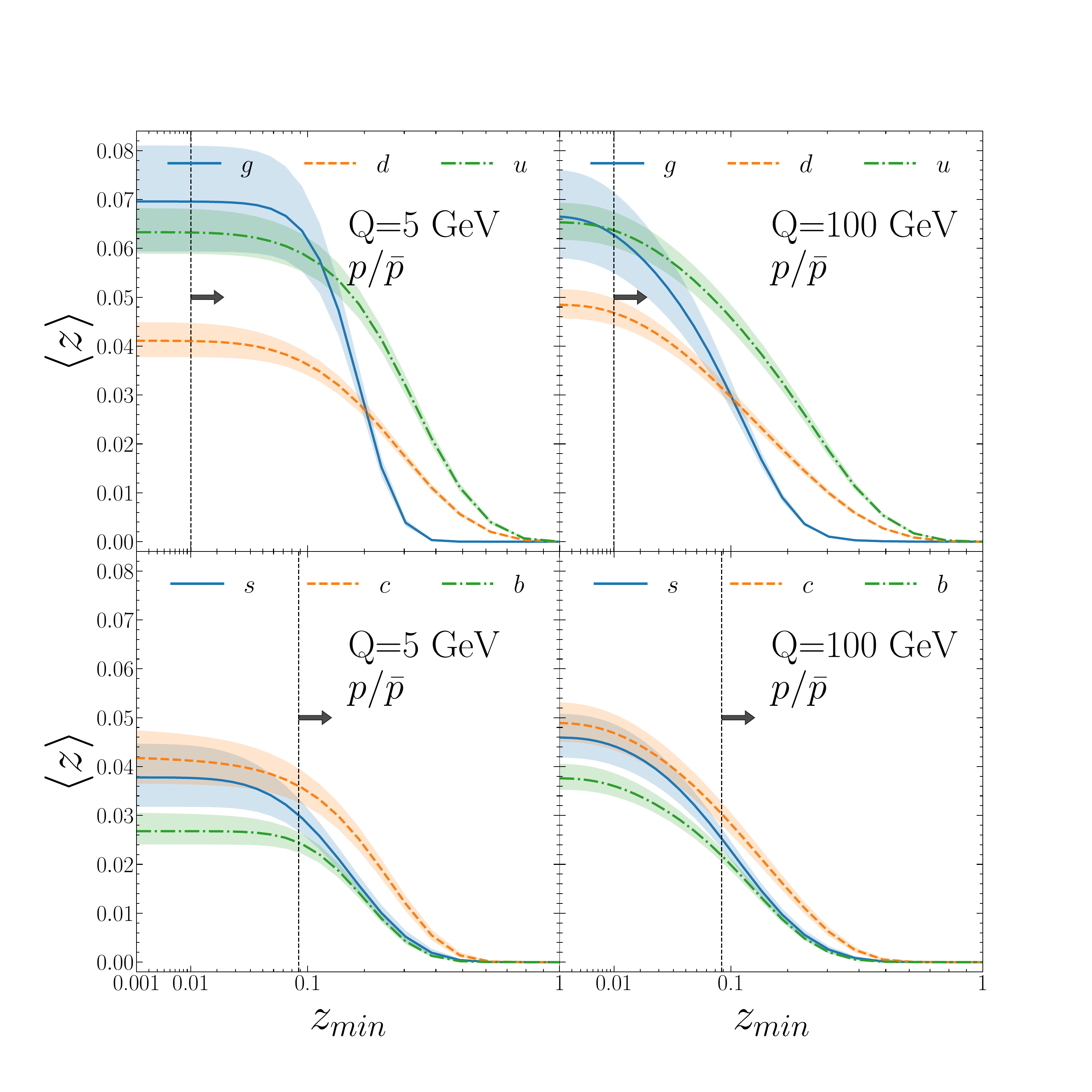}
	\caption{
  Same as \ref{Fig:ff42_moment_1} but for $K^\pm$ and $p/\bar p$.
	}
  \label{Fig:ff42_moment_2}
\end{figure}

We further show $\langle z\rangle_i^h$ as functions of $z_{min}$ in Fig.~\ref{Fig:ff42_moment_1} for the sum of light charged hadrons and $\pi^{\pm}$ at $Q=5$ and $100$ GeV, respectively.
The total momentum already saturates at $z_{min}=0.01$ for gluon and $u$, $d$ quarks at $Q=5$ GeV.
For $Q=100$ GeV, the FFs have been pushed toward small-$z$ region, and the total momentum stabilizes at a much smaller $z_{min}$.
The pion contributions are always dominant in the sum of light charged hadrons.
The total momentum continues to rise rapidly for $s$, $c$, and $b$ quarks around $z_{min}=0.088$.
They also show large uncertainties when extrapolating to even smaller $z_{min}$ values. 
We present similar results for $K^{\pm}$ and $p/\bar p$ in Fig.~\ref{Fig:ff42_moment_2}. 
The total momentum saturates much earlier in this case at about $z_{min}=0.1$ for gluon and $u$, $d$ quarks.
Also, they rise slower at small $z_{min}$ for $s$, $c$ and $b$ quarks compared to the case of pions.
The total momentum of $u$ and $d$ quarks carried by kaons and protons is similar, while the momentum carried by kaons is much larger for $s$ and heavy quarks.
A test on the momentum sum rule based on the results of charged hadrons presented here and scaling to include neutral hadrons can be found in Ref.~\cite{Gao:2024nkz}.


\subsubsection{Jet charge}

\begin{figure}[htbp]
  \centering
  \includegraphics[width=0.48\textwidth]{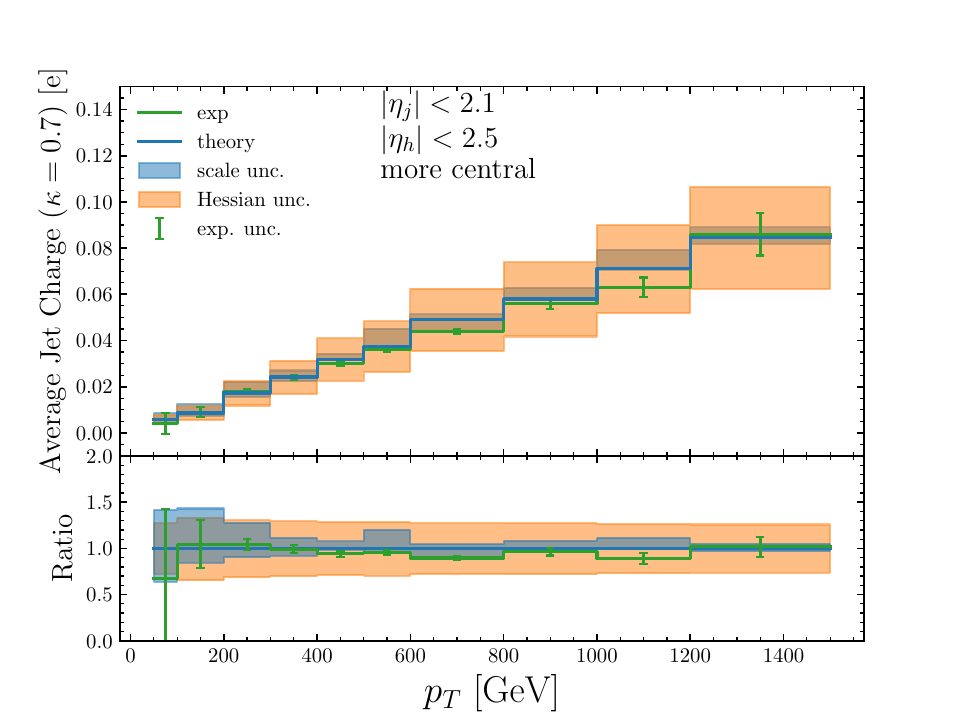}
  \includegraphics[width=0.48\textwidth]{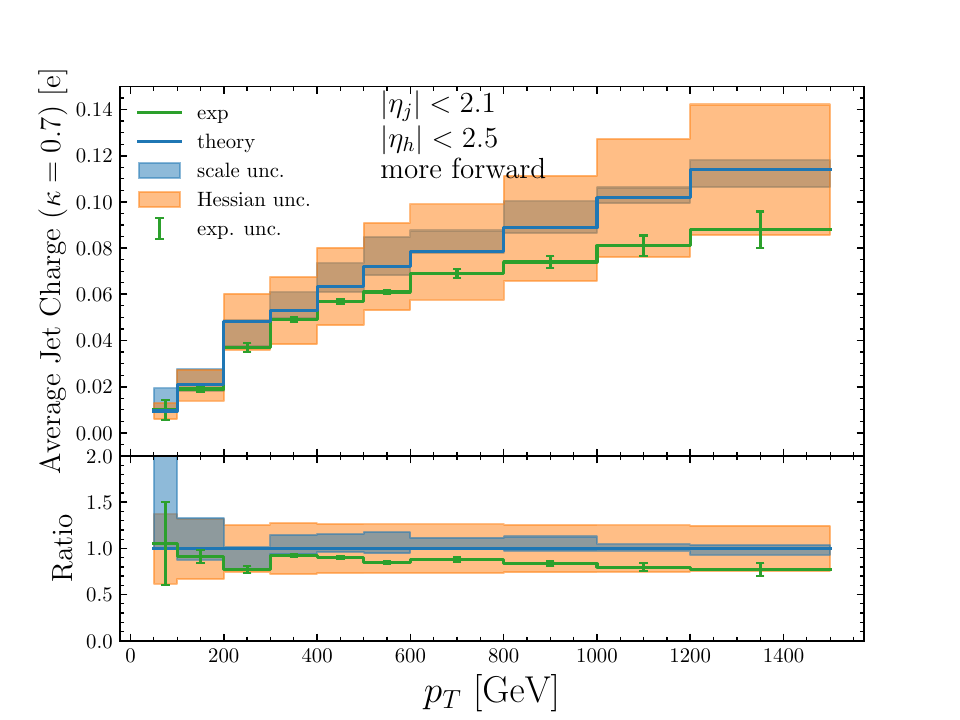}
	\caption{
 NLO predictions on average jet charge in QCD dijet production at the LHC 8 TeV, for both the more central and more forward jets, as functions of jet $p_T$, compared to the ATLAS measurements.
 The error bands represent the scale variations and Hessian uncertainties of FFs, respectively.
	}
  \label{Fig:jet_ch}
\end{figure}

Definition of jet charge begins with a clustered jet, for instance, with the anti-$k_T$ algorithm.
It involves counting the total electric charge carried by constituent hadrons inside the jet in units of the charge of the positron. 
The electric charge of the hadrons is weighted by a positive power $\kappa$ of the transverse momentum of the hadron to suppress contributions from soft particles.
Therefore, the jet charge $Q_J$ can be expressed as~\cite{Field:1977fa} 
\begin{equation}
Q_J=\sum_{i\in J}\left({p_{T,i}\over p_{T,J}}\right)^{\kappa}Q_i,
\end{equation}
where $i$ sums over all charged tracks in the jet $J$, 
$Q_i$ is the charge of particle $i$ in units of the positron charge, 
$p_{T,i}$ and $p_{T,J}$ denote the transverse momenta of the charged track and the jet, respectively.
$\kappa$ is a positive regularization parameter.
The mean value of $Q_J$ over a large sample of jets can be related to the differential cross sections of hadron production inside the jet, namely
\begin{equation}
\langle Q_J\rangle=\int_{z_0}^1dz_h z_h^{\kappa}{1\over \sigma_J}\left({d\sigma_{h^+}\over dz_h}-{d\sigma_{h^-}\over dz_h}\right),
\end{equation}
where $\sigma_{h^+}$, $\sigma_{h^-}$ and $\sigma_J$ denote the cross sections of positively and negatively charged hadrons and the jet, respectively.
The lower limit $z_0$ is determined by the experimental threshold on hadron energy or transverse momentum.
The ATLAS collaboration at the LHC has conducted measurements on jet charge over QCD dijet samples at 8 TeV~\cite{ATLAS:2015rlw}. 
Jets are clustered using the anti-$k_T$ algorithm with a jet radius of 0.4 and are required to have a pseudo-rapidity $|\eta_j|<2.1$.
The associated hadrons are required to have transverse momentum $p_{T,h}>500$ MeV and $|\eta_h|<2.5$.
The selections on the two jets are the same as the 13 TeV measurement on jet fragmentation discussed earlier~\cite{ATLAS:2019rqw}.
The two jets are classified as more central jet or more forward jet according to absolute value of their rapidities.
We have calculated NLO predictions on the average jet charge and compared them to the ATLAS measurements in Fig.~\ref{Fig:jet_ch} for both the forward jet and the central jet as functions of jet $p_T$.
We used the CT14 NLO PDFs and chose a value of $\kappa=0.7$ for which the predictions are less sensitive to FFs at small $z$. 
The error bands in Fig.~\ref{Fig:jet_ch} represent the scale variations and Hessian uncertainties of FFs, respectively.
The average jet charges are positive, as for QCD jets production in $pp$ collisions, the jets are more likely to be from $u$ quarks than $d$ quarks.
They grow with jet $p_T$ since the gluon contributions become smaller at high-$p_T$. 
Our predictions from best-fit FFs agree well with the ATLAS measurements on the more central jet and are higher by 10\%$\sim$20\% compared to the ATLAS measurements on the more forward jet. 
In both cases, the Hessian uncertainties from FFs are about 30\% for all $p_T$ ranges considered, much larger than both the experimental uncertainties and the scale variations.
This suggests that current or future data from LHC measurements on jet charges can place further stringent constraints on FFs, which we leave for future investigations.   

\section{Quality of the fit to data}
\label{sec:fit}
\subsection{Overall agreement}

\begin{table}[]
  \centering
\begin{tabular}{|c|c|c|c|}
\hline
\textbf{Experiments} & \textbf{$N_{pt}$} & \textbf{$\chi^2$} & \textbf{$\chi^2/N_{pt}$} \\ \hline
ATLAS 5.02 TeV $\gamma+j$ & 6 & 9.6 & 1.61 \\
CMS 5.02 TeV $\gamma+j$ & 4 & 11.1 & 2.78 \\
ATLAS 5.02 TeV $Z+h$ & 9 & 22.2 & 2.47 \\
CMS 5.02 TeV $Z+h$ & 11 & 6.2 & 0.56 \\
LHCb 13 TeV $Z+j$ & 20 & 30.6 & 1.53 \\
ATLAS 5.02 TeV inc. jet & 63 & 67.9 & 1.08 \\
ATLAS 7 TeV inc. jet & 103 & 91.3 & 0.89 \\
ATLAS 13 TeV dijet & 280 & 191.6 & 0.68 \\ \hline
$pp$ hadron in jet sum             & 496	&430.5	&0.87\\ \hline
ALICE 13 TeV & 49 & 45.0 & 0.92 \\
ALICE 7 TeV & 37 & 36.3 & 0.98 \\
ALICE 5.02 TeV & 34 & 37.5 & 1.10 \\
ALICE 2.76 TeV & 27 & 31.8 & 1.18 \\
STAR 200 GeV & 60 & 42.2 & 0.70 \\ \hline
$pp$ inclusive sum             &207	&192.8	&0.93\\ \hline
H1 $^\dag$              &16	&12.5	&0.78\\      
H1 (asy.) $^\dag$        &14	&12.2	&0.87\\     
ZEUS $^\dag$                &32	&65.5	&2.05\\
COMPASS 06 ($D$)             &124	&107.3	&0.87\\ 
COMPASS 16  ($p$)          &97	&56.8	&0.59\\ \hline
SIDIS sum             &283	&254.4	&0.90\\ \hline
OPAL $Z\to q\bar q$                 &20	&16.3	&0.81\\
ALEPH  $Z\to q\bar q$              &42	&31.4	&0.75\\ 
DELPHI  $Z\to q\bar q$           & 39 & 12.5 & 0.32\\   
DELPHI  $Z\to b\bar b$          & 39 & 23.9 & 0.61\\   
SLD $Z\to q\bar q$                & 66 & 53.0 & 0.8\\  
SLD $Z\to b \bar b$                & 66 & 82.0 & 1.24\\  
SLD $Z\to c \bar c$                & 66 & 76.5 & 1.16\\ 
TASSO 34 GeV inc. had.              & 3 & 2.7 & 0.9\\
TASSO 44 GeV inc. had.              &5 & 4.3 & 0.86 \\
TPC  29 GeV inc. had.                  &12	&11.6	&0.97\\
OPAL (202 GeV) inc. had. $^\dag$       &17	&24.2	&1.42\\ 
DELPHI (189 GeV) inc. had.    &9	&15.3	&1.70\\  \hline
SIA sum             &384	&353.8	&0.92\\    \hline
Global total&1370	&1231.5 &	0.90\\  \hline
\end{tabular}
	\caption{
	The number of data points, $\chi^2$, and $\chi^2/N_{pt}$ for the global datasets, groups of data from $pp$ collision, from SIA, and from SIDIS.
    The values are also shown for individual experiments.
    Datasets for production of unidentified charged hadrons are marked with a dagger. 
	}
  \label{Tab:chi2}
\end{table}

\begin{figure}[htbp]
  \centering
  \includegraphics[width=0.3\textwidth]{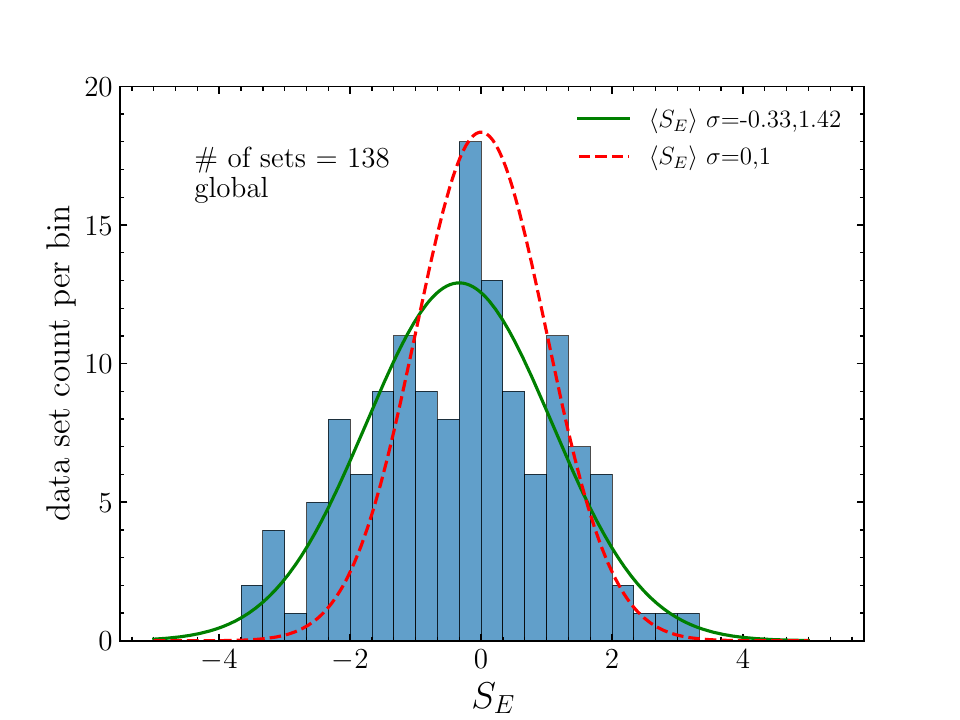}
  \includegraphics[width=0.3\textwidth]{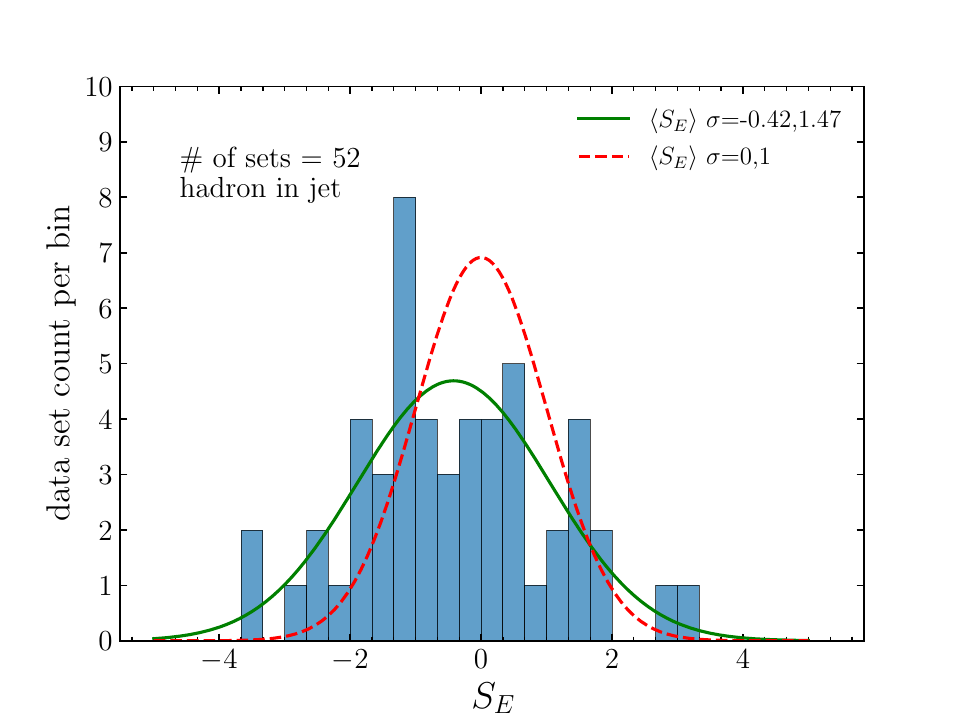}
  \includegraphics[width=0.3\textwidth]{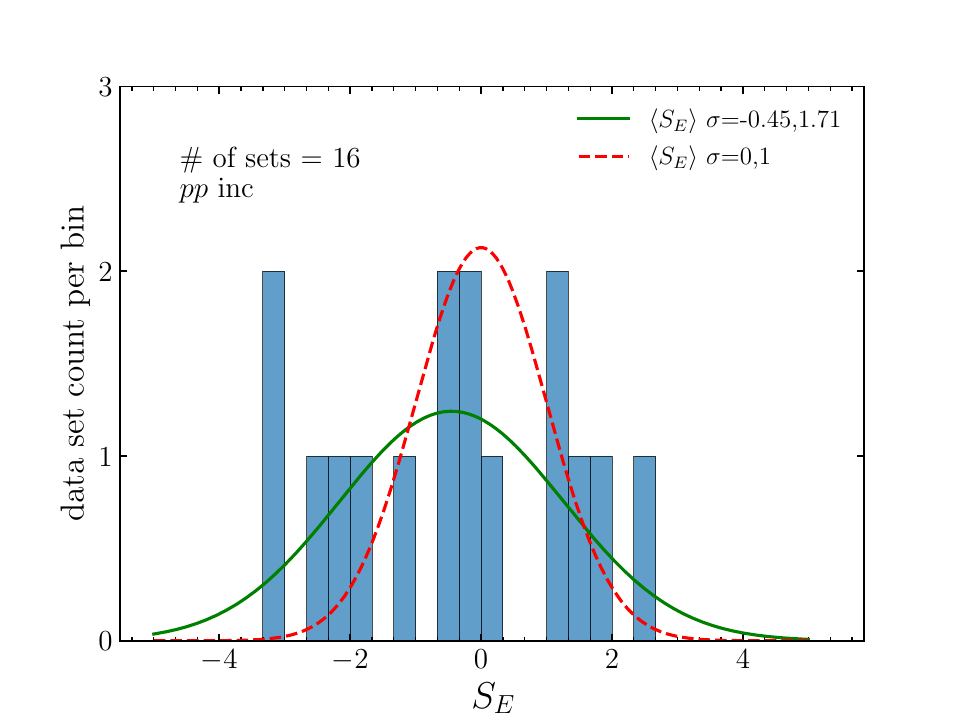}
  \includegraphics[width=0.3\textwidth]{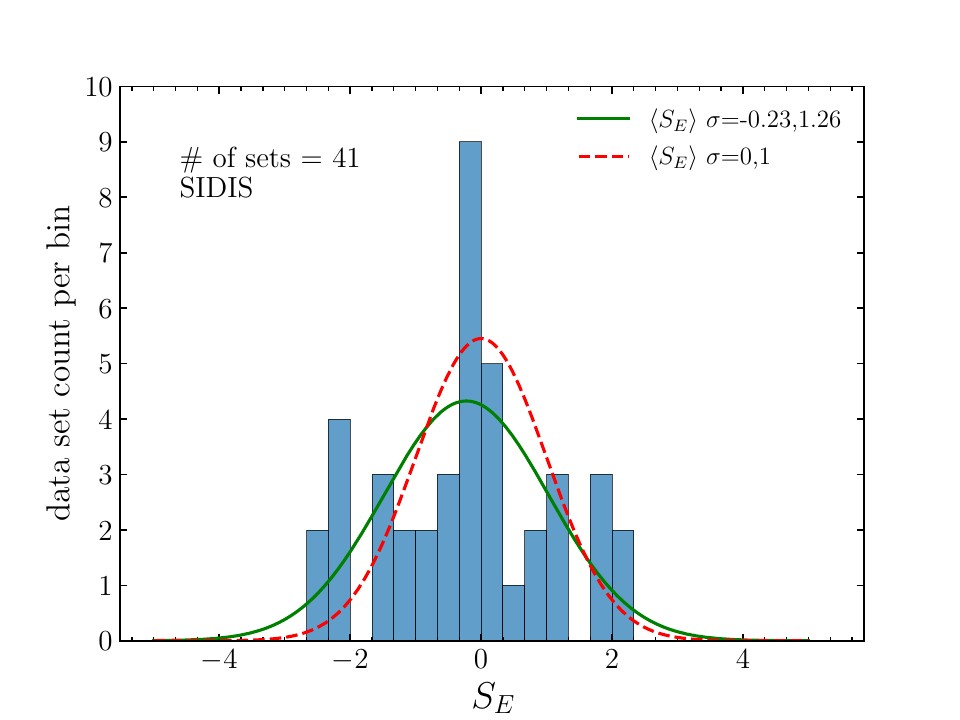}
      \includegraphics[width=0.3\textwidth]{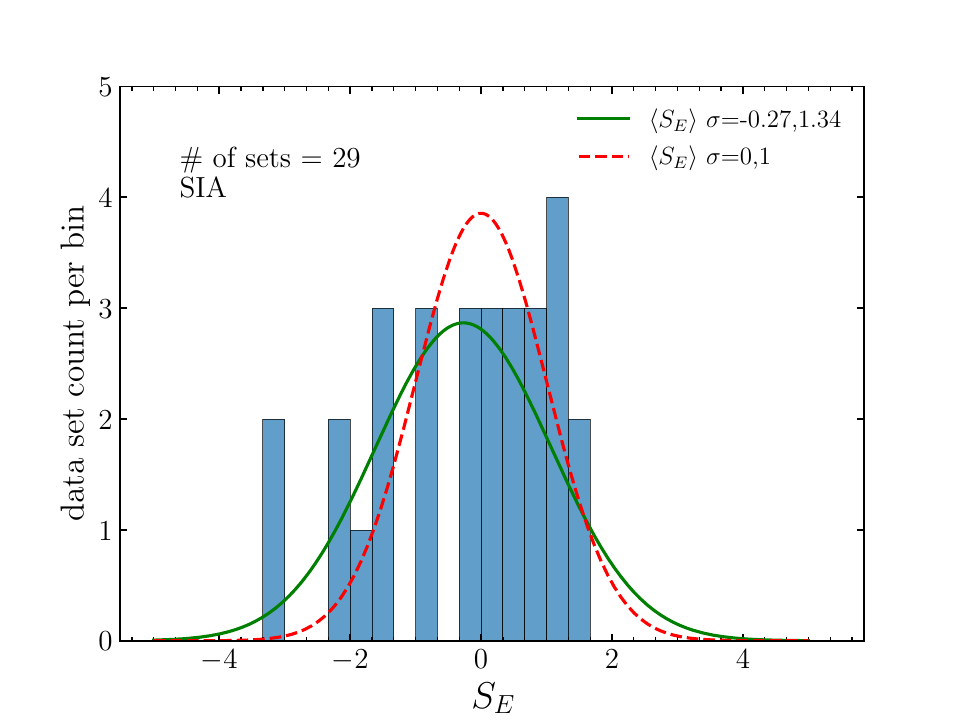}
	\caption{
    Histogram of the effective Gaussian variable $S_E$ for a total of
    138 subsets of data and subsets of SIA/hadron in jet/$pp$ inclusive/SIDIS experiments.
    The green and red curves represent a normal distribution and a Gaussian distributions with mean and standard deviation calculated from the ensemble of $S_E$.  
	}
  \label{Fig:se}
\end{figure}

We demonstrate overall agreement of our best-fit with the data by analyzing the log-likelihood functions $\chi^2$ for each of the measurements.
These are summarized in Table~\ref{Tab:chi2}, along with the sum of $\chi^2$ for each of the four groups of data and for the global data.
The global $\chi^2$ is 1231.5 units for a total number of data points of 1370, resulting in $\chi^2/N_{pt}=0.90$, indicating good agreement between theory and data.
The $\chi^2/N_{pt}$ values are 0.93, 0.87, 0.90, 0.92 for the groups of data of inclusive hadron production and jet fragmentation in $pp$ collisions, inclusive hadron production from SIA and SIDIS, respectively.
For individual measurements, only the jet fragmentation in ATLAS $Z+$~jet, CMS $\gamma+$~jet production, and the ZEUS unidentified charged hadron production show slightly worse agreement with $\chi^2/N_{pt}>2$.
We achieve very good agreement with the ALICE and STAR measurements on single inclusive hadron production because we only fit to various ratios of cross sections of different hadrons or different center-of-mass energies.
The two measurements on unidentified charged hadron production from LEP above the $Z$-pole also show a worse $\chi^2$ as compared to other SIA measurements at or below the $Z$-pole.
We also conduct a detailed investigation into the agreement of our best fit with each of the 138 subsets.
To account for the variation in the number of data points in different subsets, we introduce an effective Gaussian variable. Specifically, the sparseness is defined as:
\begin{equation}
    S_E=\frac{(18N_{pt})^{3/2}}{18N_{pt}+1}\left\{{6\over 6-\ln(\chi^2/N_{pt})}-{9N_{pt}-1\over 9N_{pt}}\right\},
\end{equation}
which follows a normal distribution if $N_{pt}$ is not too small~\cite{ct18}.
We present histograms of $S_E$ for all subsets in the global data, as well as in each of the four groups of data in Fig.~\ref{Fig:se}.
The number of subsets are 16, 52, 41 and 29 for inclusive hadron production and jet fragmentation in $pp$ collisions, inclusive hadron production from SIDIS and SIA, respectively.
The majority of the subsets (132 out of 138) have $S_E$ values smaller than 2, indicating again good agreement.
The distributions of $S_E$ from our best-fit closely resemble Gaussian distributions, with mean values and standard deviations shown in the figure and the associated Gaussian distribution represented by green curves.
For instance, for all subsets in the global data, the mean value and standard deviation are -0.33 and 1.43, respectively.
We also plot normal distributions in red curves alongside the histograms for comparisons. 
Deviation of the histogram with respect to the normal distribution indicates a possible underestimation of uncertainties by a factor of
1.43 on average.
This motivates a choice of tolerance of $\Delta\chi^2=1.43^2\approx 2$ in our determination of uncertainties of the FFs with the
Hessian method~\cite{ct18}, as introduced earlier.

\subsection{Description of individual datasets}

In this section, we present comparisons of our theoretical predictions to all data subsets included in our global analysis.
The data have been separated into groups of inclusive hadron production and jet fragmentation in $pp$ collisions, as well as inclusive hadron production from SIDIS and SIA.
For each of them, we show our NLO predictions on the differential cross sections based on the best-fit FFs and their Hessian uncertainties.
\subsubsection{Hadron collisions}
\begin{figure}[htbp]
  \centering
  \includegraphics[width=0.83\textwidth]{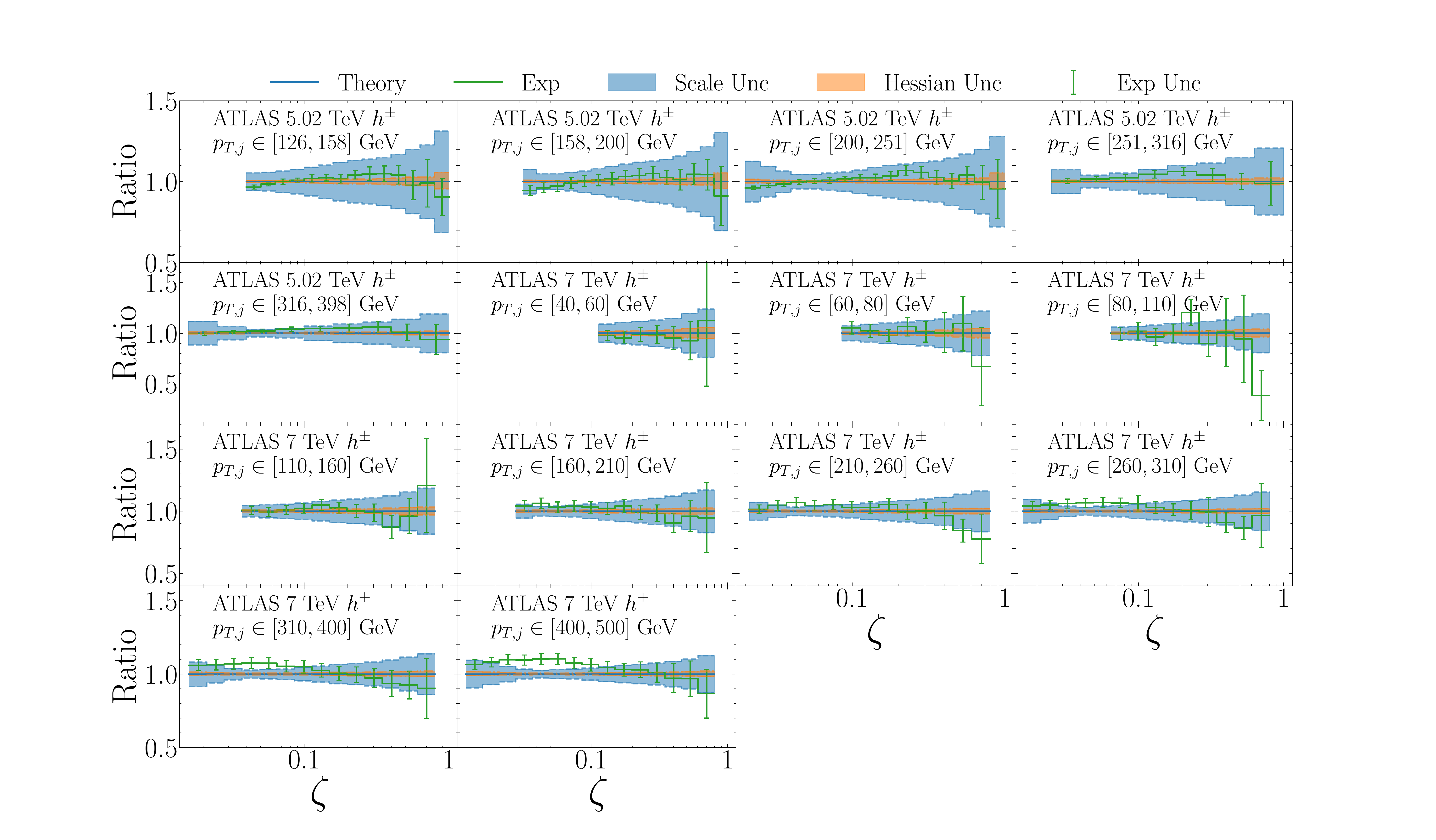}
	\caption{
 Comparison of predictions and experimental data from ATLAS 5 TeV and 7 TeV measurements on jet fragmentation. The results are normalized to theoretical predictions. The error bars indicate experimental uncertainty. The light shaded bands indicate scale uncertainty, obtained by taking the envelope of theory predictions with the 9 scale combinations of $\mu_F/\mu_{F,0}$ = $\mu_R/\mu_{R,0} = \{1/2,~1,~2\}$ and $\mu_D/\mu_{D,0} = \{1/2,~1,~2\}$. The dark shaded bands indicate Hessian uncertainty obtained with Hessian error sets.
	}
  \label{Fig:atlas57}
\end{figure}

\begin{figure}[htbp]
  \centering
  \includegraphics[width=0.83\textwidth]{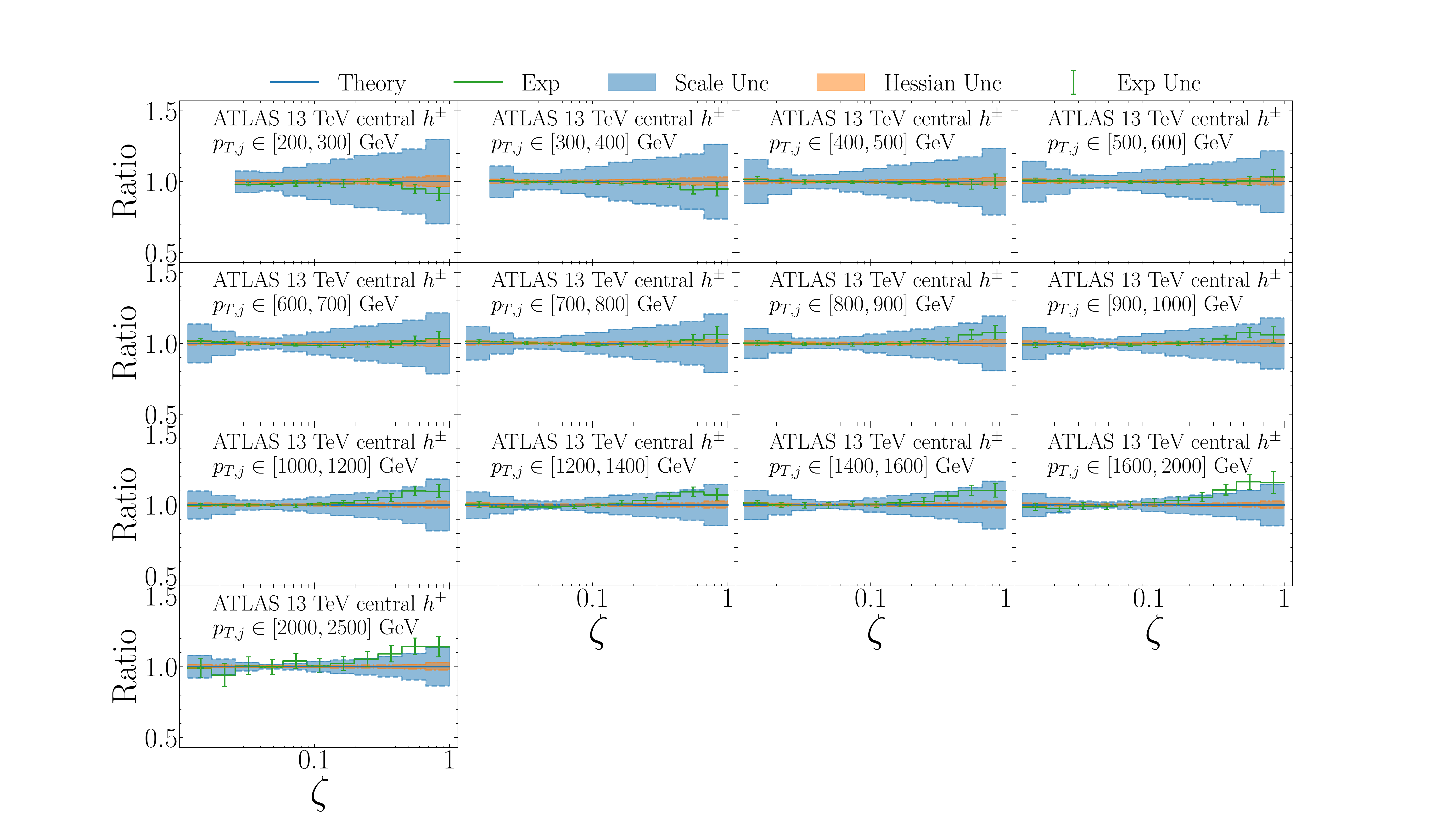}
	\caption{
 Similar to Fig. \ref{Fig:atlas57} but for data from ATLAS 13 TeV central jets.
	}
  \label{Fig:atlas_cen}
\end{figure}

\begin{figure}[htbp]
  \centering
  \includegraphics[width=0.83\textwidth]{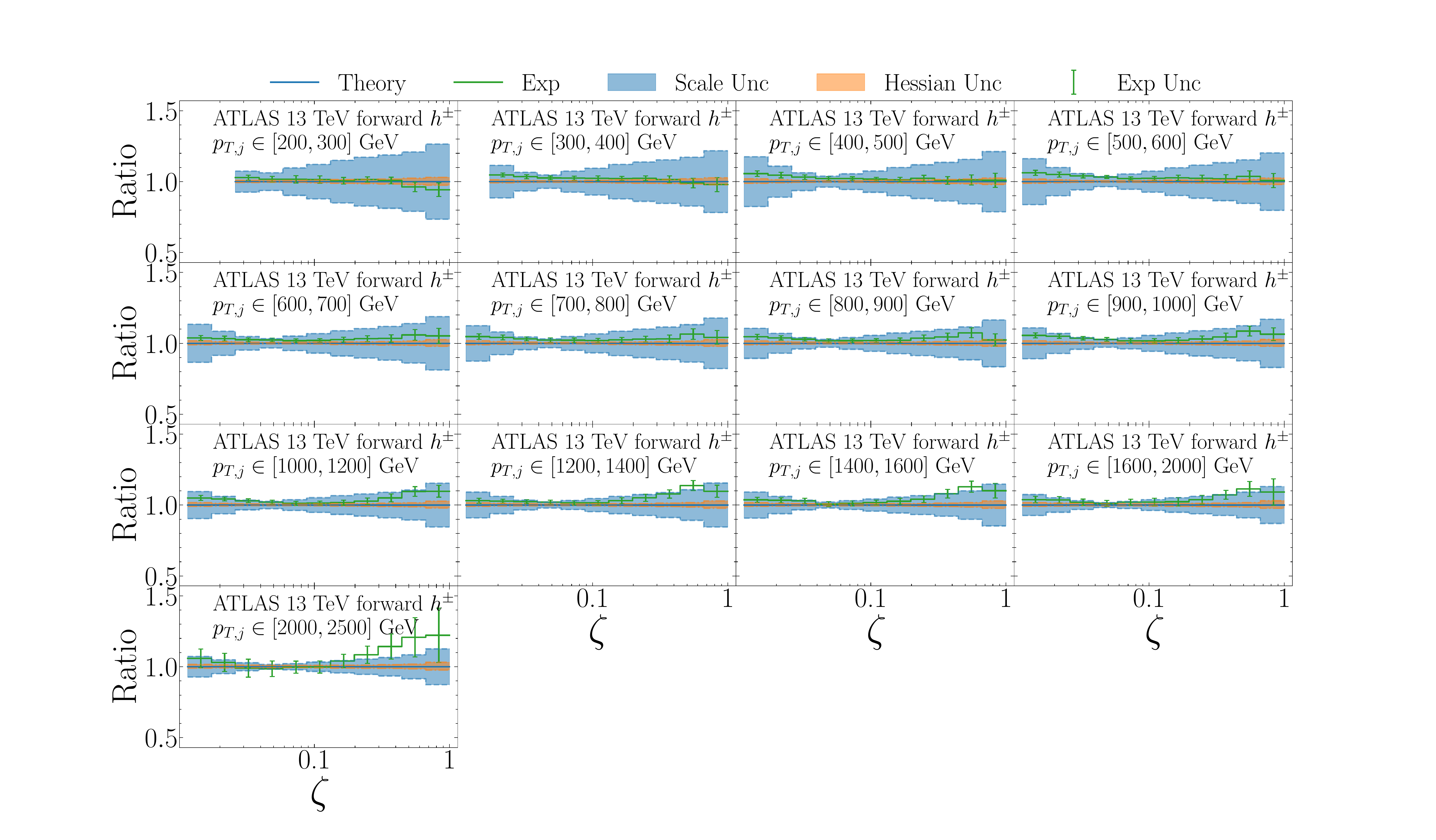}
	\caption{
 Similar to Fig. \ref{Fig:atlas57} but for data from ATLAS 13 TeV forward jets.
	}
  \label{Fig:atlas_for}
\end{figure}

Fig.~\ref{Fig:atlas57} shows a comparison of production cross sections as functions of hadron momentum fraction, for unidentified charged hadron from ATLAS inclusive jet fragmentation measurements at center-of-mass energies of 5.02 and 7 TeV for typical transverse momenta of the jet. 
In each panel of the plots, all results including data are normalized to our central predictions from the best-fit, and the colored bands represent scale variations and Hessian uncertainties from our error sets, respectively.
Error bars represent data central values together with total uncertainties, namely statistical uncertainties and uncorrelated systematic uncertainties adding in quadrature.
Experimental normalization uncertainties, which are fully correlated among different data points, are not shown.
We find very good agreement between theory and data within uncertainties, except for slight excesses of data at $z\sim 0.05$ for the high jet $p_T$ subset of 7 TeV measurements.
The Hessian uncertainties are generally smaller than the experimental uncertainties.
The scale variations are much larger and can reach 20\%$\sim$30\% in the highest $z$ region.
However, we assume the theoretical uncertainties to be fully correlated among different data points in each subset of data.
Figs.~\ref{Fig:atlas_cen} and~\ref{Fig:atlas_for} show similar comparisons but for the fragmentation of the central and forward jet in QCD dijet production at ATLAS 13 TeV.
We observe a deficit of theory predictions at large $z$, especially in forward jet production at large $p_T$, presumably due to the fragmentation of $u$ quarks.  

\begin{figure}[htbp]
  \centering
  \includegraphics[width=0.83\textwidth]{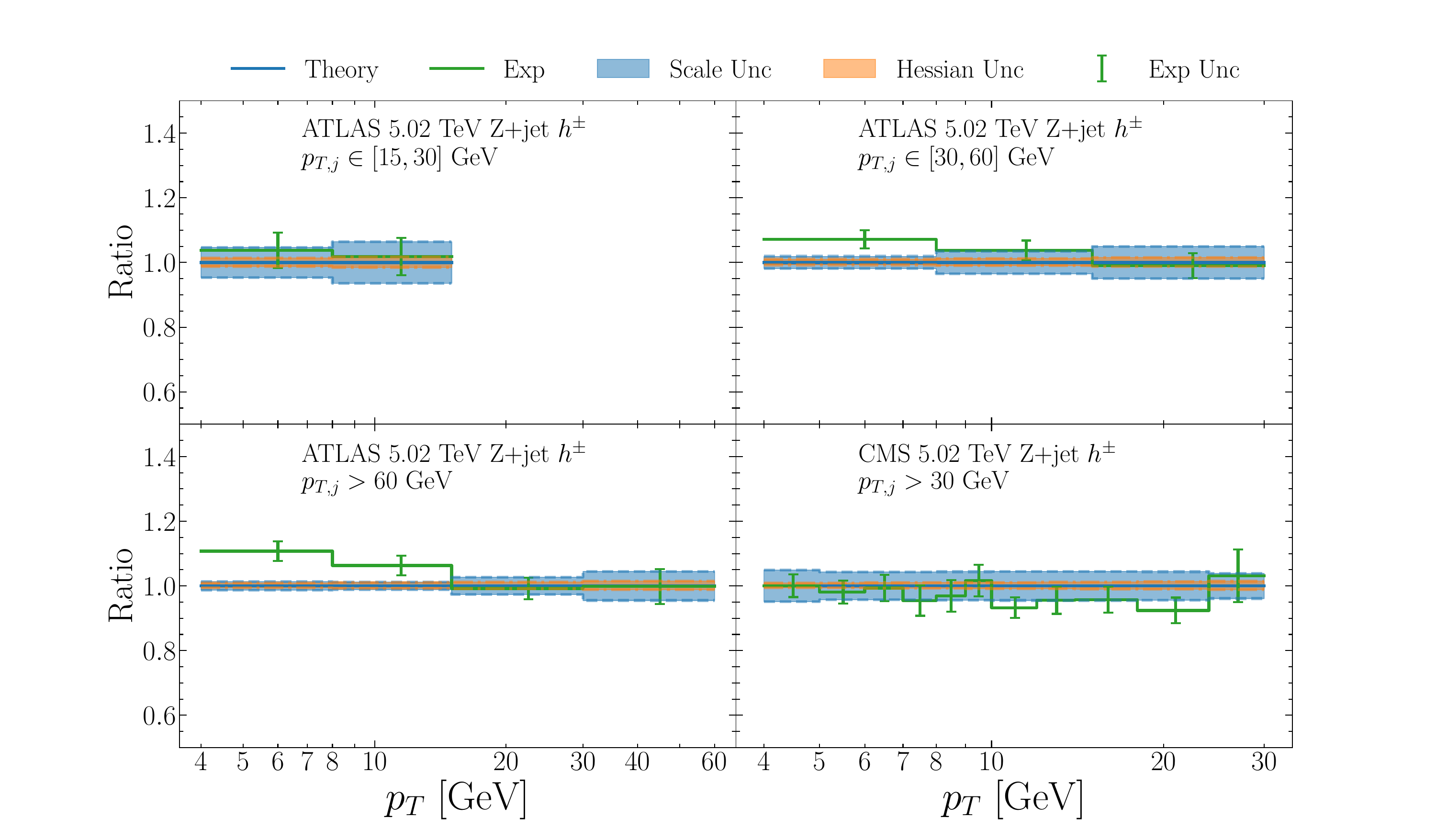}
	\caption{
  Similar to Fig. \ref{Fig:atlas57} but for data from ATLAS and CMS Z tagged jets.
	}
  \label{Fig:z_jet}
\end{figure}

\begin{figure}[htbp]
  \centering
  \includegraphics[width=0.83\textwidth]{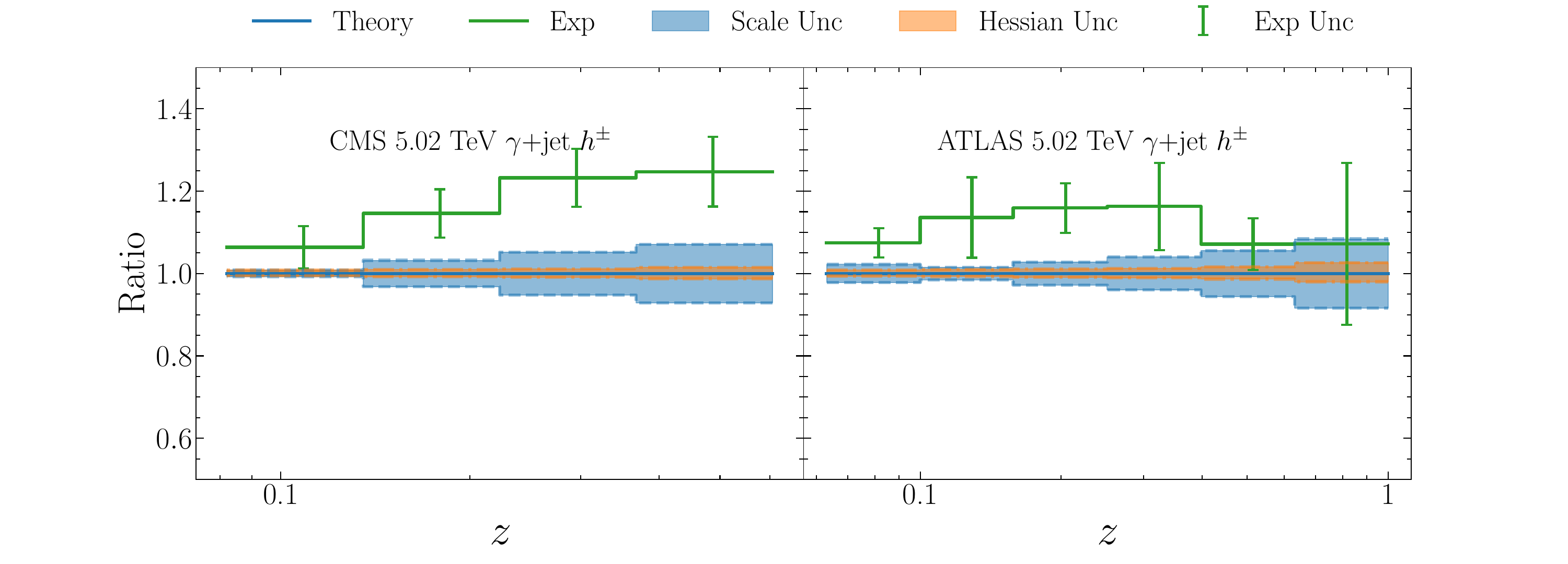}
	\caption{
  Similar to Fig. \ref{Fig:atlas57} but for data from ATLAS and CMS $\gamma$ tagged jets.
	}
  \label{Fig:photon_jet}
\end{figure}

\begin{figure}[htbp]
  \centering
  \includegraphics[width=0.83\textwidth]{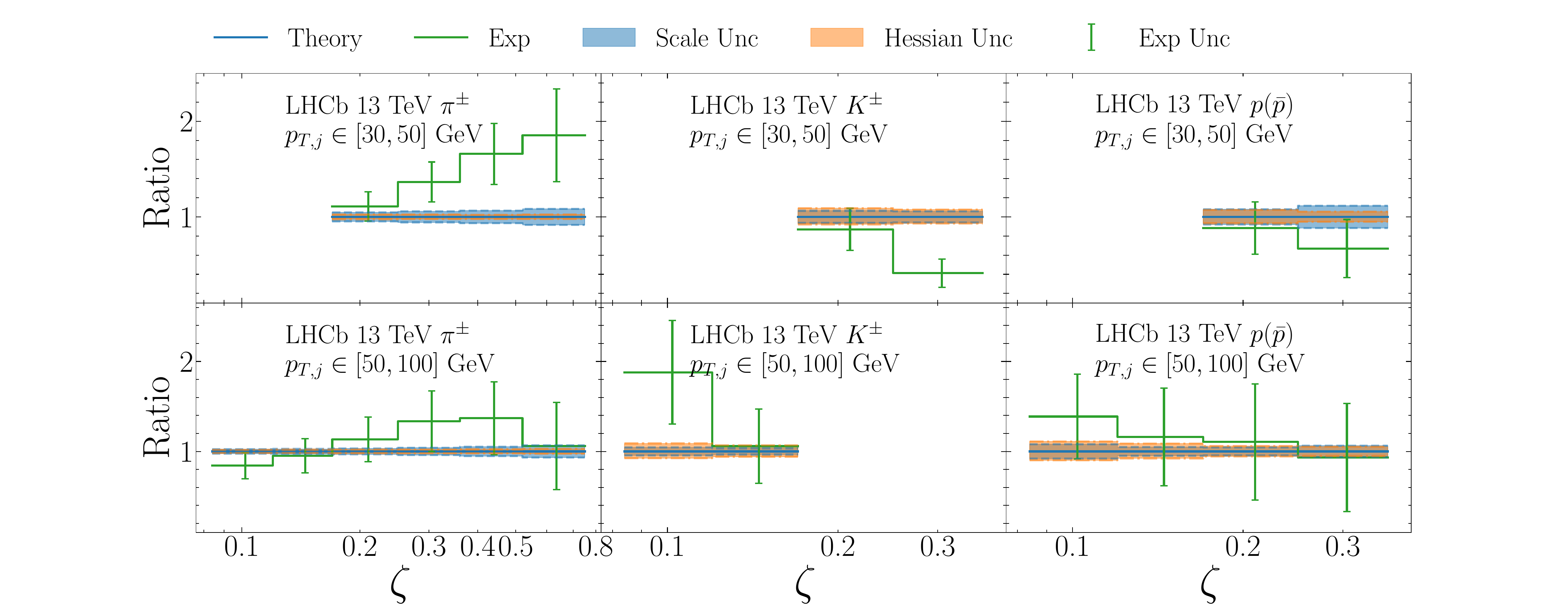}
	\caption{
  Similar to Fig. \ref{Fig:atlas57} but for data from LHCb 13 TeV  $Z$-tagged jets.
	}
  \label{Fig:lhcb}
\end{figure}

We now compare jet fragmentation measurements in photon or $Z$ boson tagged events, which are mostly sensitive to fragmentation from $u$ and $d$ quarks. 
The comparisons are shown in Fig.~\ref{Fig:z_jet} for unidentified hadrons from $Z+$jet production at ATLAS and CMS and in Fig.~\ref{Fig:photon_jet} for unidentified hadrons from $\gamma+$jet production at CMS and ATLAS.
We find reasonable agreement between theory and data, except for the CMS measurement from $\gamma+$jet production where the theory predictions are about 20\% lower at large $z$. 
The scale variations are much smaller compared to inclusive jet productions and are at most 10\%.
Lastly, in Fig.~\ref{Fig:lhcb}, we show a comparison to measurements on identified charged hadrons from LHCb $Z+$jet production at 13 TeV.
For subsets with low jet $p_T$ ($30\sim 50$ GeV), we observe poor agreement of our predictions with LHCb data.
However, the experimental uncertainties can be larger than 30\% in the high-$z$ bins. 

\begin{figure}[htbp]
  \centering
  \includegraphics[width=0.83\textwidth]{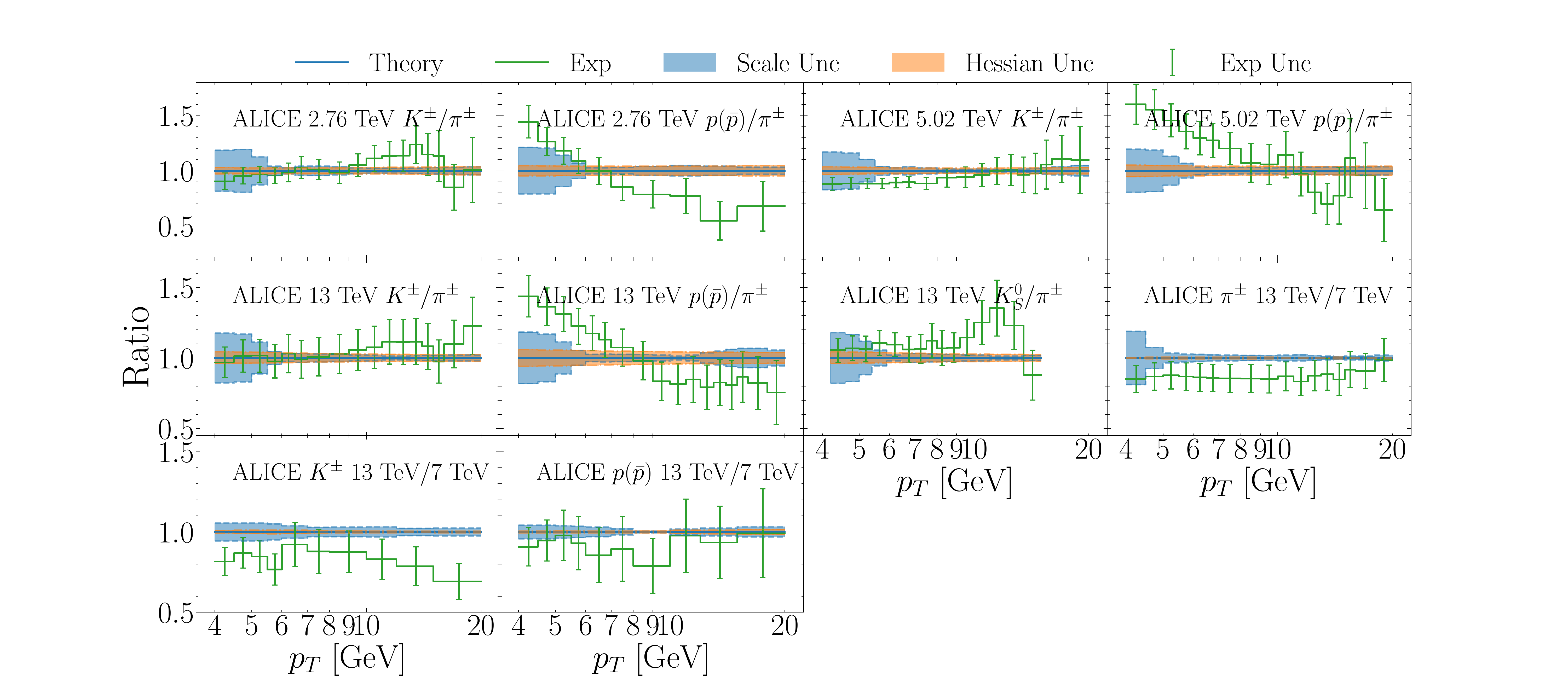}
	\caption{
 Similar to Fig. \ref{Fig:atlas57} but for data from ALICE.
	}
  \label{Fig:alice}
\end{figure}

\begin{figure}[htbp]
  \centering
  \includegraphics[width=0.83\textwidth]{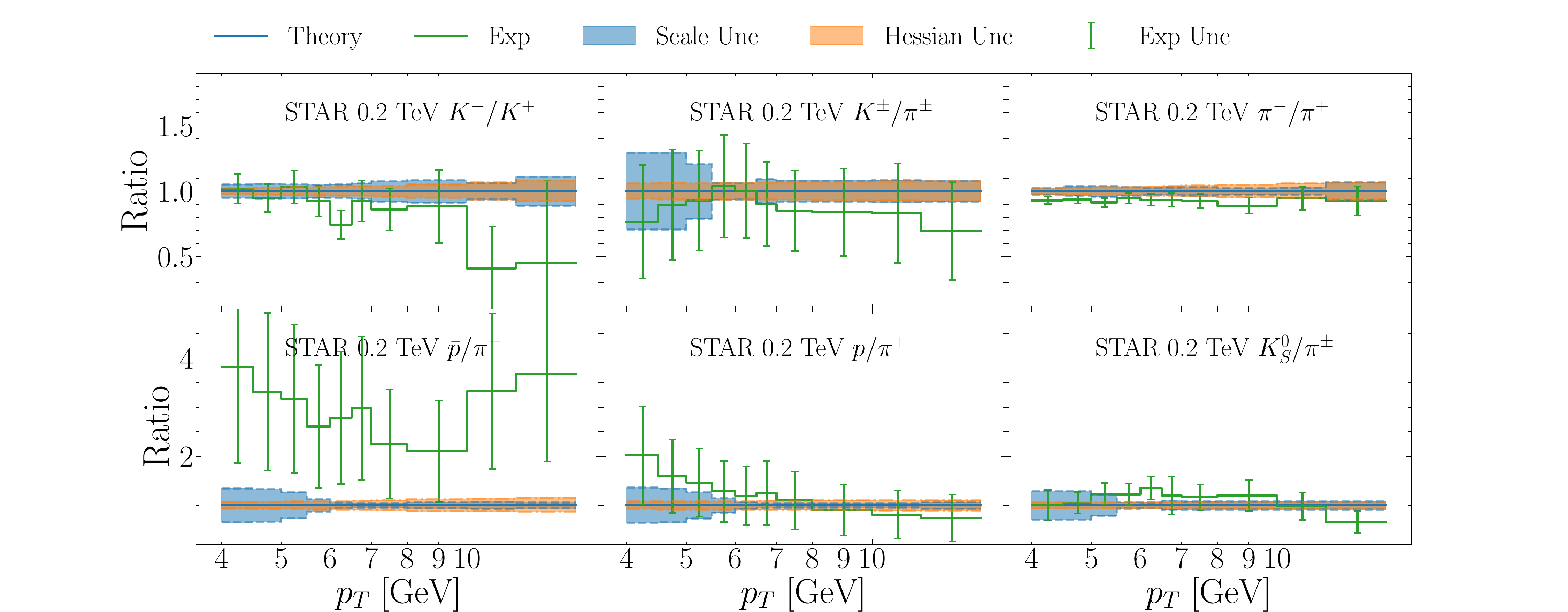}
	\caption{
  Similar to Fig. \ref{Fig:atlas57} but for data from STAR 0.2 TeV datasets.
	}
  \label{Fig:star}
\end{figure}

Fig.~\ref{Fig:alice} shows a comparison to data on ratios of inclusive production cross sections of different charged hadrons as functions of hadron transverse momentum, from ALICE measurements at center-of-mass energies of 2.76, 5.02, and 13 TeV.
We find good agreement for ratios of kaons to pions.
The scale variations are small for such ratio observable and are only significant at low-$p_T$ region of hadrons. 
There are notable slopes in data normalized to theory predictions for ratios of protons to pions at all three energies, especially at low-$p_T$ region of hadrons.
We also show a comparison to data on ratios of inclusive production cross sections at 13 TeV to those at 7 TeV.
There is a trend of normalization difference for all three charged hadrons, though within experimental uncertainties in general.
Fig.~\ref{Fig:star} shows a similar comparison to STAR measurements on inclusive hadron production at a center-of-mass energy of 0.2 TeV, which are also sensitive to separation on electric charges of hadrons.
Our theory predictions agree well with the data, except for the measurement on the ratio $\bar p/\pi^-$.
The central values of data are $2\sim 4$ times the theory predictions but with rather large uncertainties of about 50\%.

\subsubsection{SIDIS}

\begin{figure}[htbp]
  \centering
  \includegraphics[width=0.83\textwidth]{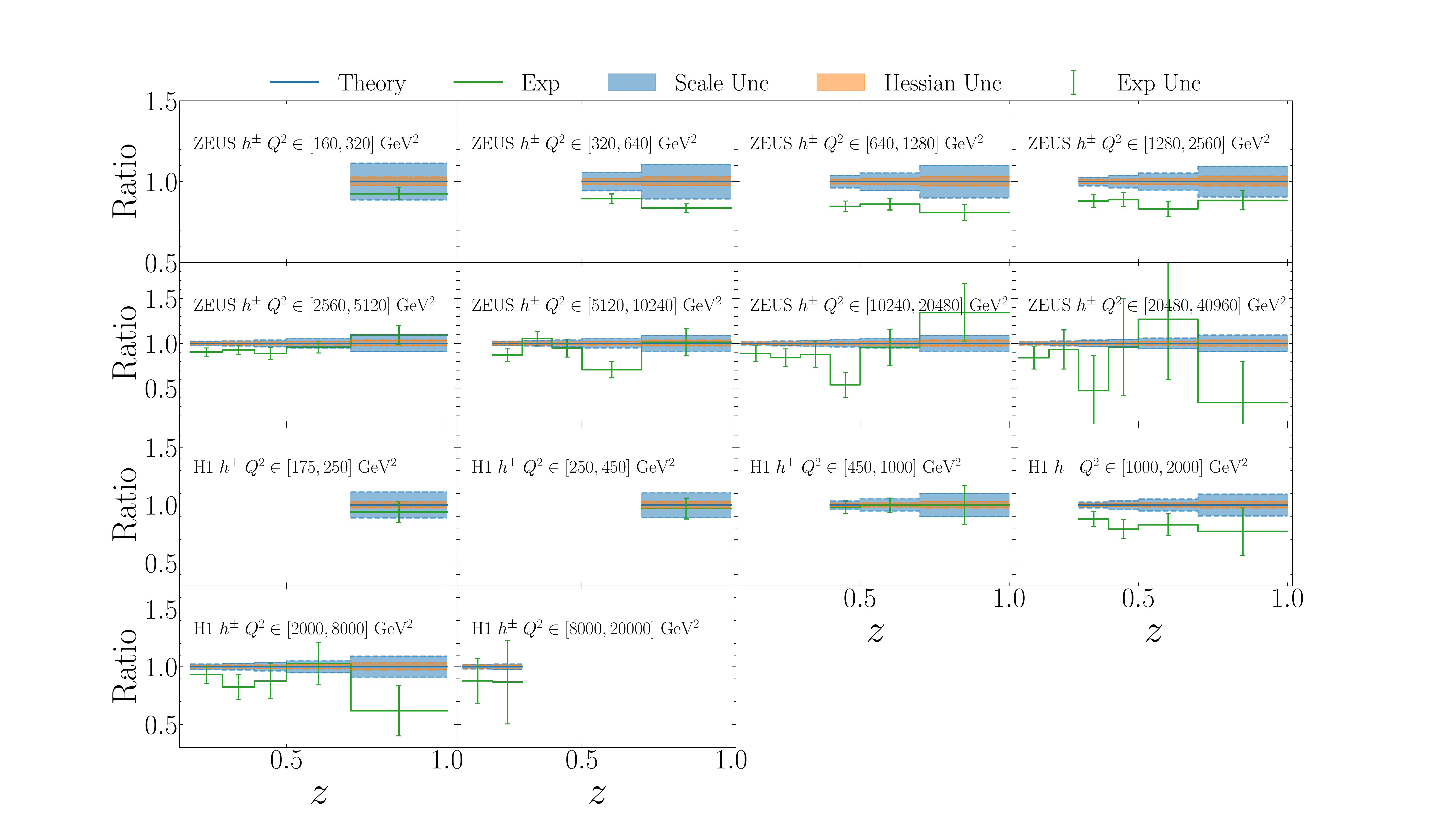}
	\caption{
   Similar to Fig. \ref{Fig:atlas57} but for data from ZEUS and H1 datasets.
	}
  \label{Fig:hera}
\end{figure}

\begin{figure}[htbp]
  \centering
  \includegraphics[width=0.83\textwidth]{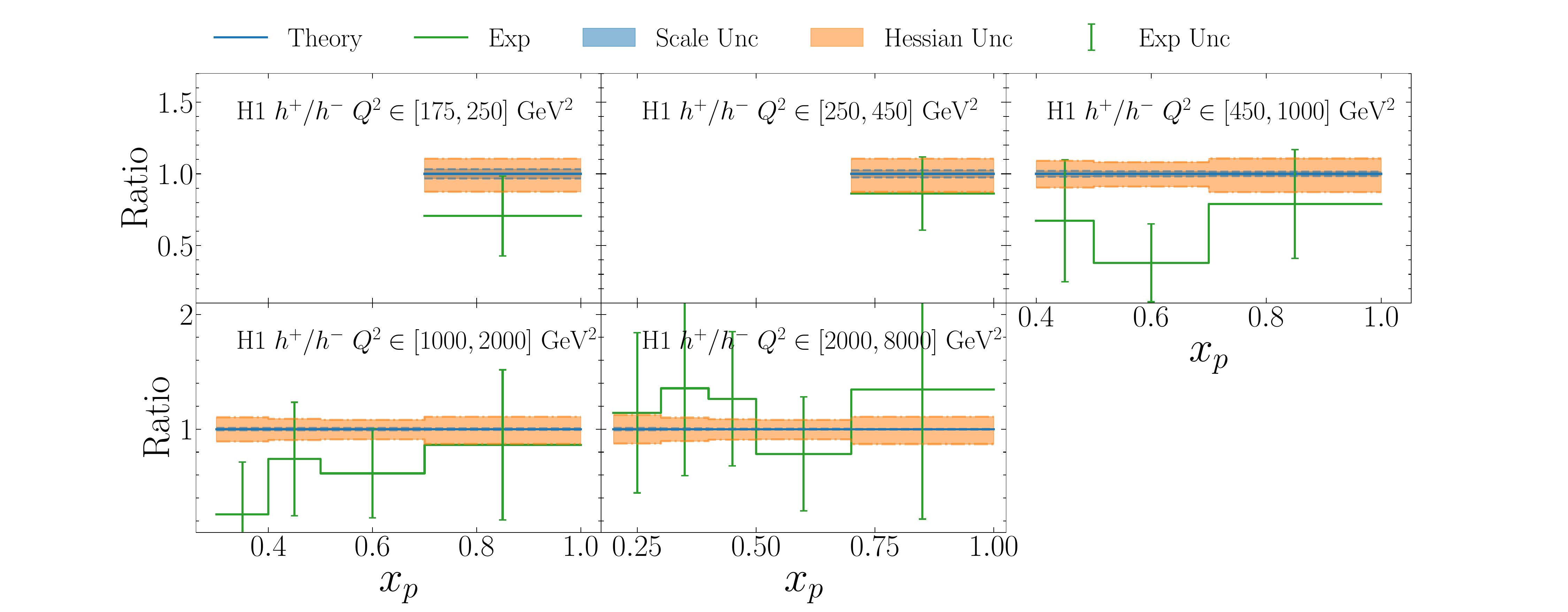}
	\caption{
   Similar to Fig. \ref{Fig:atlas57} but for data from H1 charge asymmetry datasets.
	}
  \label{Fig:heraasy}
\end{figure}

Fig.~\ref{Fig:hera} shows a comparison to production cross sections of unidentified charged hadrons measured by ZEUS and H1 at various $Q^2$ bins as functions of hadron momentum fraction $z$.
Our theory predictions overshoot the ZEUS data by about 20\% at relatively low $Q^2$ region, but they agree with the H1 data for similar $Q^2$ values.
Scale variations can reach 10\% for large $z$ values.
Additionally, we compare our predictions to the charge asymmetry data from H1 in Fig.~\ref{Fig:heraasy}, and they agree within the large experimental uncertainties.

\begin{figure}[htbp]
  \centering
  \includegraphics[width=0.83\textwidth]{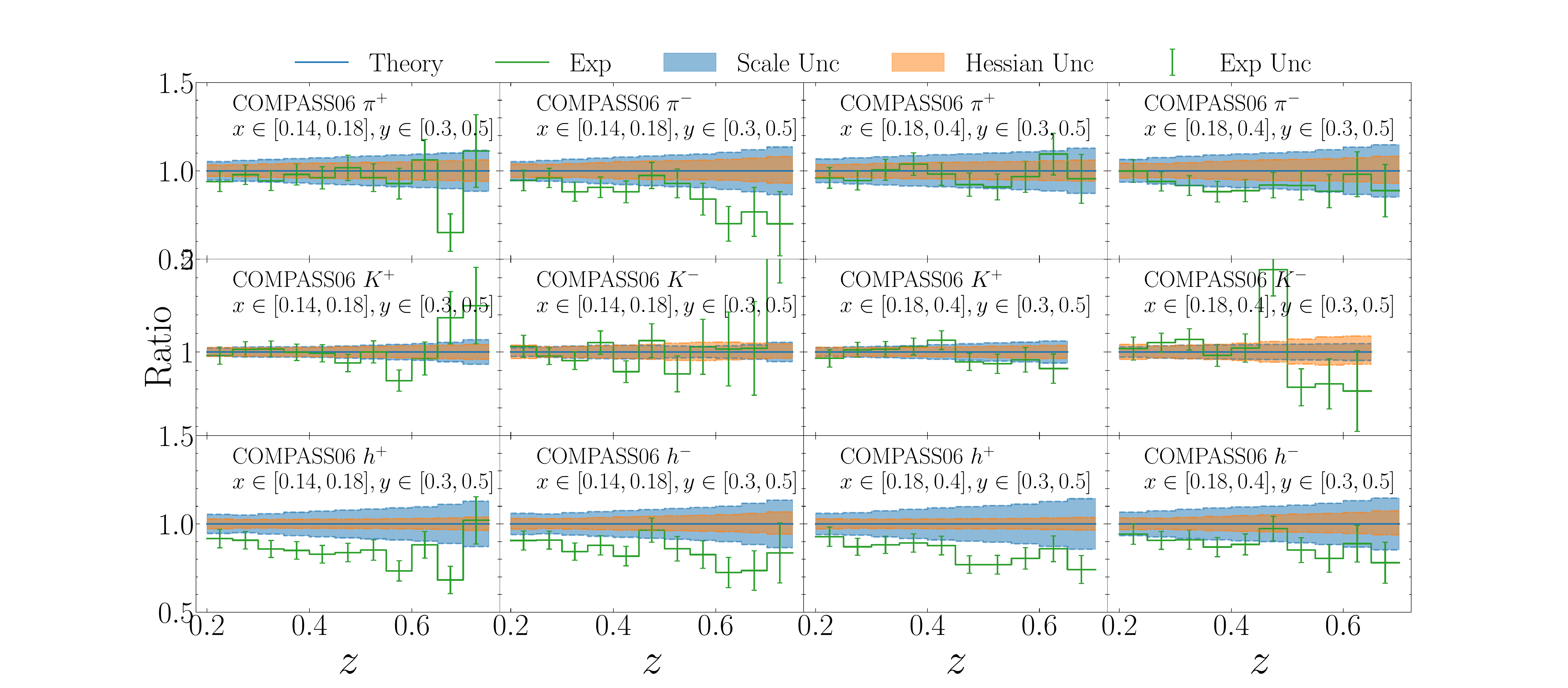}
	\caption{
   Similar to Fig. \ref{Fig:atlas57} but for data from COMPASS06 datasets.
	}
  \label{Fig:compass06}
\end{figure}

\begin{figure}[htbp]
  \centering
  \includegraphics[width=0.83\textwidth]{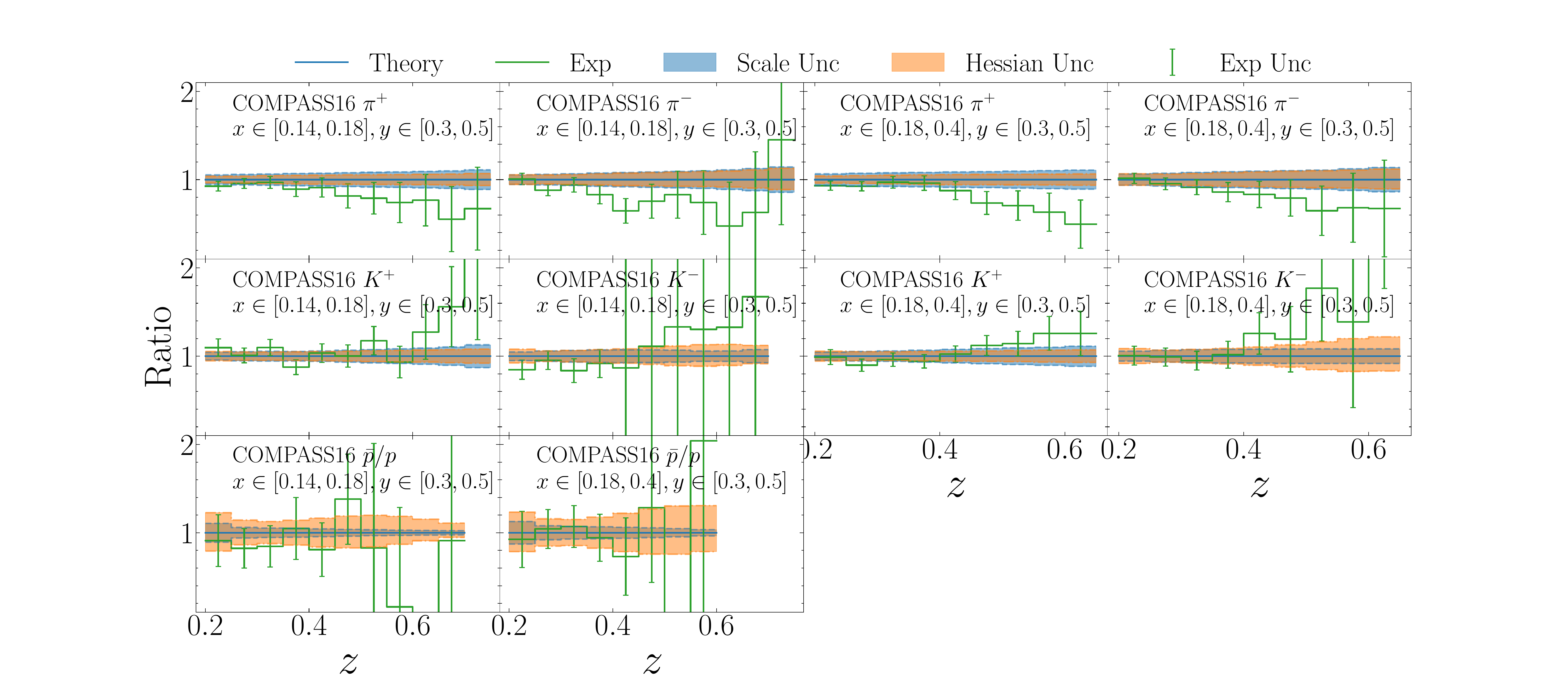}
	\caption{
   Similar to Fig. \ref{Fig:atlas57} but for data from COMPASS16 datasets.
	}
  \label{Fig:compass16}
\end{figure}

Comparison to the COMPASS measurements of SIDIS on isoscalar and proton targets are shown in Figs.~\ref{Fig:compass06} and~\ref{Fig:compass16}, respectively. 
We find good agreement with the COMPASS06 data on the production of pions and kaons for both two bins in Bjorken-$x$.
However, for the unidentified charged hadrons, our theory predictions consistently overshoot the data by $10\sim 20$\%.
This discrepancy can be attributed to our predictions on the cross sections of proton production, which are overestimated due to the absence of hadron mass corrections and the low-$Q^2$ values of COMPASS measurements. 
For the comparison to COMPASS16 data, we observe a slight slope in data normalized to theory predictions for pion production.
Our theory predictions on ratios of production cross sections of anti-proton to proton agree well with the data since the effects of hadron mass largely cancel in ratios.
The Hessian uncertainties are also large for proton ratios due to our poor constraint on FFs to proton at large $z$ values.  

\subsubsection{SIA}

\begin{figure}[htbp]
  \centering
  \includegraphics[width=0.83\textwidth]{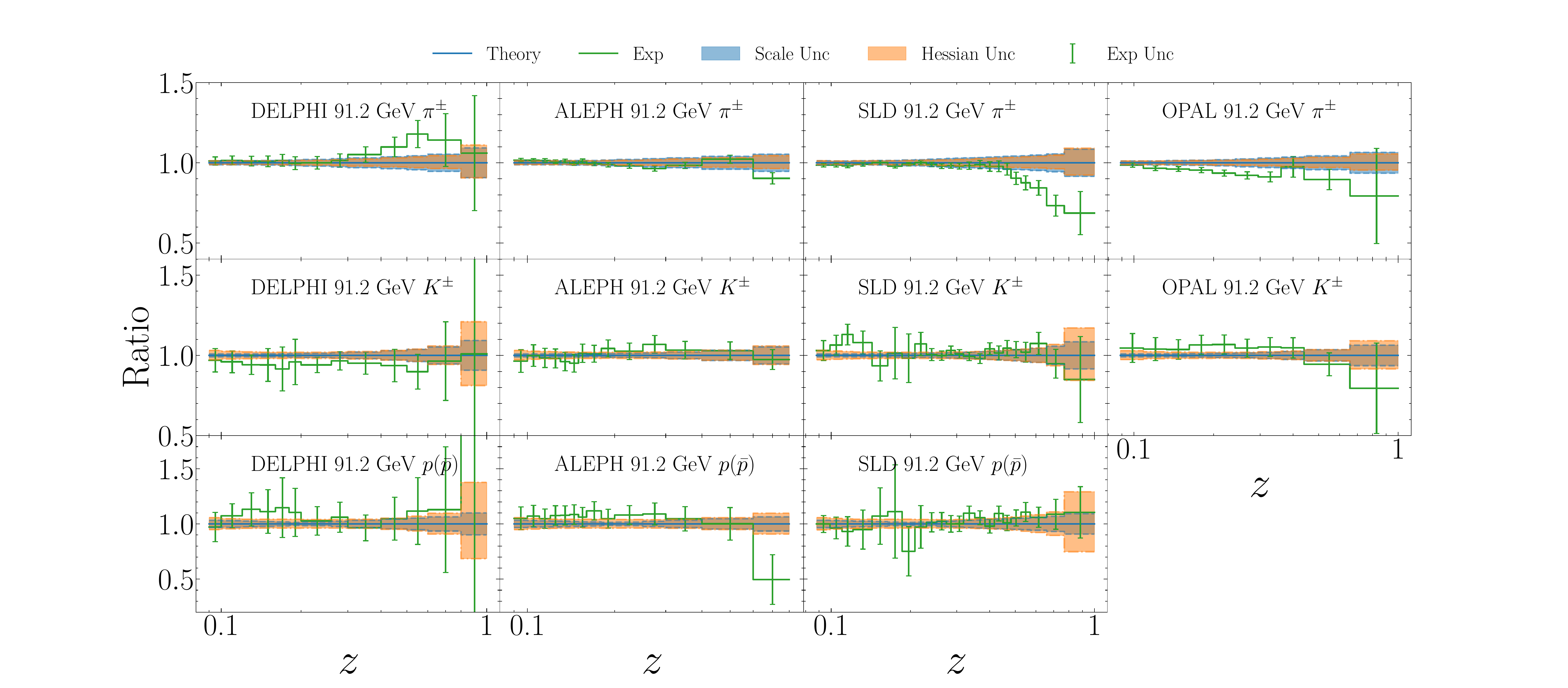}
	\caption{
   Similar to Fig. \ref{Fig:atlas57} but for data from SIA experiments at Z pole including DELPHI, ALEPH, SLD, OPAL.
	}
  \label{Fig:delphi}
\end{figure}

Fig.~\ref{Fig:delphi} shows a comparison to the majority of the SIA data on identified charged hadron production at the $Z$-pole from DELPHI, ALEPH, SLD, and OPAL.
The theory predictions are fully correlated for the above measurements on the same charged hadron.
Our predictions agree well with data in general, especially for lower $z$ values where the data are very precise.
However, significant discrepancies are seen for the comparison to SLD pion measurements at the large-$z$ region.
This could be due to the underestimation of experimental uncertainties, as the predictions agree well with data from other measurements in the same region.
For kaons and protons, our predictions agree very well with the SLD data in the entire kinematic region.

\begin{figure}[htbp]
  \centering
  \includegraphics[width=0.83\textwidth]{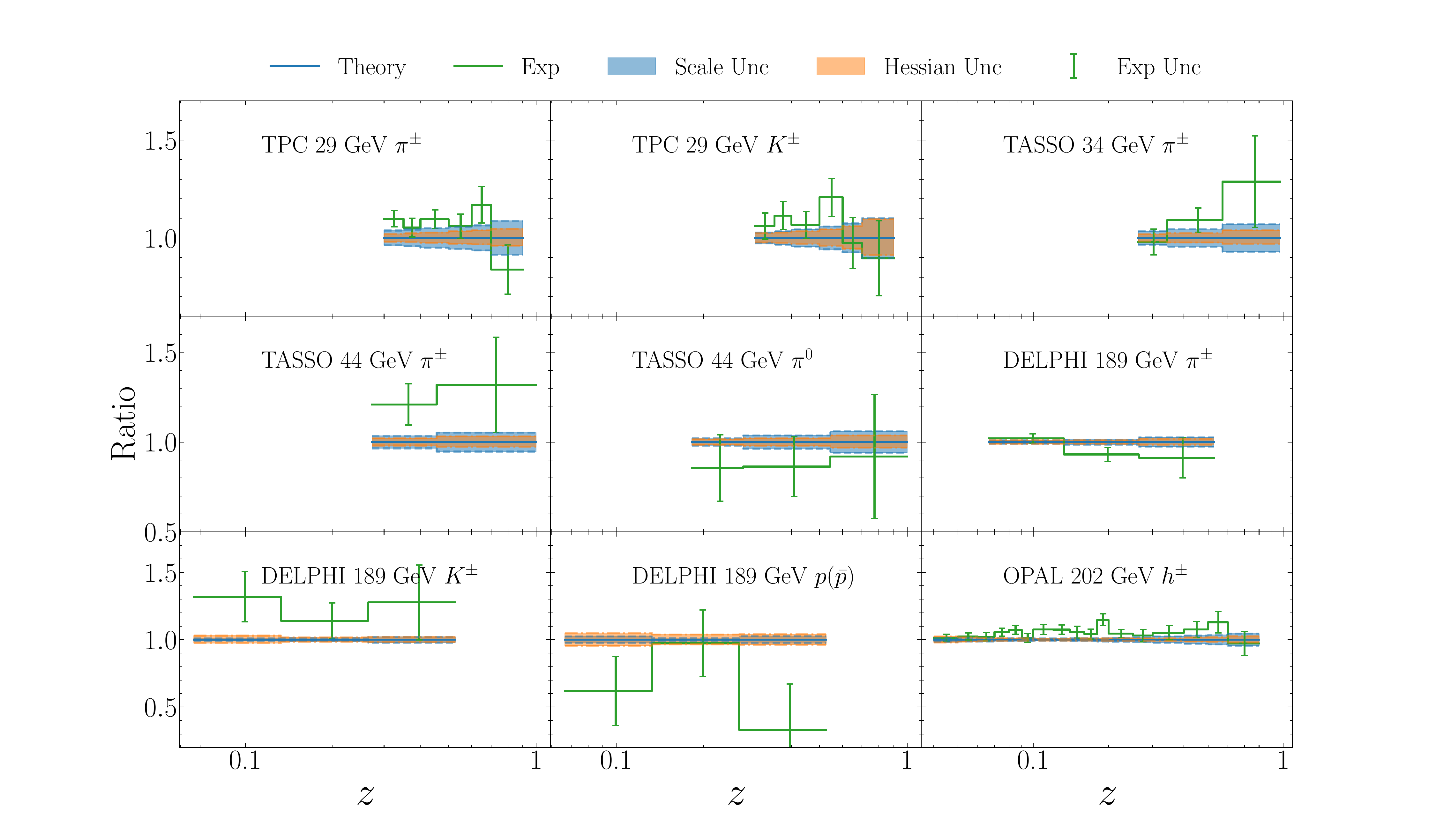}
	\caption{
   Similar to Fig. \ref{Fig:atlas57} but for data from SIA experiments from different energy scale including TPC, TASSO, DELPHI, OPAL.
	}
  \label{Fig:tass}
\end{figure}

\begin{figure}[htbp]
  \centering
  \includegraphics[width=0.83\textwidth]{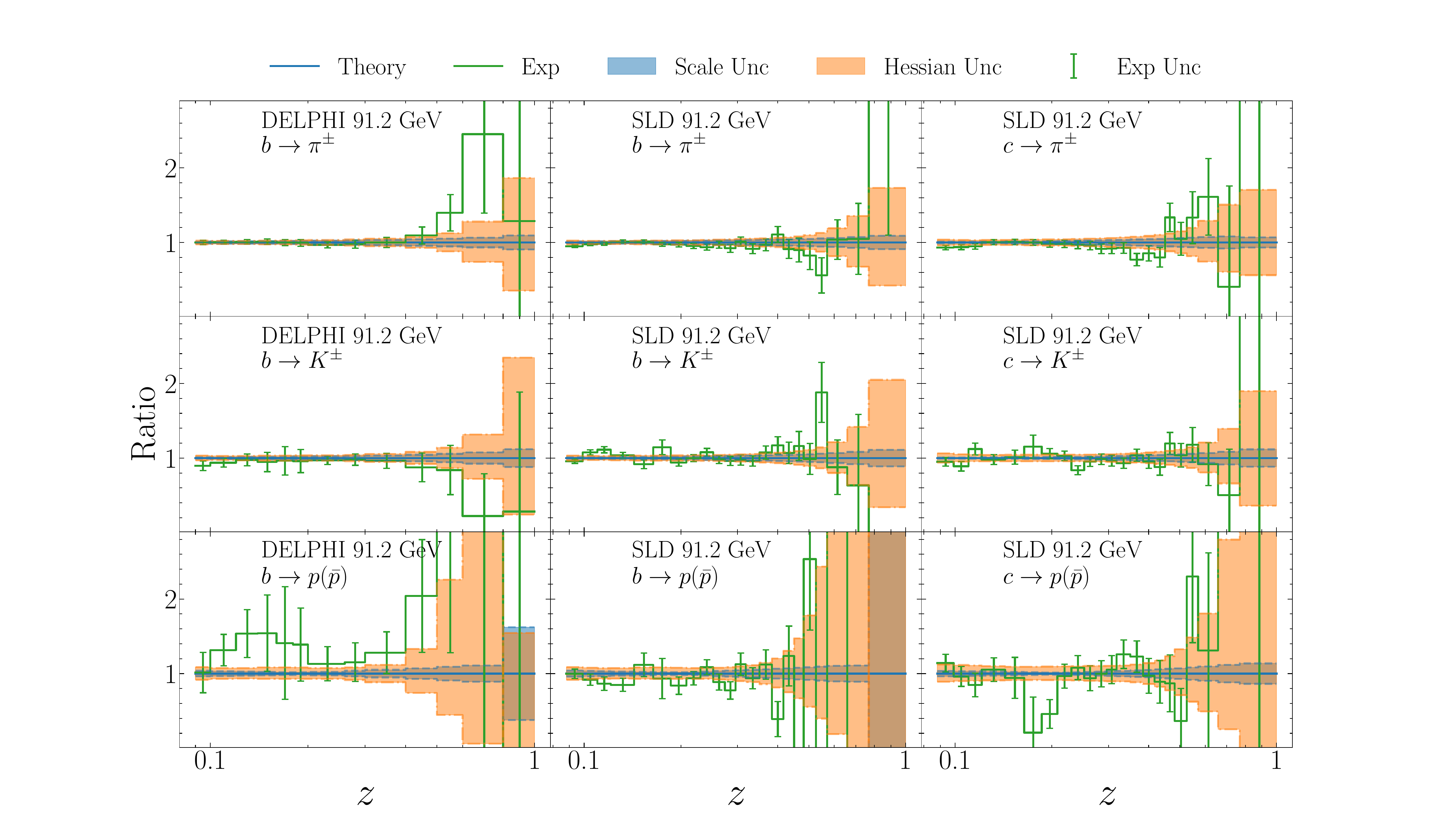}
	\caption{
   Similar to Fig. \ref{Fig:atlas57} but for data from SIA experiments of $b$, $c$ quark fragmentation on DELPHI and SLD.
	}
  \label{Fig:cb}
\end{figure}

Fig.~\ref{Fig:tass} shows a comparison to SIA data below the $Z$-pole on identified hadron production from TPC and TASSO, and above the $Z$-pole on identified hadron production from DELPHI and unidentified hadron production from OPAL.
The experimental uncertainties are large, except for the OPAL measurement.
Our theory predictions generally agree with data within uncertainties.
Finally, Fig.~\ref{Fig:cb} shows a comparison to SIA data at the $Z$-pole on light charged hadron production with heavy-flavor tagging from DELPHI ($b$ quark) and SLD ($b$ or $c$ quark).  
We find very good agreement between theory and data, and also consistency between DELPHI and SLD measurements.
The Hessian uncertainties blow up at large $z$ values due to the rapid increase of experimental uncertainties, as these data provide the only direct constraint on fragmentation from heavy quarks. 

\section{Alternative fits}
\label{sec:alter_fit}

Our best fit of the global analysis presented so far is based on a combination of choices regarding the experimental datasets and theoretical predictions. 
In this section, we explore various alternative fits to assess the impact of individual datasets and the effects of different choices for kinematic cuts and theoretical uncertainties on the extracted FFs.
By comparing these alternative fits to our baseline best-fit, we aim to reveal possible tension between different datasets and potential systematic uncertainties in addition to Hessian uncertainties presented earlier.  

\subsection{Dataset subtraction}

\begin{figure}[h!]
\centering
\includegraphics[width=0.9\textwidth]{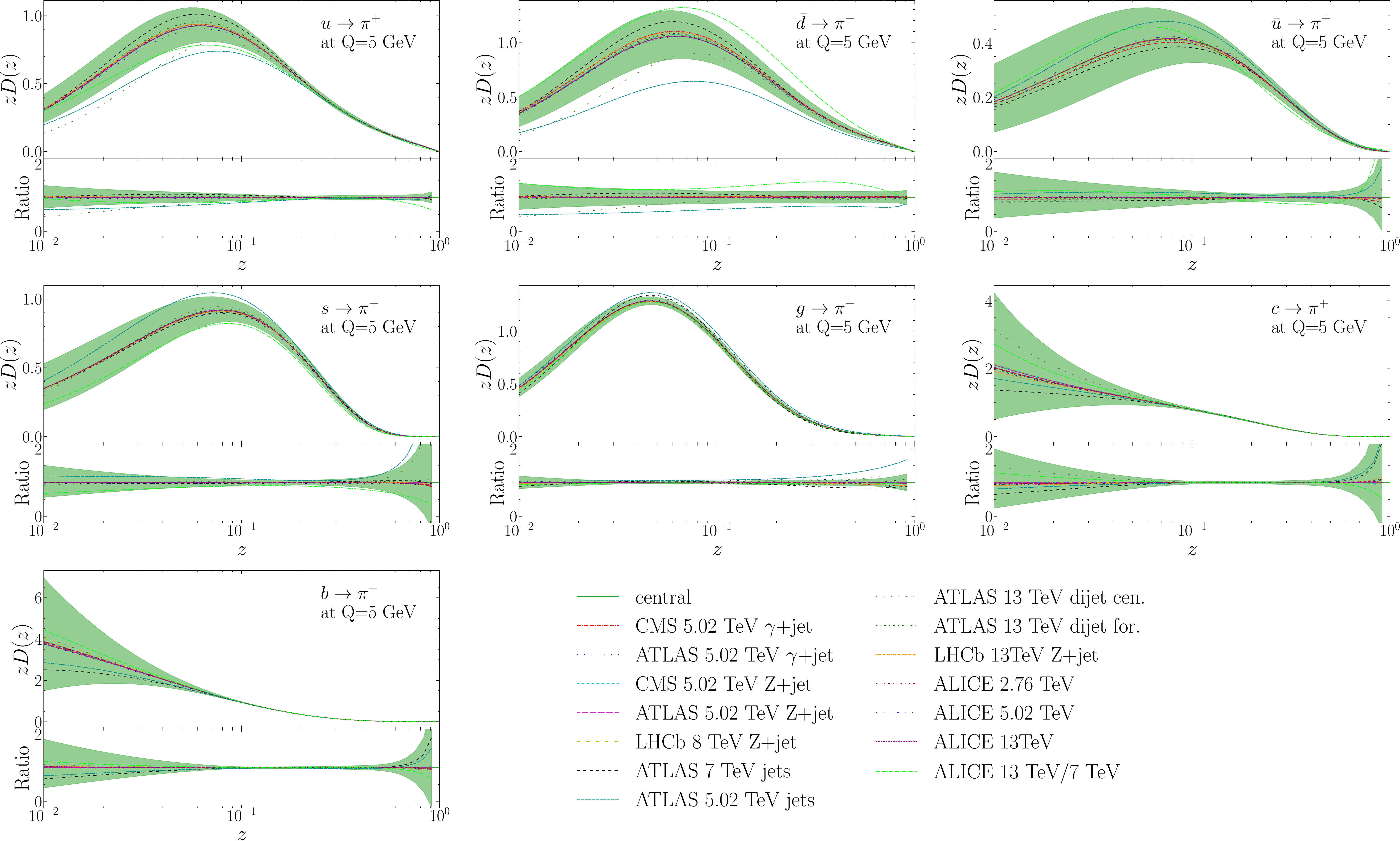}
\caption{
  Comparison of the global fit for the $\pi^+$ fragmentation function and a refitted version by excluding a specific group of subsets from pp collisions. The legend indicates the subsets that were removed for the refit. The colored band represents the Hessian uncertainty at a confidence level of 68\%. 
}
\label{Fig:pi_pp}
\end{figure}

\begin{figure}[h!]
\centering
\includegraphics[width=0.9\textwidth]{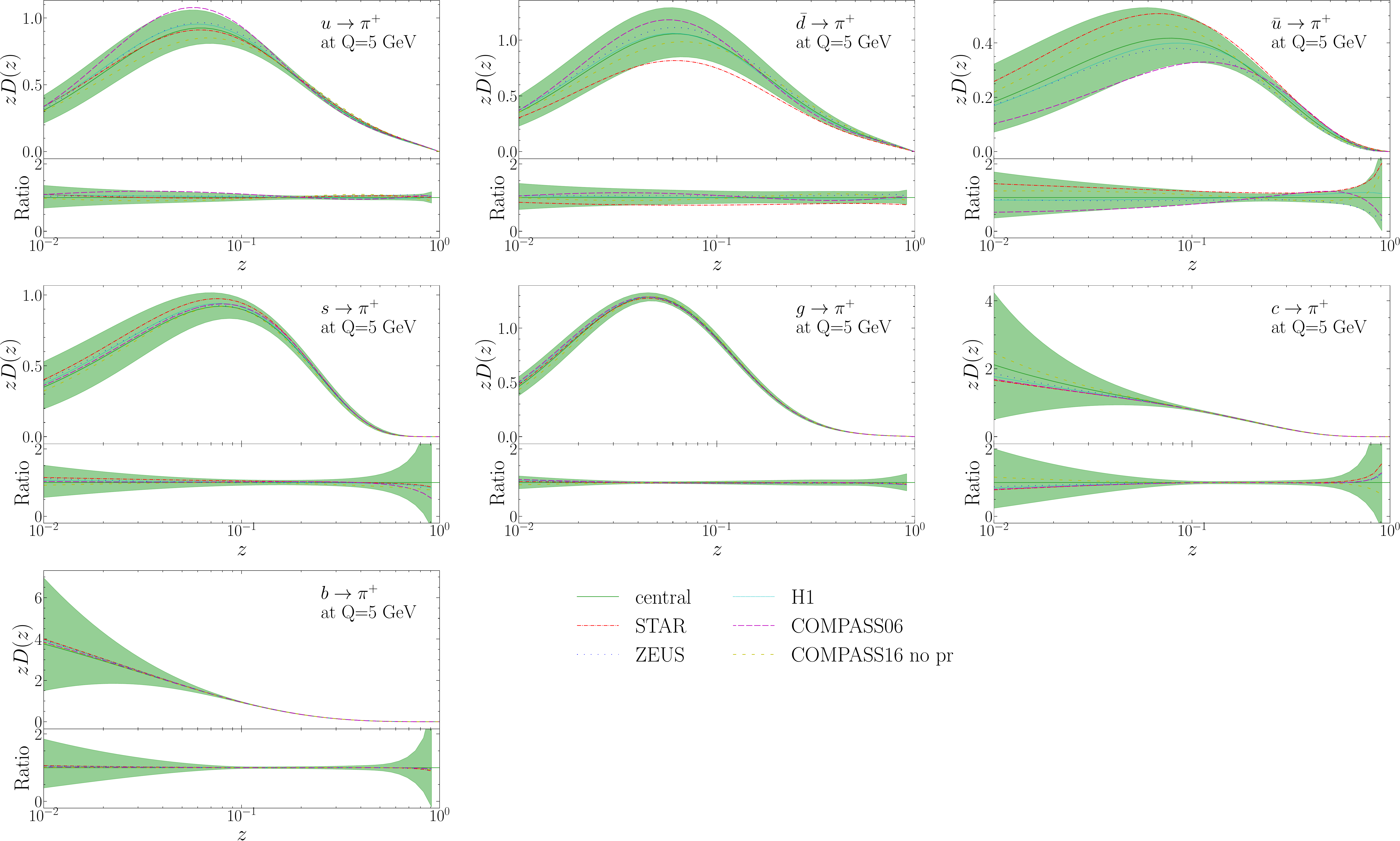}
\caption{
  Similar to Fig. \ref{Fig:pi_pp} but for FF of $\pi^+$ and subtractions from SIDIS processes.
}
\label{Fig:pi_sidis}
\end{figure}

\begin{figure}[h!]
\centering
\includegraphics[width=0.9\textwidth]{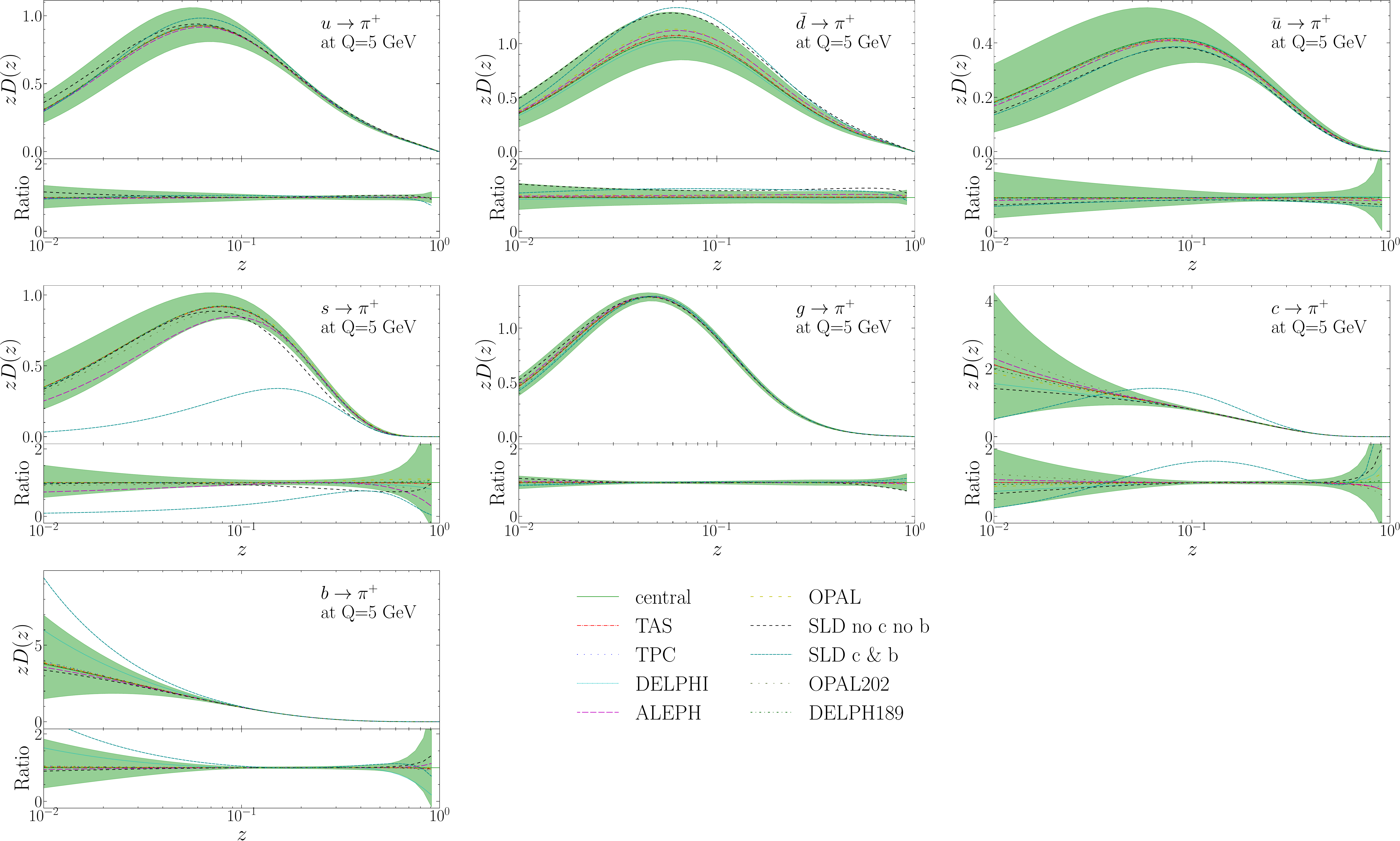}
\caption{
  Similar to Fig. \ref{Fig:pi_pp} but for FF of $\pi^+$ and subtractions from SIA processes.
}
\label{Fig:pi_sia}
\end{figure}

To assess the influence of specific datasets on distinct fragmentation processes, we conduct alternative fits by systematically excluding one dataset at a time and re-fitting the FFs at NLO. 
When referring to a single dataset, we consider the full dataset listed in Table~\ref{Tab:chi2}, including measurements on all possible hadrons.
These alternative fits solely consider the central values, with Hessian sets disabled. 
Throughout the fitting process, we maintain all other variables consistent with the baseline fit, including the kinematic cuts and the treatment of theoretical uncertainties.
The preference of a single dataset can thus be seen as opposite to the shift of the resulting FFs compared with the baseline fit, namely the best-fit of the global data.
The comparison between alternative fits and the baseline fit at $Q=5$ GeV is depicted in Figs.~\ref{Fig:pi_pp}-\ref{Fig:pi_sia}. 
For brevity, we only include a comparison of the FFs to pions here, while those for kaons and protons can be found in appendix~\ref{sec:kpsub}.
The figures are organized based on the processes in which the respective datasets are involved: hadron collisions, SIDIS, and SIA. 
Each plot illustrates FFs of different partons fragmenting into $\pi^+$. 
The momentum fraction spans from 0.01 to 1. 
In the upper panel, $zD(z)$ is presented for various subtractions, while the lower panel displays all results normalized to the baseline fit value. 
Labels indicate the removed dataset, and the colored band represents the Hessian uncertainty at 68\% confidence level for the baseline fit.
The figures reveal that the majority of the alternative fits lie within or close to the error bands of the baseline fit, suggesting no significant bias towards particular sets. 
In the case of processes in hadron collisions, the FFs to $\pi^+$ from un-favored quarks, namely $\bar u$, $s$, $c$, and $b$, are rather stable after removal of a single dataset.
That stability is expected since they are mostly constrained by datasets from SIDIS and SIA.
The ATLAS 13 TeV data on jet fragmentation and the STAR data on inclusive hadron production show a large impact on FFs from favored quarks.
After the removal of either of the three datasets, the FFs of $\pi^+$ from $u$ quark decrease for $z$ around 0.1 or smaller.
The FFs from $\bar d$ quark decrease (increase) for the removal of the ATLAS (STAR) datasets. 
For the FFs from the gluon, apart from the ATLAS 13 TeV data, the ALICE 13 TeV data on inclusive hadron production and ATLAS 5 TeV data on jet fragmentation also show strong pulls.
Both of the ATLAS 13 TeV data prefer smaller FFs from the gluon, while the ATLAS 5 TeV data prefers smaller (larger) FFs from the gluon for $z$ below (above) 0.1. 
We observe different preferences on the best-fit FFs from the ATLAS 13 and 5 TeV measurements.
However, it is important to note that the ATLAS 13 TeV data place a dominant constraint on FFs from the gluon at the large $z$ region.
After their removal, we expect much larger Hessian uncertainties on FFs from the gluon as well.
The ALICE 13 TeV data also prefer smaller FFs from the gluon for $z$ in the range 0.1$\sim$0.6.
For datasets from SIDIS, their preference for FFs to pions is mostly consistent with our best-fit of the global data.
In the dataset subtraction, we further split the 2016 COMPASS measurements into a subset on the proton to anti-proton ratios and a subset on pion and kaon production.
The 2016 COMPASS data prefer larger FFs to $\pi^+$ from the $u$ quark in the region with $z>0.2$ covered by the measurement.
The FFs from the $u$ quark at small $z$ values also increase after the removal of the 2016 COMPASS data.
The trends are opposite for the same data on FFs to $\pi^+$ from the $\bar u$ quark.
The ZEUS data on unidentified hadron production also show large pulls with a preference for larger FFs from the $\bar d$ quark and smaller FFs from the $\bar u$ quark.
The FFs from $s$, $c$, and $b$ quarks, or gluon are almost unchanged since they are not directly constrained by SIDIS data. 
Finally, the impact of SIA data is shown in Fig.~\ref{Fig:pi_sia}.
As mentioned before, the SIA data place constraints on FFs from heavy quarks either indirectly via the inclusive measurements or directly from the heavy-flavor tagged measurements.  
Especially, the $c$-tagged measurement from SLD is unique in the separation of FFs from $c$ and $s$ quarks.
We thus further split the SLD data into a subset for $c$ and $b$-tagged measurements and a subset for the rest from SLD. 
Notably, the SLD $c\&b$ subset exerts a substantial impact on the FFs to pions from $c$ and $s$ quarks. 
Their removal leads to a large increase in the FFs from the $c$ quark and an associated decrease in the FFs from the $s$ quark.
Due to the lack of constraints on flavor separation, the fits become unstable and have large uncertainties.   
The FFs from the $b$ quark are less affected because of the constraints placed by a similar measurement from DELPHI.
The rest subset from SLD also show significant pulls on FFs from the $\bar d$ and $s$ quarks due to the large number of data points and high precision.
The remaining dataset from SIA is very much consistent with the global data.

\subsection{Kinematic cuts and theoretical uncertainties}

\begin{figure}[h!]
\centering
\includegraphics[width=0.9\textwidth]{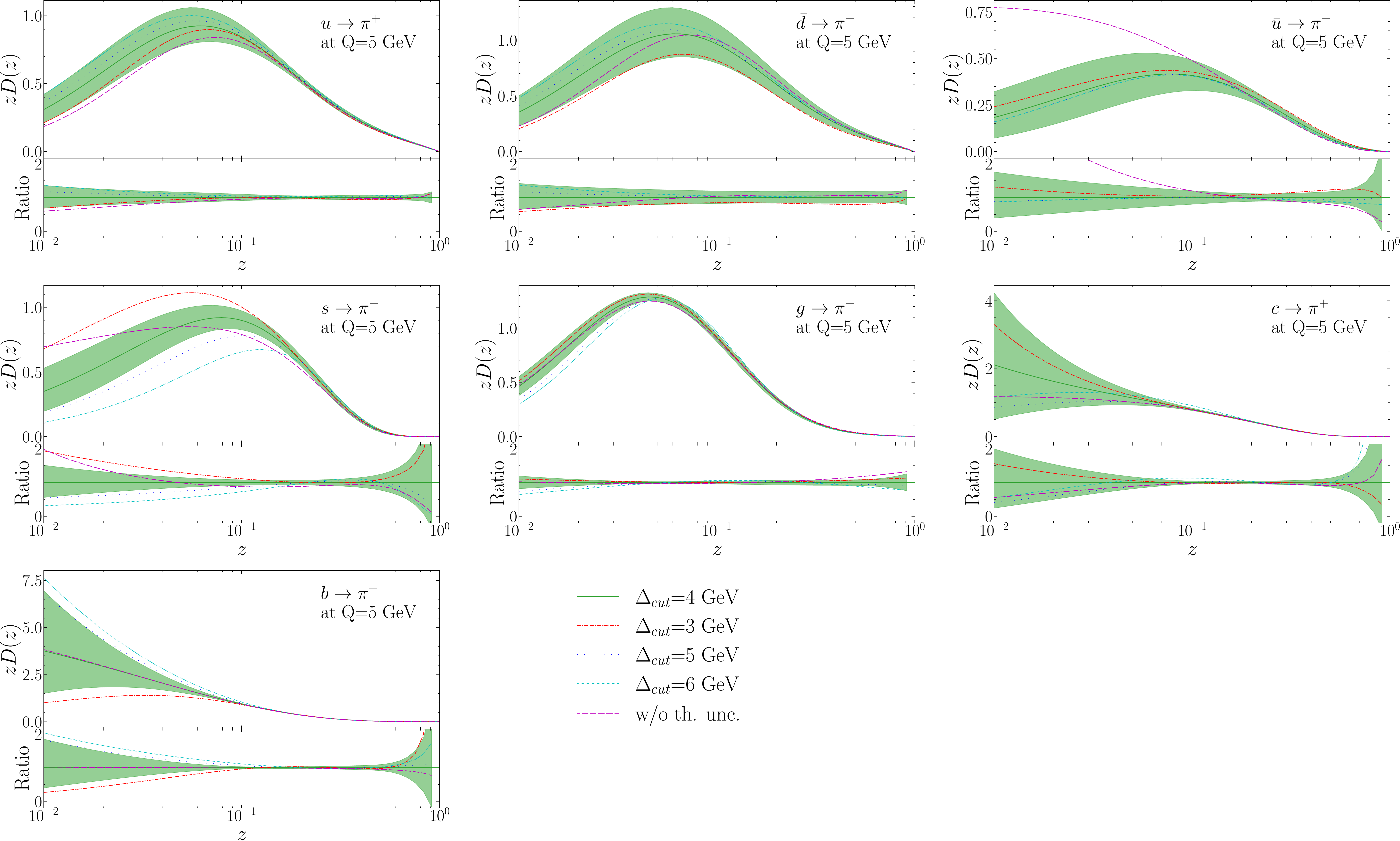}
  \caption{
  Comparison of the global fit for the $\pi^+$ fragmentation function with different choices of $p_T$ cut. The colored bands indicate the 68\% confidence level Hessian uncertainty of this work ($p_T$ = 4 GeV).
  }
  \label{Fig:pt_pi}
\end{figure}

\begin{figure}[h!]
\centering
\includegraphics[width=0.9\textwidth]{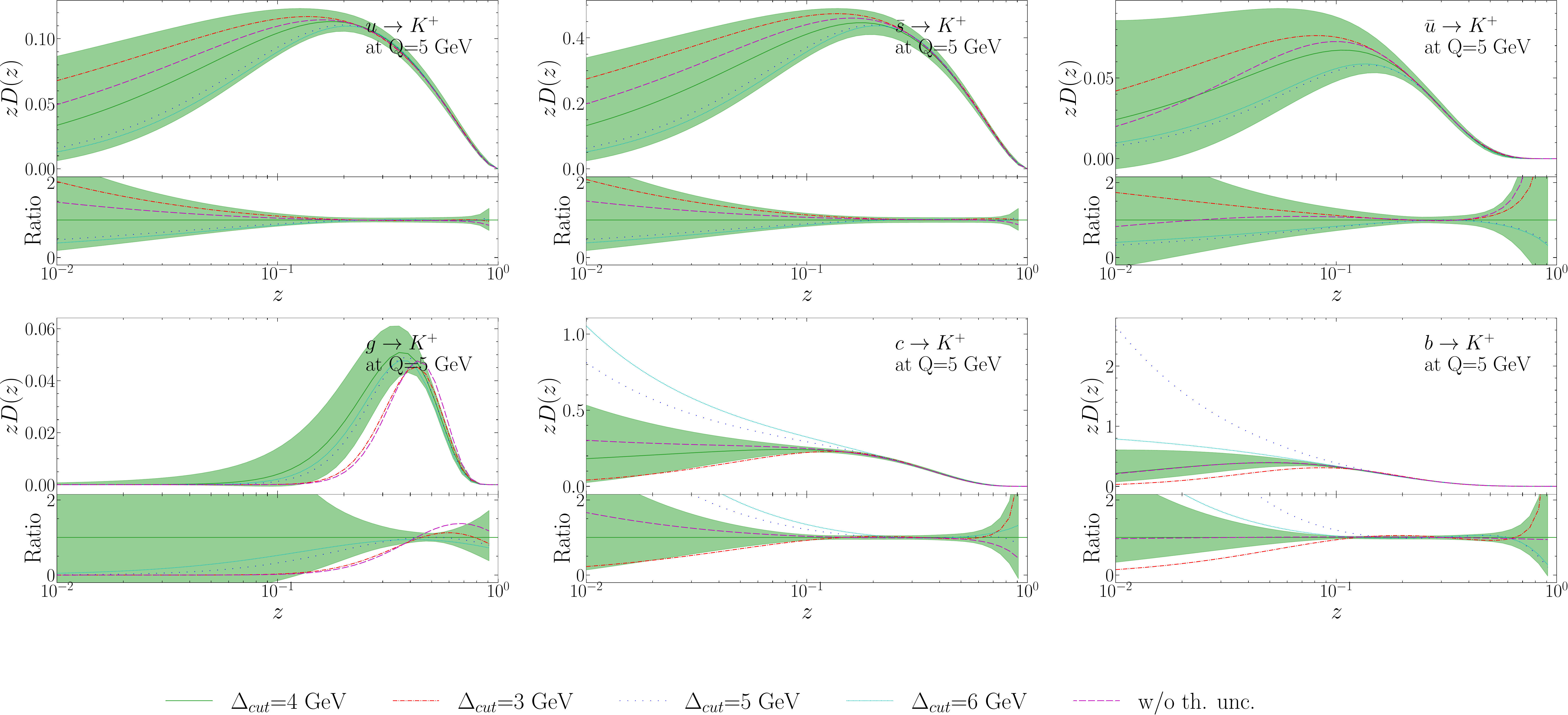}
  \caption{
  Similar to Fig. \ref{Fig:pt_pi} but for $K^+$ fragmentation function.
  }
  \label{Fig:pt_ka}
\end{figure}

\begin{figure}[h!]
\centering
\includegraphics[width=0.9\textwidth]{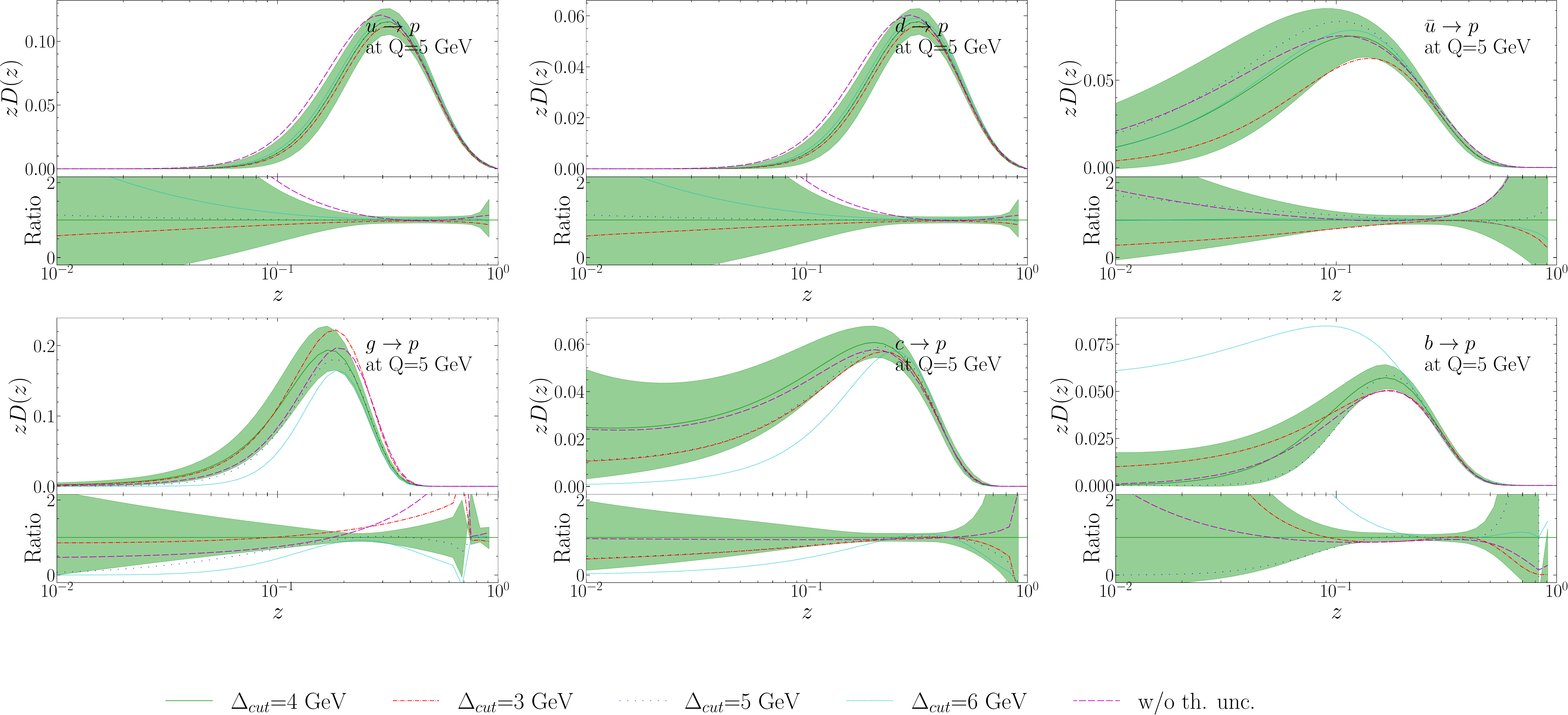}
  \caption{
  Similar to Fig. \ref{Fig:pt_pi} but for $p$ fragmentation function.
  }
  \label{Fig:pt_pr}
\end{figure}

In our nominal fit, we apply a uniform kinematic cut of $4$~GeV on either the transverse momentum or energy of the hadrons. 
We perform alternative fits with cutoff choices of $3$, $5$, and $6$~GeV to investigate potential bias by our default choice.
Furthermore, we conduct alternative fits without including the theoretical uncertainties.
All resulting FFs are compared to the baseline fit and summarized in Figs.~\ref{Fig:pt_pi}-\ref{Fig:pt_pr} for FFs to $\pi^+$, $K^+$ and $p$, respectively.

We begin with the FFs to $\pi^+$ shown in Fig.~\ref{Fig:pt_pi}.
The FFs from $u$, $\bar d$ and $b$ quarks to $\pi^+$ increase monotonically with the increasing cutoff for most of the $z$ values shown.
Conversely, the FFs from gluon and other quarks exhibit the opposite trend.
The variations are generally within the Hessian uncertainties of the baseline fit, except for the FFs at small $z$ values from gluon, $s$, and $b$ quarks.
For $s$ and $b$ quarks, the large variations are mostly due to the shrink of kinematic coverage of the heavy-flavor tagged measurements.
Excluding theoretical uncertainties can also have a certain impact, especially for the $\bar u$ quark, but this would significantly increase the $\chi^2$ of the global data.
On the other hand, the FFs from $u$, $\bar s$, and $\bar u$ quarks to $K^+$ decrease monotonically with the increasing cutoff for most of the $z$ values shown.
The variations in FFs to $K^+$ due to the kinematic cutoff are well within uncertainties for gluon and light quarks.
The variations in FFs to $p$ due to the kinematic cutoff are small, considering the large Hessian uncertainties for gluon and light quarks.
The FFs to $K^+$ or $p$ from $c$ and $b$ quarks are again unstable for the same reason as for FFs to pions.
The impact of the theoretical uncertainties is generally within Hessian uncertainties for FFs to kaons and protons.
\newpage

\clearpage

\section{Discussion and Conclusions}
\label{sec:summary}

Determining various non-perturbative inputs of QCD is essential for precision programs at the LHC and upcoming Electron-Ion colliders, as well as for understanding QCD confinement. 
For the simplest form that includes the unpolarized collinear fragmentation functions and parton distribution functions within the framework of QCD factorization.
We present a joint determination of fragmentation functions to light charged hadrons from global analysis at next-to-leading order in QCD (NPC23), including estimations of uncertainties.
We find very good agreement between our best-fit predictions and various measurements of SIA, SIDIS and hadron collisions after careful selection of the kinematics of the measurements. 
A variety of precision measurements on fragmentation from the LHC have been included in the global analysis and have led to significant constraints on FFs especially of gluons.  
Our work introduces several advances, apart from the new data and selections of kinematics.
Our joint determination allows a flexible parametrization form while taking into account correlations between experimental measurements and theoretical predictions on different hadron species.
We are able to include theoretical uncertainties as estimated with residual scale variations into the analysis due to the development of associated theoretical tools and improvements in the efficiency of the calculations. 
Furthermore, measurements on jet fragmentation at the LHC have been included in the global analysis of FFs on light charged hadrons for the first time.
A tolerance condition is applied in the estimation of the Hessian uncertainties based on the investigation of the agreement of theoretical predictions to individual datasets.   
The FFs to charged pions are well constrained in general for momentum fractions of hadrons $z$ from 0.01 to 0.8, especially for gluon and constituent quarks. 
The FFs to charged kaons and protons are also determined with reasonable precision for $z$ between 0.1 and 0.7 for most flavors of partons.
We calculate the total momentum fraction of each parton carried by various charged hadrons as functions of the lower limit on the momentum fractions, which are relevant for testing the momentum sum rule once FFs to neutral hadrons can be determined with similar precision. 
We also conduct a series of alternative fits to investigate pulls from each dataset in the global analysis and the impact of various choices of parameters in the global analysis.
The extracted FFs are found to be stable, with variations within Hessian uncertainties of our nominal fit in general. 
Comparing our results with previous determinations, as shown in appendix~\ref{sec:dsscom}, we find significant differences, especially in the fragmentation functions to kaons and protons.
Discrepancies are also observed for the fragmentation functions of non-constituent quarks and gluon to charged pions.
In the future, benchmark exercises involving different groups of global analysis will be needed for clarification on the differences observed.
The grids for NPC23 FFs in LHAPDF6 format are publicly available~\footnote{\url{https://fmnlo.sjtu.edu.cn/\~fmnlo/data/NPC23_CH.tar.gz}}, with details given in appendix~\ref{sec:lha6}.
There are several important observations and questions raised by our analysis of the experimental measurements.
So far, we have not distinguished between the experimental definitions of unidentified charged hadrons and charged tracks.
In order to maximize available experimental data, we simply assume both of them equal to the sum of the three light charged hadrons, which in principle should be a good approximation.
However, the exact differences between them are usually not explained and quantified in publications of experimental analyses.
Even for the measurements of identified charged hadron production, we notice that their exact definition can still differ between different experiments.
For instance, contributions from decays of short-lived hadrons, e.g., $K^0_S$ and $\Lambda$, are excluded in the ALICE measurements but very likely not for all the SIA measurements.
This inconsistency between experiments can have an impact and also lead to ambiguities in the extracted FFs of strange quarks to charged pions and to protons. 
We hope that all of these valid points can be clarified in future experimental analyses.
Additionally, we noticed significant impact of the heavy-flavor tagged measurements on quark flavor separation of FFs.
More measurements of such kind, in addition to the only few from SLD and DELPHI, would be desirable.
Further more, we find that the LHC measurements on jet fragmentation have a wide kinematic coverage and great potential for further improvements. 
We would encourage future LHC analyses dedicated to identified hadrons or distinguishing the sign of charges of unidentified hadrons. 
We demonstrate the broad applications of the NPC23 FFs when combined with the FMNLO program.
For instance, we calculate the average jet charges for QCD inclusive dijet production at the LHC and find good agreement with the experimental measurements.
Additionally, we present predictions on $Z$-tagged jet fragmentation in $PbPb$ collisions at the LHC while neglecting final state medium effects, with more information given in appendix~\ref{sec:addcom}.
Meanwhile, we have published version {\tt 2.0} of the {\tt FMNLO} program~\footnote{\url{https://fmnlo.sjtu.edu.cn/\~fmnlo/data/FMNLOv2.0.tar.gz}}, which now includes a module for calculations of hadron production in SIDIS at next-to-leading order in QCD.
Details of the module and benchmark comparisons to existing results are also provided in appendix~\ref{sec:fmnlo}.
All theoretical predictions used in this analysis, including those for hadron collisions, SIDIS or SIA, can be reproduced with this program. 
Upon the completion of NPC23 for charged hadrons, an immediate follow-up study would be a global analysis of FFs to neutral hadrons.
In a longer term, we are working on implementation of hard coefficient functions at next-to-next-to-leading order in QCD in the FMNLO program, especially given recent progresses on calculations of SIDIS coefficient functions~\cite{Goyal:2023xfi,Bonino:2024qbh}.
This will pave the way towards a global analysis of FFs at NNLO in QCD. 

\quad \\
\noindent \textbf{Acknowledgments.} 
We thank Ignacio Borsa for providing the latest DSS
pion fragmentation function grids. The work of J. G. is
supported by the National Natural Science Foundation
of China (NSFC) under Grant No. 12275173 and open
fund of Key Laboratory of Atomic and Subatomic
Structure and Quantum Control (Ministry of Education).
H. X. is supported by the NSFC under Grant No. 12475139
and No. 12035007, and by the Guangdong Major
Project of Basic and Applied Basic Research Grant
No. 2020B0301030008 and No. 2022A1515010683. Y. Z.
is supported by the NSFC under Grant No. U2032105 and
CAS Project for Young Scientists in Basic Research Grant
No. YSBR-117. X. S. is supported by the Helmholtz-
OCPC Postdoctoral Exchange Program under Grant
No. ZD2022004.

\appendix

\section{Comparison to other groups}
\label{sec:dsscom}

\begin{figure}[htbp]
    \centering
    \includegraphics[width=0.83\linewidth]{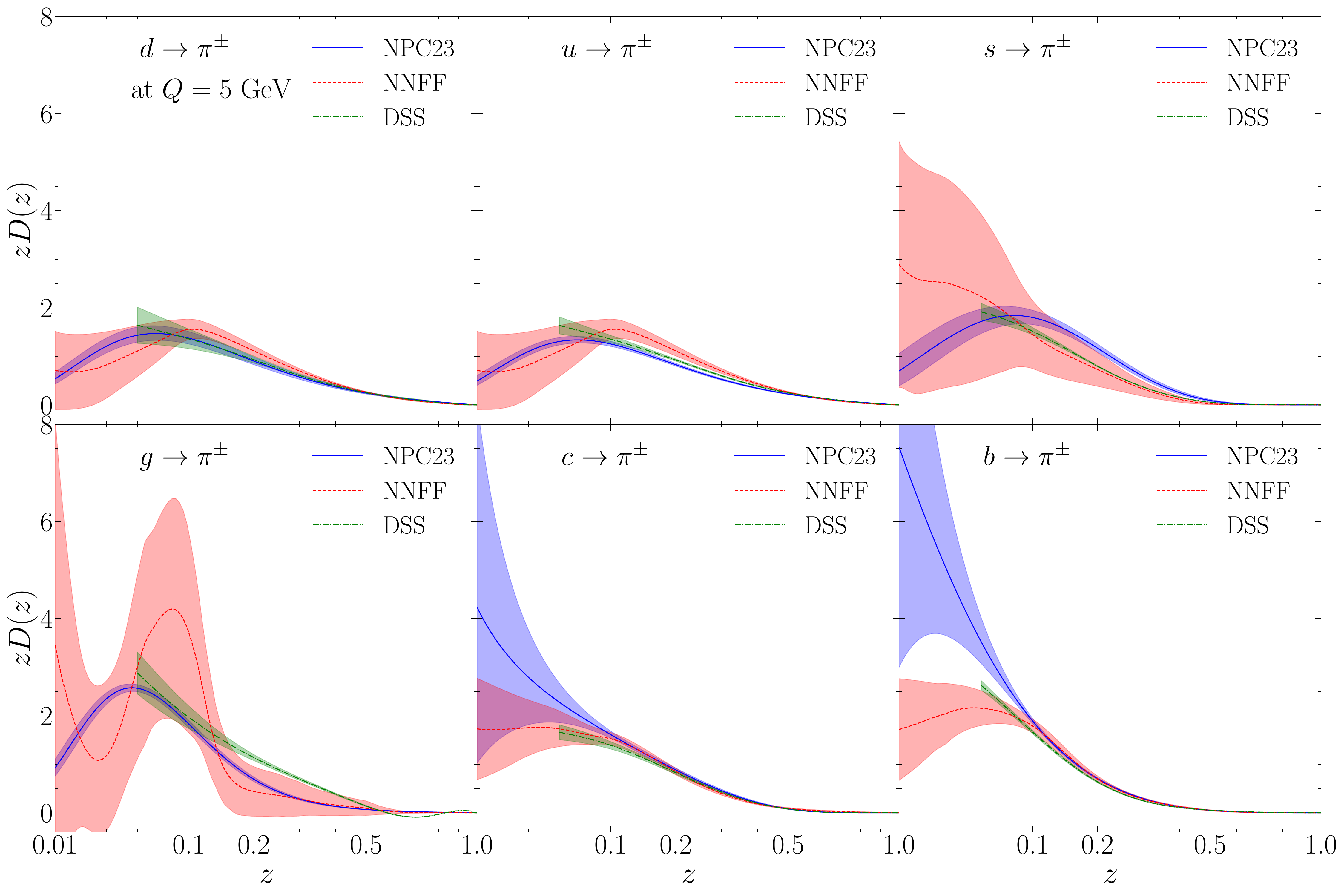}
    \caption{
 	Comparison of our NLO fragmentation functions to pion with those from NNFF and DSS at $Q=5$~GeV.
The DSS results are calculated by using 
 DSS21 sets~\cite{Borsa:2021ran}. 
         The NNFF results are from NNFF1.0~\cite{Bertone:2017tyb}.
    }
    \label{fig:ffsum_pi}
\end{figure}

\begin{figure}[htbp]
    \centering
    \includegraphics[width=0.83\linewidth]{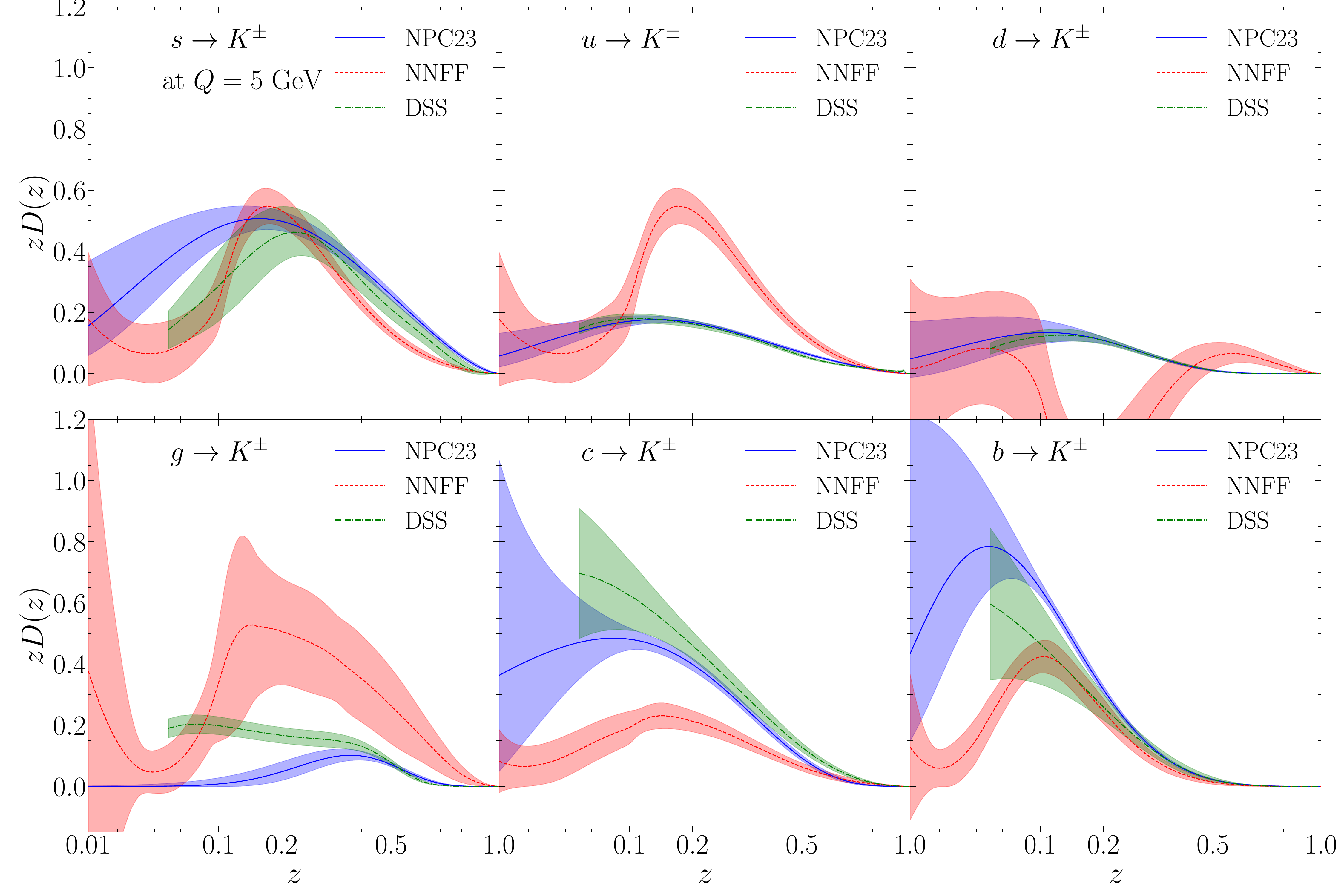}
    \caption{Same as Fig.~\ref{fig:ffsum_pi} but for FFs to $K^\pm$.
    The DSS results are calculated by using DSS17 sets \cite{deFlorian:2017lwf}.}
    \label{fig:ffsum_ka}
\end{figure}

\begin{figure}[htbp]
    \centering
    \includegraphics[width=0.83\linewidth]{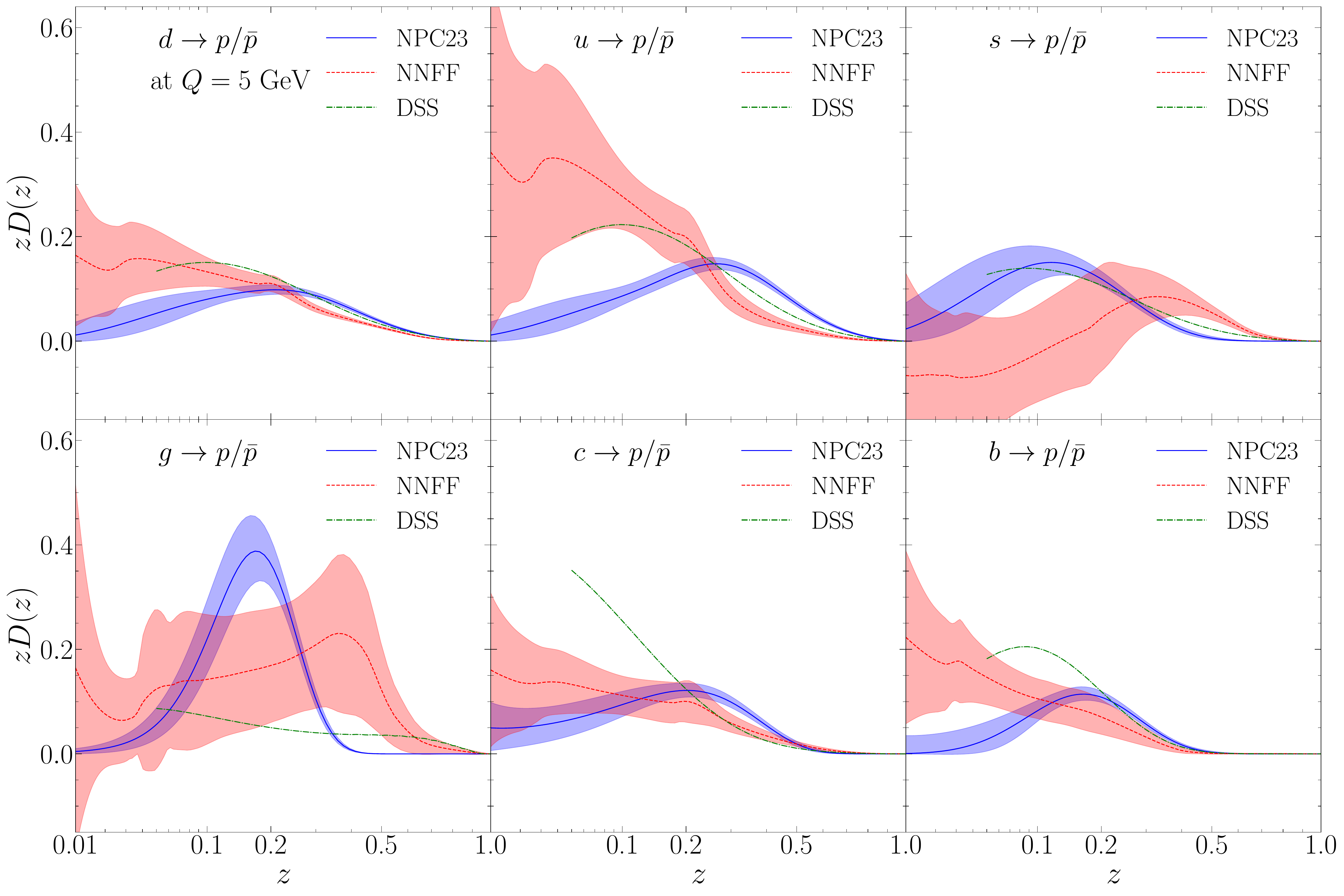}
    \caption{Same as Fig.~\ref{fig:ffsum_pi} but for FFs to $p/\bar p$.
    The DSS results are calculated by using DSS07 sets~\cite{deFlorian:2007ekg}.}
    \label{fig:ffsum_pr}
\end{figure}

In this section, we present a comparative analysis of our fit results with those fragmentation functions from the DSS, NNFF, JAM and MAP groups.
The comparisons of NLO fragmentation functions at 5 GeV summed over charges are depicted in Figs.~\ref{fig:ffsum_pi}-\ref{fig:ffsum_pr}.
To illustrate the difference between fragmentation from favored and
unfavored quarks, we also show the FFs from different groups for positively charged hadrons in Figs.~\ref{fig:ffplus_pi} and~\ref{fig:ffplus_ka}.

The DSS fits, sourced from DSS21 \cite{Borsa:2021ran}, DSS17 \cite{deFlorian:2017lwf}, DSS14\cite{deFlorian:2014xna},and DSS07 \cite{deFlorian:2007ekg}, provide fragmentation functions for $\pi^\pm, K^\pm$ and $p/\bar p$. 
Note that DSS only publishes FFs with momentum fraction $z>0.05$. For the uncertainties, the DSS21 release includes Monte Carlo uncertainties, while the DSS17 and DSS14 release incorporate Hessian uncertainties.
Correspondingly, the NNFF sets used are from NNFF1.0 \cite{Bertone:2017tyb}.
For NNFF sets, the uncertainties are estimated by the Monte Carlo method with FF replicas.
For JAM\cite{Moffat:2021dji} and MAP\cite{Khalek:2021gxf} FF sets, uncertainties are estimated with Monte Carlo method as well.
The error band of NPC23 is estimated using the Hessian method at a 68\% confidence level. 
As evident from Figs.~\ref{fig:ffsum_pi}-\ref{fig:ffsum_pr}, the uncertainties of NPC23 are notably smaller than those from NNFF due to the inclusion of a large variety of LHC data as well as the SIDIS data, which impose significant constraints on the FFs.

\subsection{Fragmentation to $\pi^\pm$}
For fragmentation of pions, NNFF assumes isospin symmetry, resulting in coinciding results for $\pi^{\pm}$ from $u$ and $d$ quarks. 
Both DSS and our fit allow for different normalization of the two constituent quarks, and our results show a moderate violation of isospin symmetry, though within uncertainties. 
Overall, our results agree well with those of DSS for constituent quarks to pions FFs but differ with those of NNFF, especially in high-$z$ regions, where the differences can be larger than the uncertainties.
The NNFF determination exhibits notably larger uncertainties compared to both NPC23 and DSS, particularly in the gluon and strange quark channels at low-$z$ region. Both DSS and NPC23 show relatively smaller and comparable uncertainty bands, suggesting more constrained determinations.
For $s$ quark fragmentation, FFs from NNFF and DSS are close, while central value of NPC23 is much larger, with a discrepancy of about 50\%$\sim$ 60\%. 
This difference becomes less significant when uncertainties are taken into account.
In low-$z$ regions, our result and NNFF show different trends, with NPC23 showing an increasing-then-decreasing pattern and NNFF displaying a continuously decreasing trend.
For gluon fragmentation, FFs differ significantly in low-$z$ regions, especially where NNFF shows large oscillations with a substantial uncertainty that even dips below zero.
However, at $z > 0.3$, NNFF and NPC23 agree well, while DSS exhibits larger values.
For heavy quark fragmentations ($c, b$), FFs from different fits agree well for $z > 0.1$ and differ notably at low- $z$ regions, accompanied by large uncertainties.

\subsection{Fragmentation to $K^\pm$}

Fragmentation to $K^{\pm}$ from the $u$ quark is similarly described by fits from NPC23 and DSS, while NNFF displays a large discrepancy. The error bands of NPC23 and DSS overlap in intermediate-$z$ region ($0.1<z<0.7$).
Note that NNFF assumes symmetry between $D_u^{K^\pm}$ and $D_s^{K^\pm}$.
Fragmentation from the $s$ quark slightly deviates, but NPC23 and DSS show a similar trend with a peak in the middle, and the peak locations are close.
For $u$ quark fragmentation to $K^{\pm}$, DSS and NPC23 analyses show agreement in most regions. 
In contrast, NNFF diverges significantly in intermediate-$z$ regions, displaying a much larger peak in the middle.
For fragmentation from the $d$ quark, NPC23 aligns well with DSS and the error bands encompass DSS curve in most regions.
NPC23 and NNFF are close in the low-$z$ region ($z < 0.05$), after which NNFF starts oscillating and even becomes negative. 
For gluon fragmentation, DSS and NPC23 agree at $z \sim 0.5$, with significant discrepancies in other regions, and NNFF consistently shows a large discrepancy.
For fragmentation from $c$ quark, the three results are quite different, with DSS being the largest, NPC23 in between, and NNFF being the smallest. But there are still slight overlaps between NPC23 and DSS in region $z<0.4$.
For fragmentation from $b$ quark, results from three analyses are close and the error bands overlap in region $z>0.2$. But in low-$z$ region discrepancies persist. 

\subsection{Fragmentation to $p/\bar p$}

For fragmentation to $p/\bar p$ from the $d$ quark, the three results are close at $z > 0.2$ , but error bands of NPC23 and NNFF do not overlap.
For fragmentation from $u, s$ quarks and $g$, the three results differ significantly in all regions.
For fragmentation from the $c$ quark, DSS and NNFF are close at $z > 0.2$, while NPC23 is much larger. 
For fragmentation from the $b$ quark, NPC23 and NNFF are close at $z > 0.2$, with NNFF differing significantly.
Overall, fragmentation functions to $p/\bar p$ from different fits show the least agreement.
\\

\begin{figure}[htbp]
    \centering
    \includegraphics[width=0.83\linewidth]{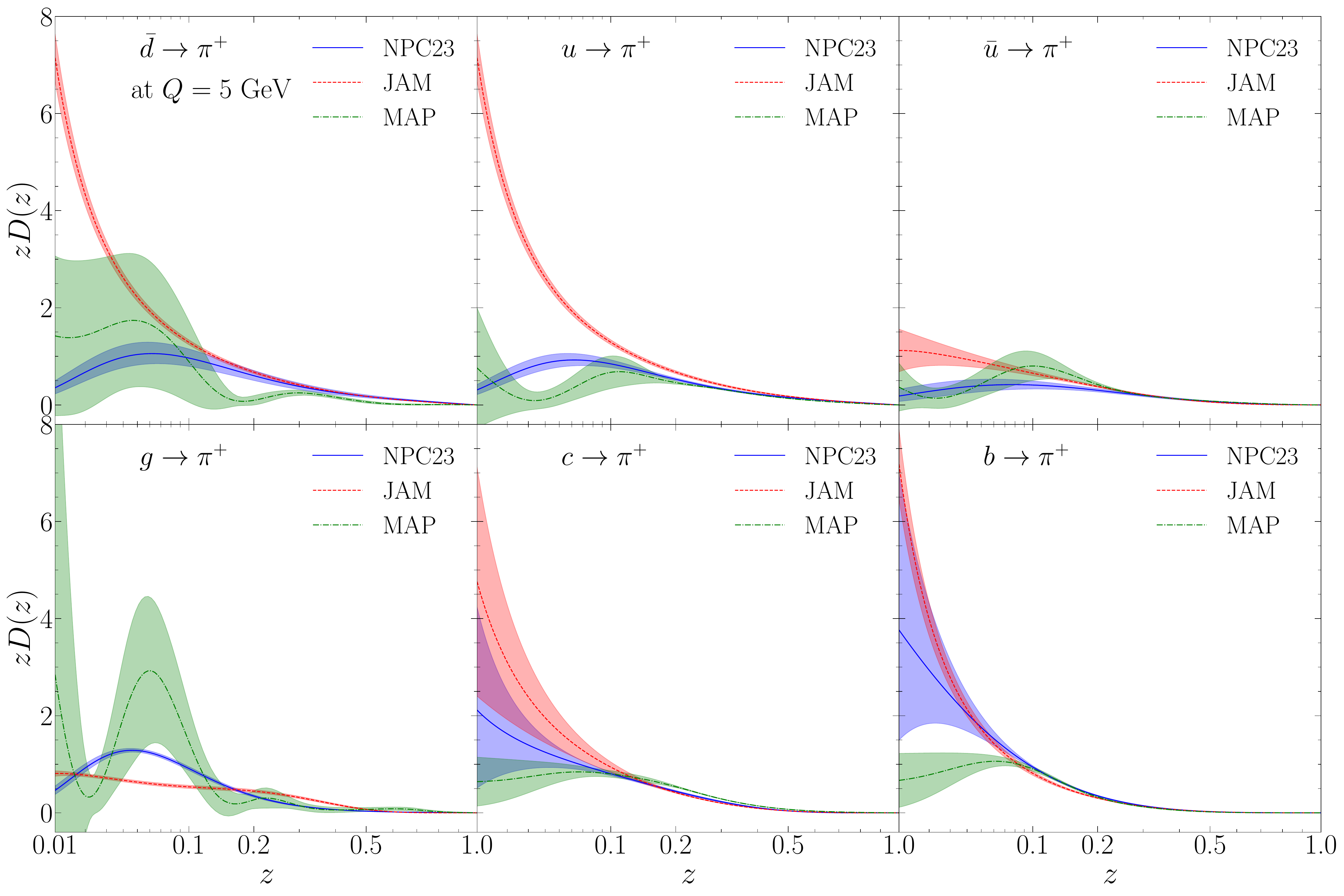}
    \caption{Same as Fig.~\ref{fig:ffsum_pi} but for FFs to $\pi^+$. The JAM20\cite{Moffat:2021dji} and MAPFF10\cite{Khalek:2021gxf} sets are used respectively.}
    \label{fig:ffplus_pi}
\end{figure}

\begin{figure}[htbp]
    \centering
    \includegraphics[width=0.83\linewidth]{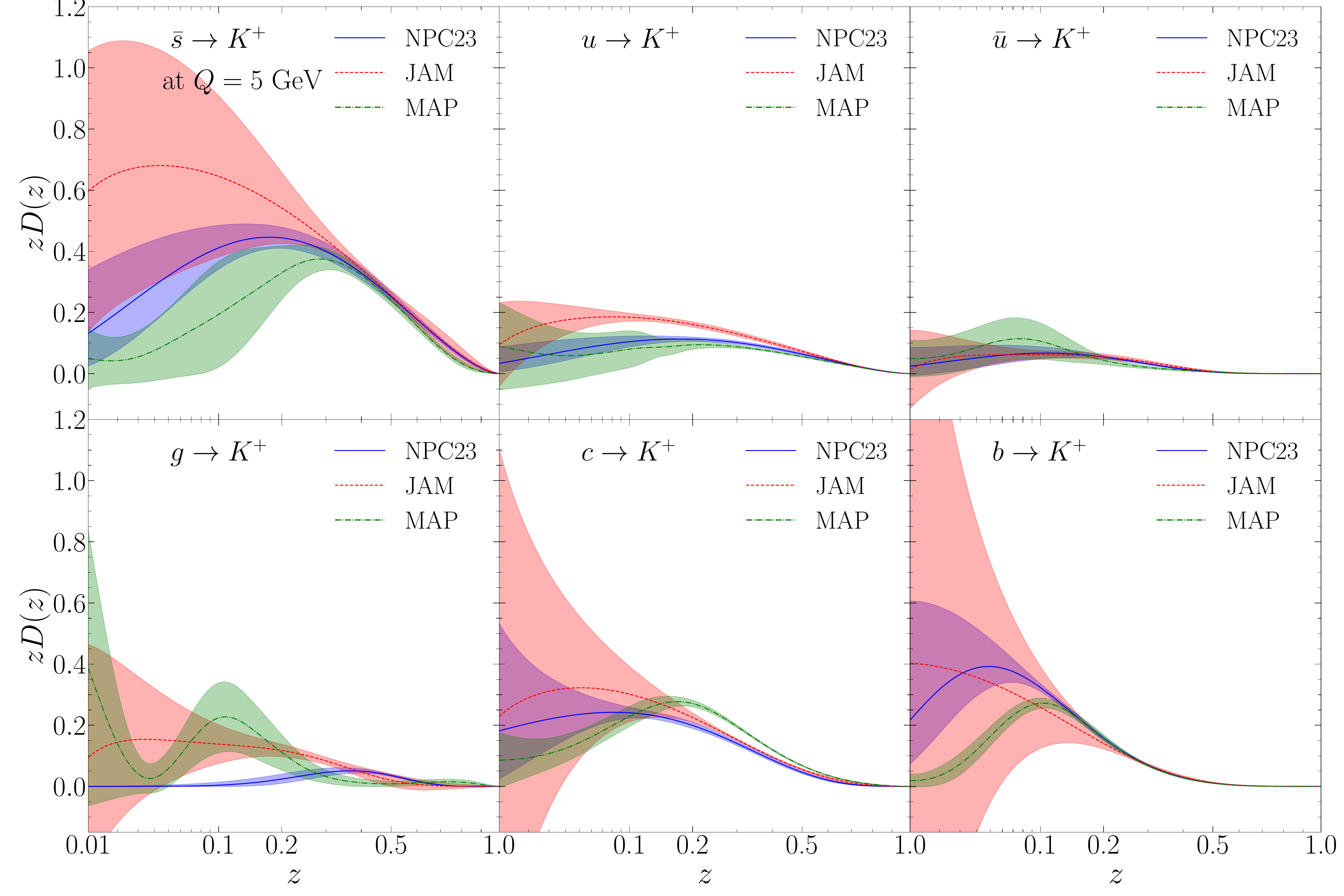}
    \caption{Same as Fig.~\ref{fig:ffsum_pi} but for FFs to $K^+$. The JAM20\cite{Moffat:2021dji} and MAPFF10\cite{Khalek:2021gxf} sets are used respectively.}
    \label{fig:ffplus_ka}
\end{figure}

\subsection{Fragmentation to $\pi^+$, $K^+$}

In Figs.~\ref{fig:ffplus_pi} and~\ref{fig:ffplus_ka} a comparative analysis of fragmentation functions to positively charged hadrons is presented. 
The fragmentation functions derived from NPC23, JAM~\cite{Moffat:2021dji}, and MAP~\cite{Khalek:2021gxf} analyses exhibit varying patterns of agreement and tension across different parton flavors for both $\pi^+$ and $K^+$ at $Q = 5$ GeV. 
For favored quarks, JAM consistently predicts higher fragmentation probabilities compared to both NPC23 and MAP, with this feature being particularly pronounced in the pion sector. 
In contrast, the unfavored quark channels show better agreement among the three analyses, especially notable in the $\bar u\to K^+$ fragmentation. 
The gluon fragmentation functions demonstrate significant differences across all analyses for both mesons in terms of shape and magnitude, with MAP showing more pronounced fluctuations in the intermediate-$z$ region. 
For heavy quarks, all three analyses converge well within their uncertainties, particularly in the high-$z$ region, although some tension persists in the low-$z$ domain where JAM typically exhibits larger uncertainty bands. 
While the analyses show reasonable agreement in the high-$z$ region across most parton species, significant differences in both central values and uncertainty estimates remain in the low-$z$ region.

\section{Dataset subtraction for $K^{\pm}$ and $p/\bar p$}
\label{sec:kpsub}
\begin{figure}[h!]
\centering
\includegraphics[width=0.9\textwidth]{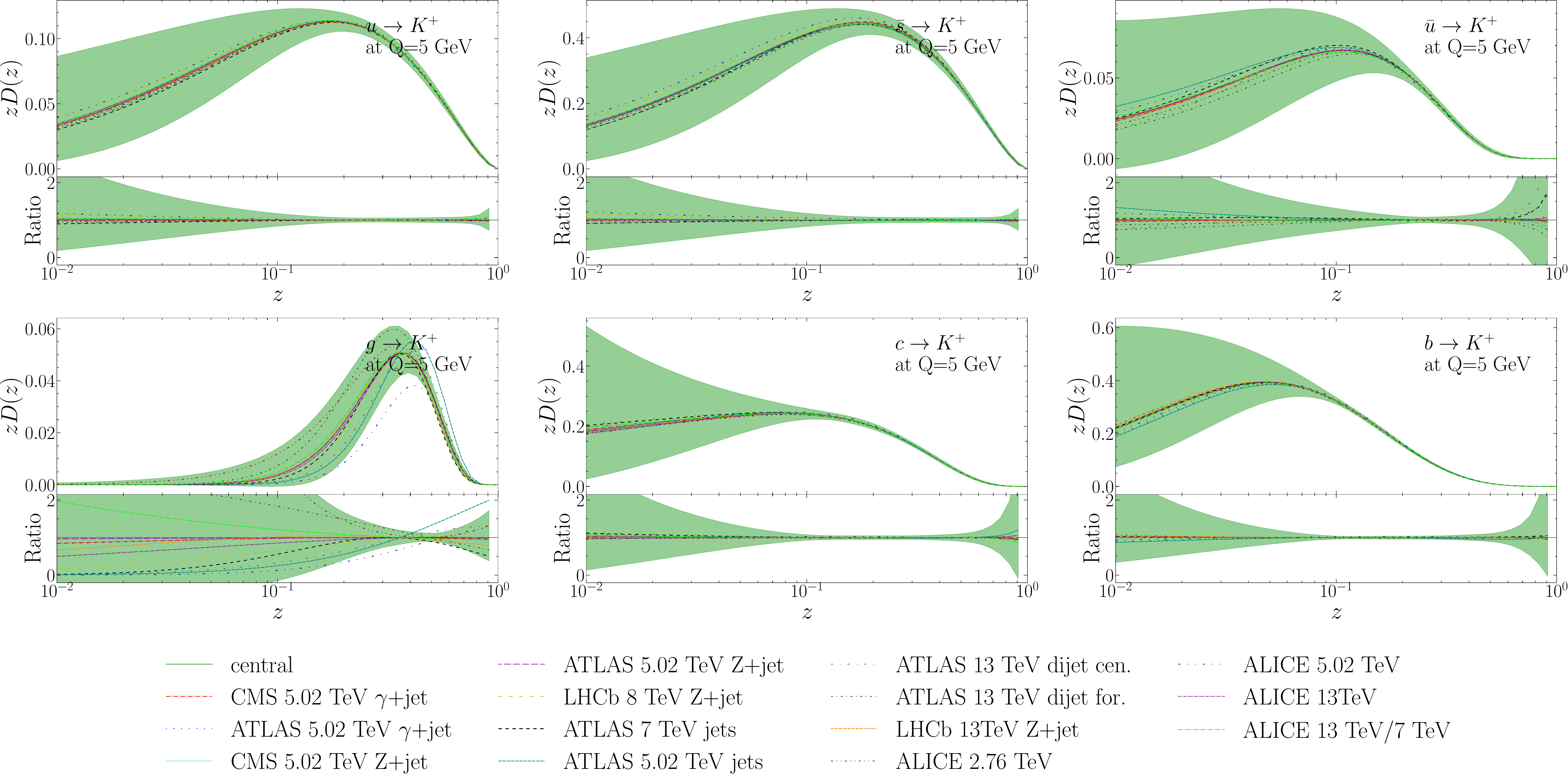}
\caption{
  Similar to Fig. \ref{Fig:pi_pp} but for FF of $K^+$ and subtractions from $pp$ collisions.
}
\label{Fig:ka_pp}
\end{figure}

\begin{figure}[h!]
\centering
\includegraphics[width=0.9\textwidth]{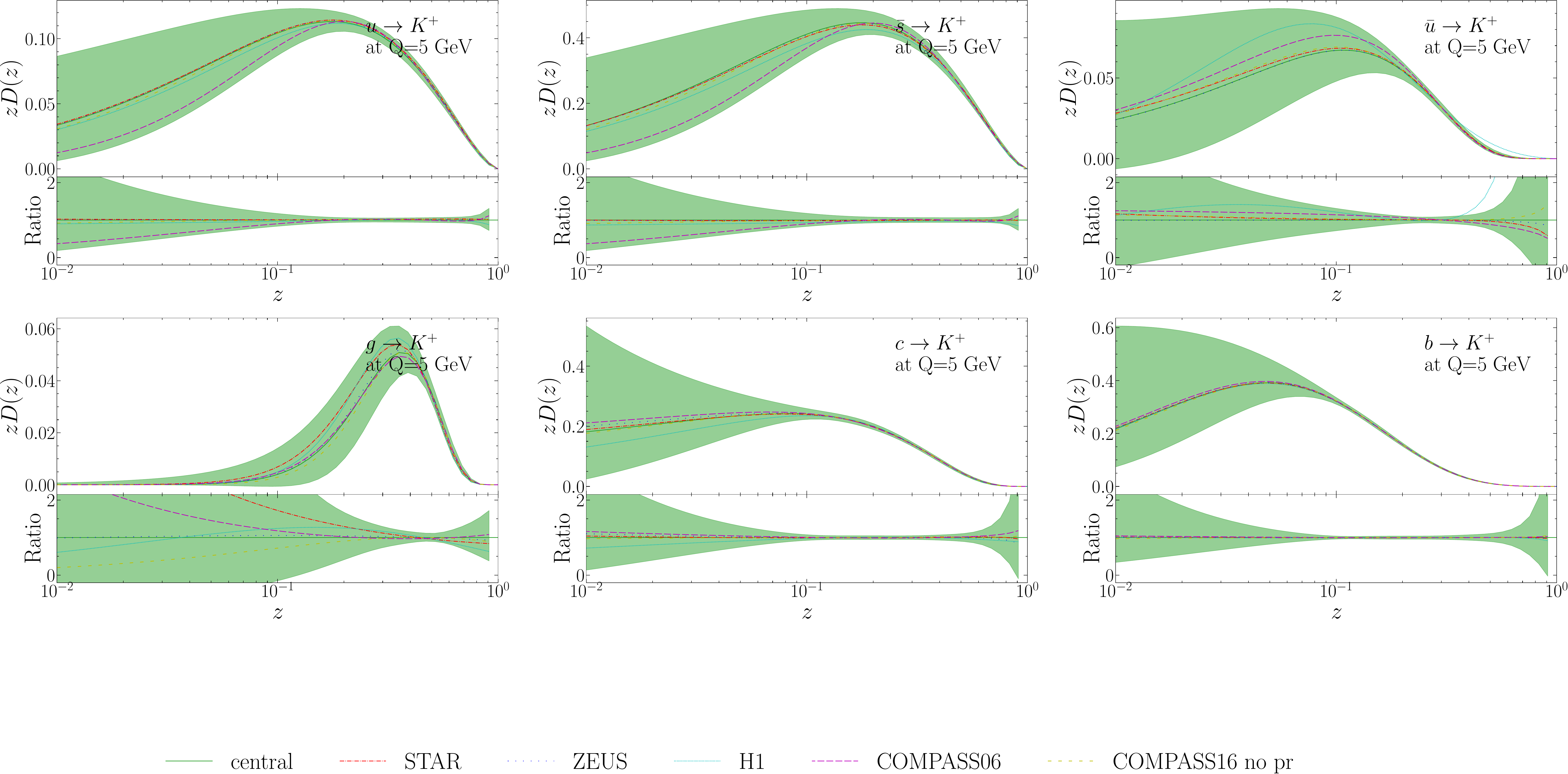}
\caption{
  Similar to Fig. \ref{Fig:pi_pp} but for FF of $K^+$ and subtractions from SIDIS processes.
}
\label{Fig:ka_sidis}
\end{figure}

\begin{figure}[h!]
\centering
\includegraphics[width=0.9\textwidth]{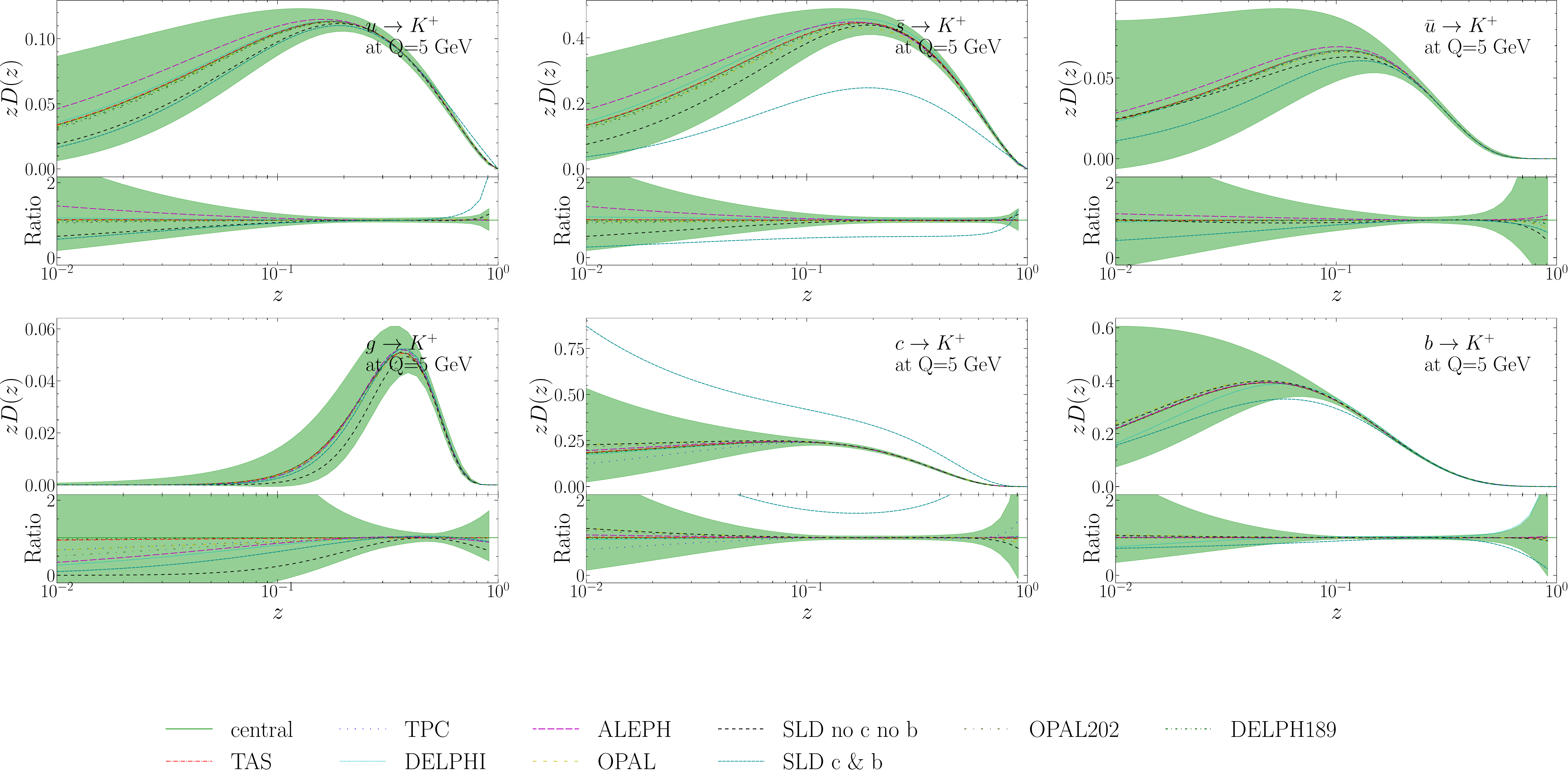}
\caption{
  Similar to Fig. \ref{Fig:pi_pp} but for FF of $K^+$ and subtractions from SIA processes.
}
\label{Fig:ka_sia}
\end{figure}

\begin{figure}[h!]
\centering
\includegraphics[width=0.9\textwidth]{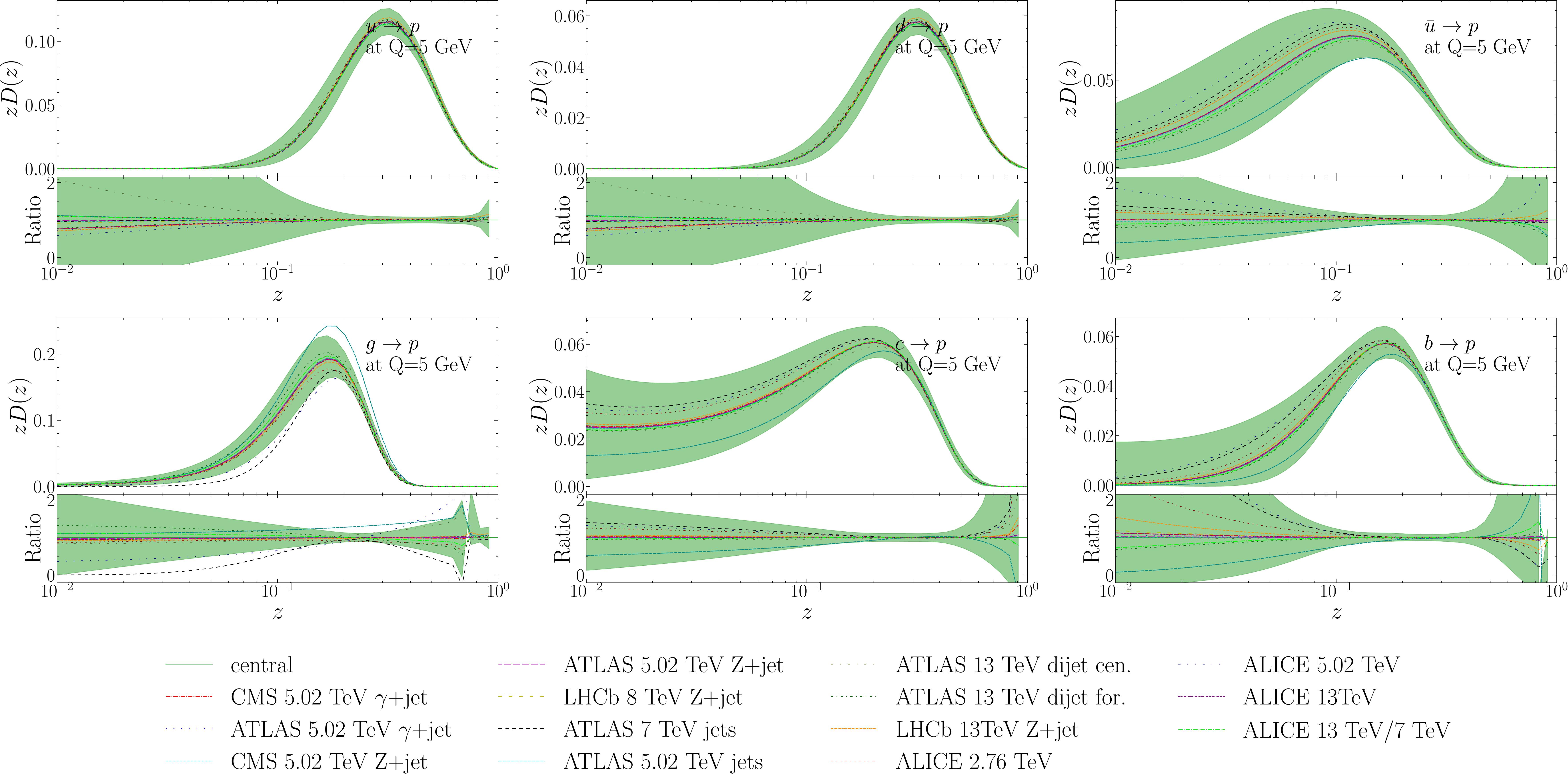}
\caption{
  Similar to Fig. \ref{Fig:pi_pp} but for FF of $p$ and subtractions from $pp$ collisions.
}
\label{Fig:pr_pp}
\end{figure}

\begin{figure}[h!]
\centering
\includegraphics[width=0.9\textwidth]{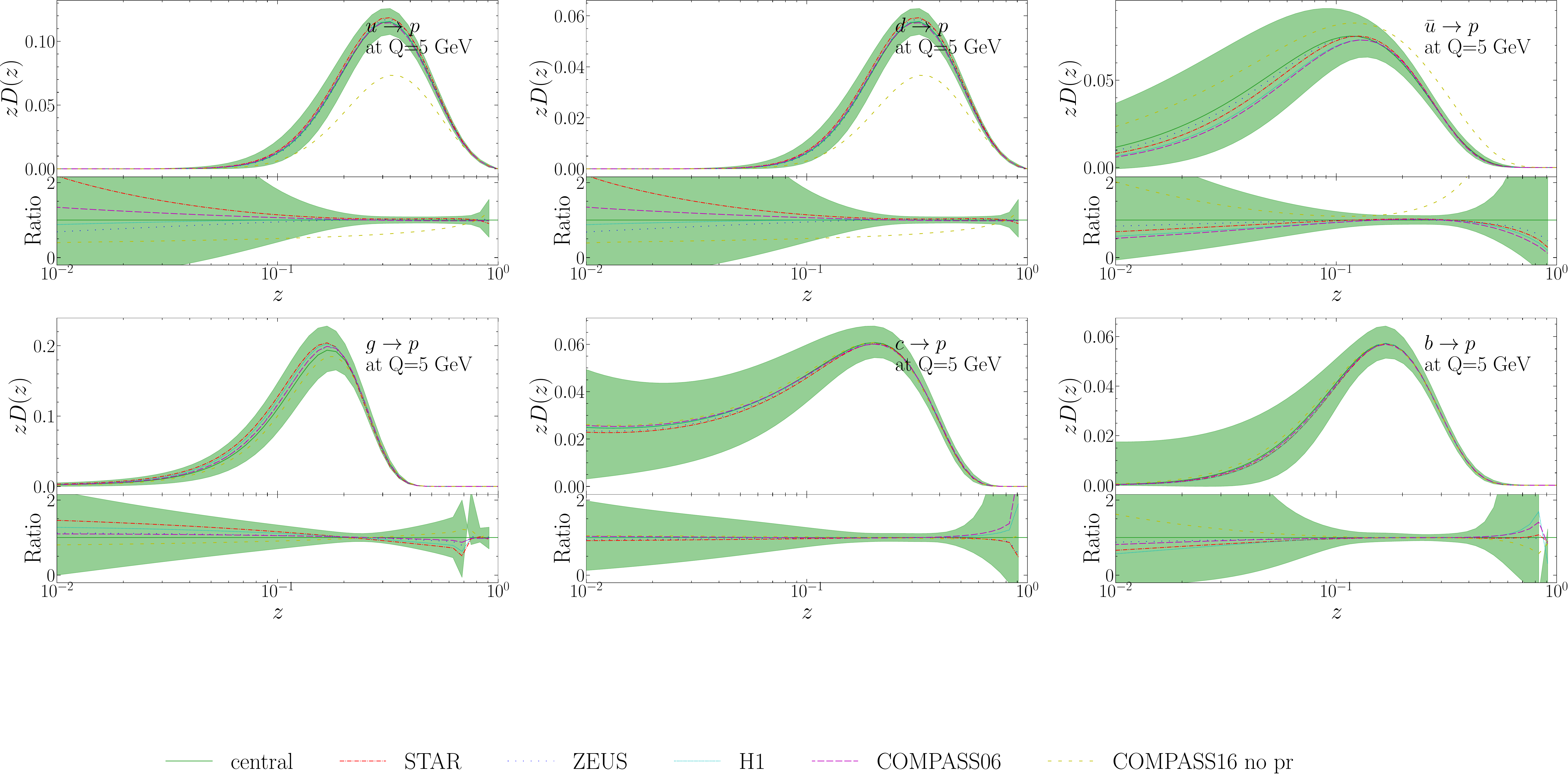}
\caption{
  Similar to Fig. \ref{Fig:pi_pp} but for FF of $p$ and subtractions from SIDIS processes.
}
\label{Fig:pr_sidis}
\end{figure}

\begin{figure}[h!]
\centering
\includegraphics[width=0.9\textwidth]{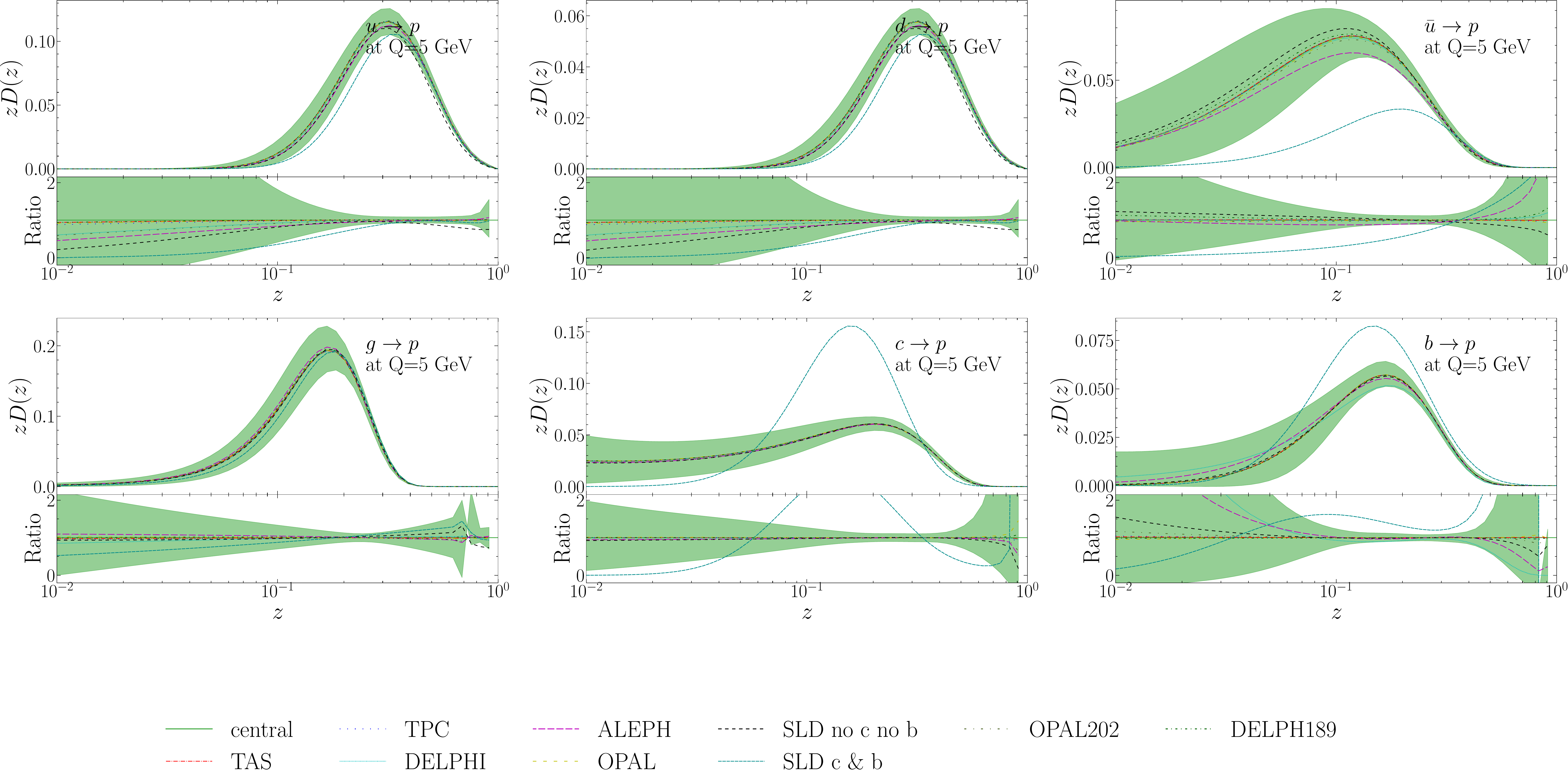}
\caption{
  Similar to Fig. \ref{Fig:pi_pp} but for FF of $p$ and subtractions from SIA processes.
}
\label{Fig:pr_sia}
\end{figure}

In this section, we continue the description of the effects of data subtraction on FFs.
The comparison between alternative fits and the baseline fit for kaons is depicted in Figs.~\ref{Fig:ka_pp}-\ref{Fig:ka_sia}, and for protons in Figs.~\ref{Fig:pr_pp}-\ref{Fig:pr_sia}.
The figures, labels, observables, etc., are organized similarly to those for pions.
For subtractions of datasets from $pp$ collisions, FFs to $K^+$ closely align with the baseline fit except for FFs from gluons.
After removal of the ALICE 13 TeV dataset, the gluon FF goes slightly outside the error band for $0.2<z < 0.5$. 
In the large $z$ region ($>0.4$), after removal of the ATLAS 13 TeV dijets data the gluon FF increases considerably similar to that for pions, since they contain most data points and provides dominant constraints for gluon FF.
In the case of SIDIS datasets, for FFs to kaons from favored quarks, only the COMPASS16 dataset shows a pull for $z$ at 0.1 or below which is well within the uncertainties. 
After removal of COMPASS06 data, FFs from $\bar u$ quark increase significantly at $z>0.4$ as the SIDIS data provides dominant constraints for fragmentation from unfavored quarks.
The impact of subtraction of SIDIS data are negligible for kaon FFs from gluon and heavy quarks as expected.
In the case of SIA processes, the most significant pulls are from heavy flavor tagged data similar to the case for pions discussed earlier. 
Notably, after removal of SLD $c\&b$ dataset FFs to kaons from $c$ quark increase largely and become almost unconstrained.
As a result FFs from $s$ quarks decrease accordingly in order to maintain description of the inclusive SIA data.
Subtraction of other SIA datasets only lead to small variations within uncertainties.
For FFs to protons, in the case of $pp$ collisions, after dataset removal, variations for FFs from quarks are within uncertainties in general. 
Variations due to subtraction of ATLAS 13 TeV dijet data are slightly larger, since they contain most data points. 
For FFs from gluon, variations for ATLAS 5.02 TeV jet data, 13 TeV dijet data, or ALICE 13 TeV data can fall outside the error bands. 
After removal of ATLAS 5.02 TeV jet data, FFs are lowered. 
But for ALICE 13 TeV data, ratios to baseline fit continuously increase, from less than 1 in small $z$ region to larger than 1, and the turning point is between 0.2 and 0.3. 
For removal of SIDIS datasets, FFs from alternative fits lie very close with the exception of COMPASS16 proton to antiproton ratio dataset. 
It lowers the FFs from $u$ and $d$ quarks while raising the FFs from $\bar u$ quark, and the variations gradually fall outside of the error bands after $z > 0.1$. 
For SIA datasets, SLD $c\&b$ dataset still have the strongest impact. 
After its removal, the FFs from $c$ and $b$ quarks raise significantly.
Meanwhile, FFs from $u$ and $d$ quarks decrease in the full $z$ region, while FFs from $\bar u$ quark decrease for $z < 0.3$ and increase for $z > 0.4$. .

\section{Additional comparison to data}
\label{sec:addcom}

\begin{figure}[htbp]
  \centering
  \includegraphics[width=0.47\textwidth]{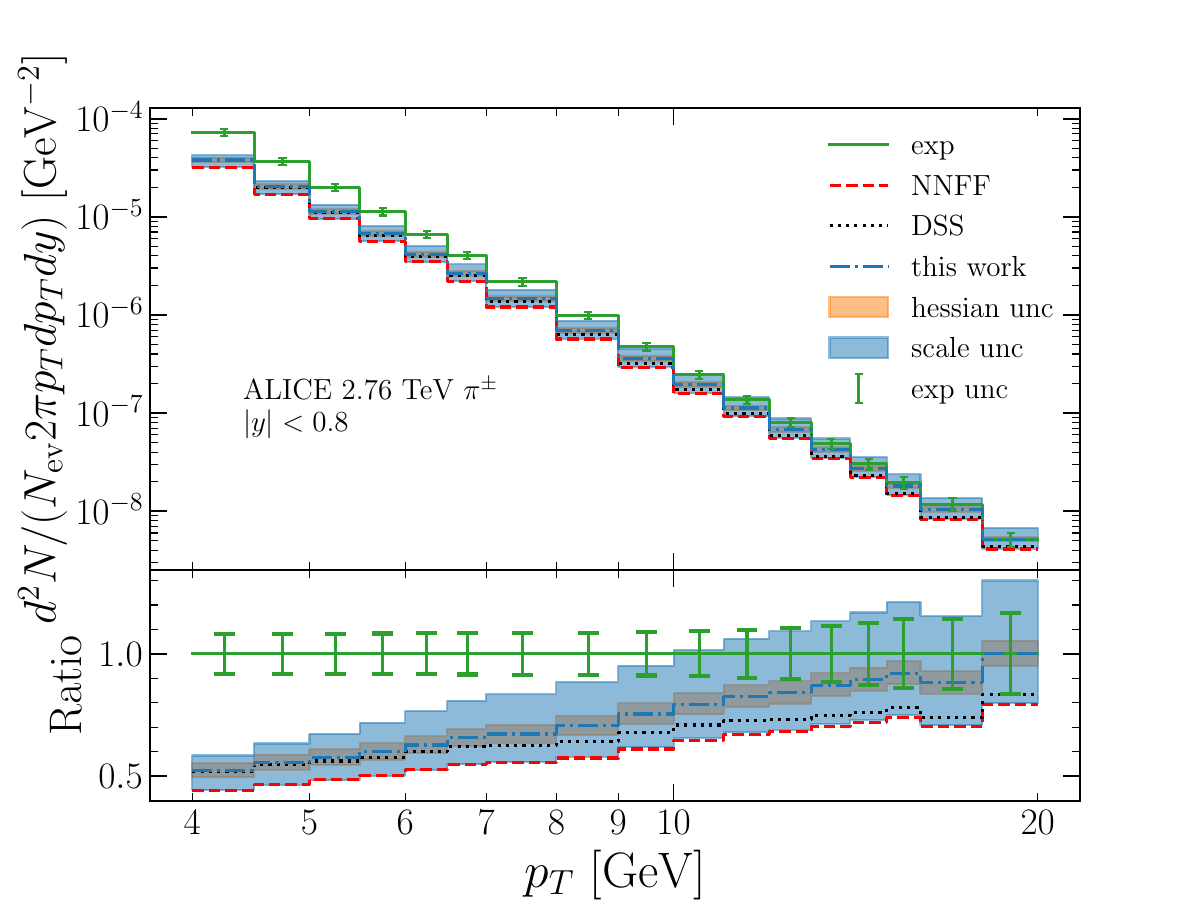}
    \includegraphics[width=0.47\textwidth]{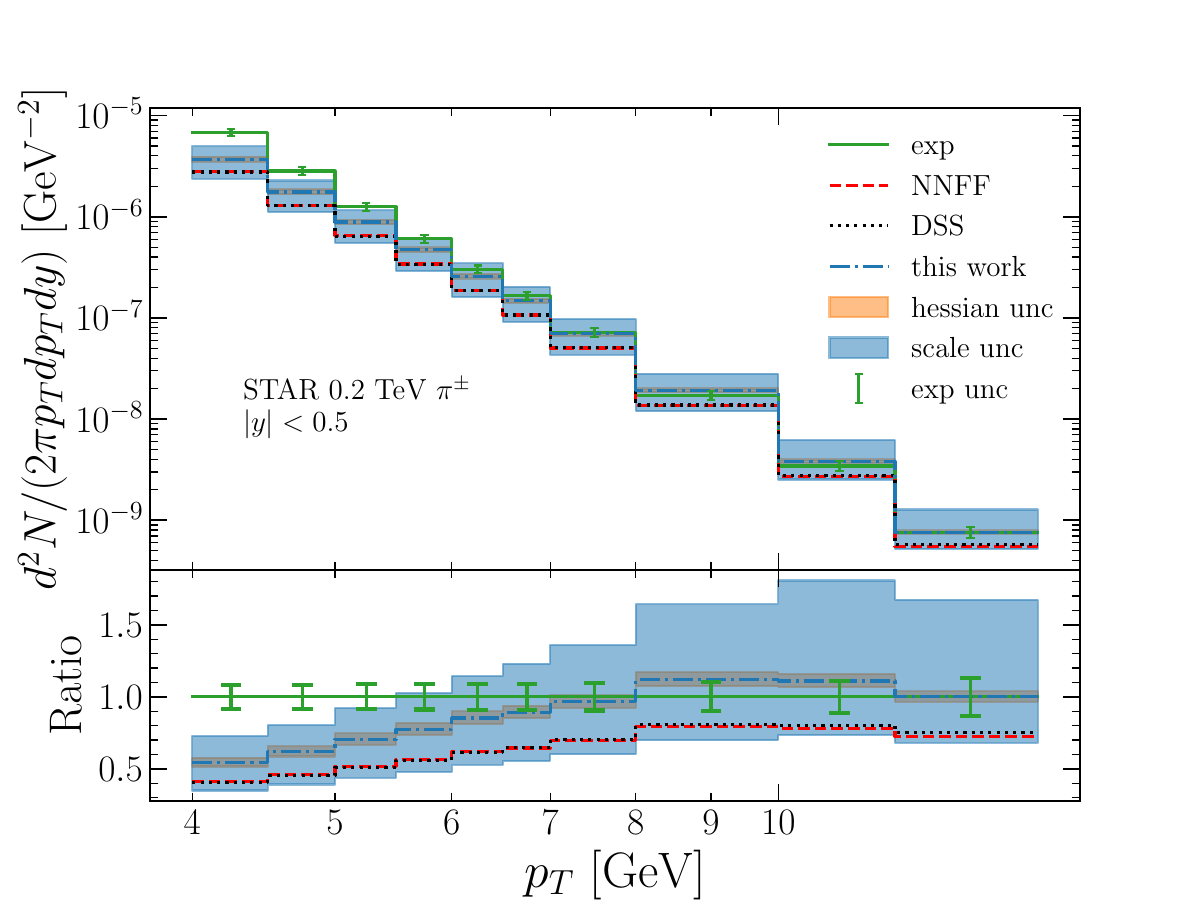}
	\caption{
 Comparison of predictions from various fragmentation functions (FFs) for $\pi^\pm$ and experimental data at ALICE with a center of mass energy of 2.76 TeV. The theoretical results are renormalized by a factor to align with the experimental data in the last bin. In the lower panel, the ratio to experimental data is presented. The green bars represent experimental errors, while the light and dark green bands indicate scale uncertainties and Hessian uncertainties, respectively.
 }
  \label{Fig:alice_2t}
\end{figure}
\begin{figure}[htbp]
  \centering
  \includegraphics[width=0.83\textwidth]{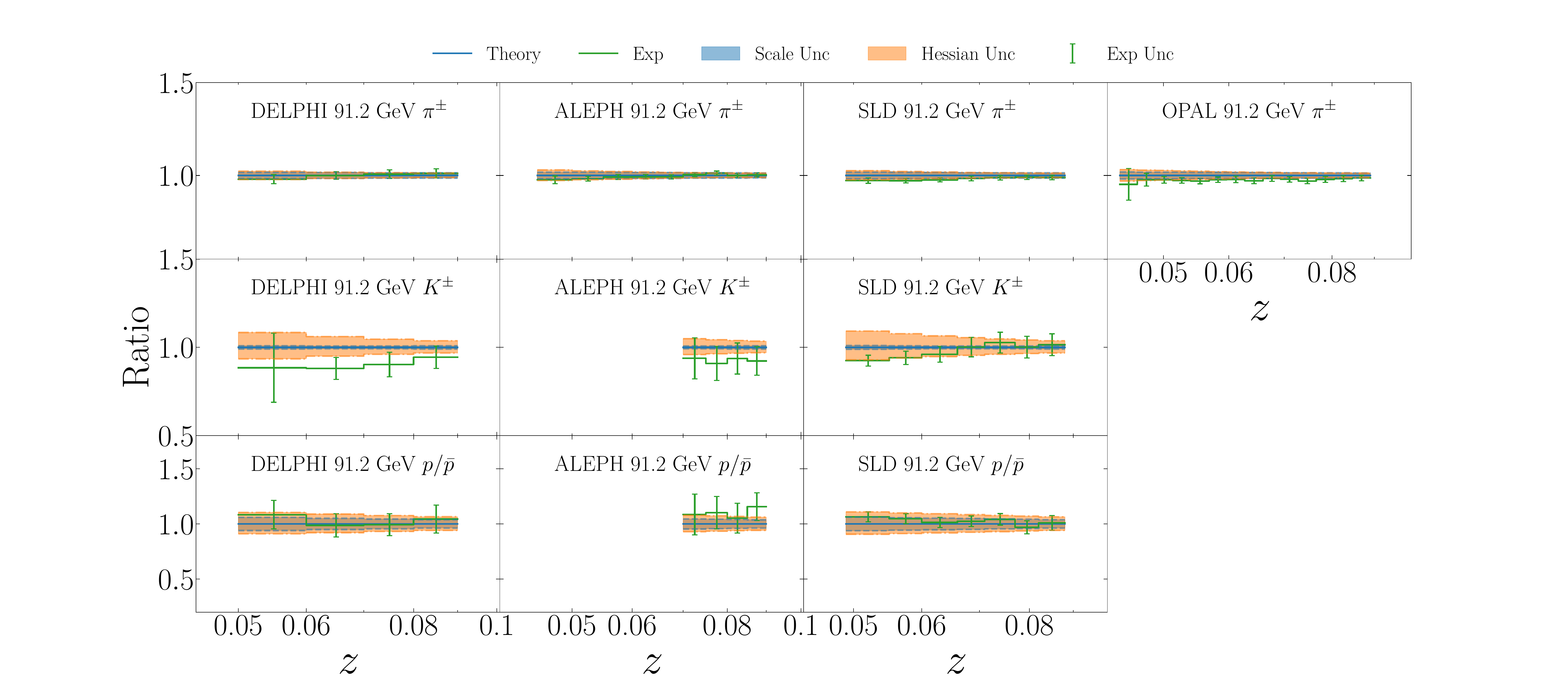}
	\caption{
 NLO predictions on cross sections of single inclusive hadron production in SIA comparing to various measurements at $Z$-pole from DELPHI, ALEPH, SLD and OPAL, for kinematic region with $z\sim 0.5-0.088$.
	}
  \label{Fig:siasmallz}
\end{figure}
\begin{figure}[htbp]
  \centering
  \includegraphics[width=0.83\textwidth]{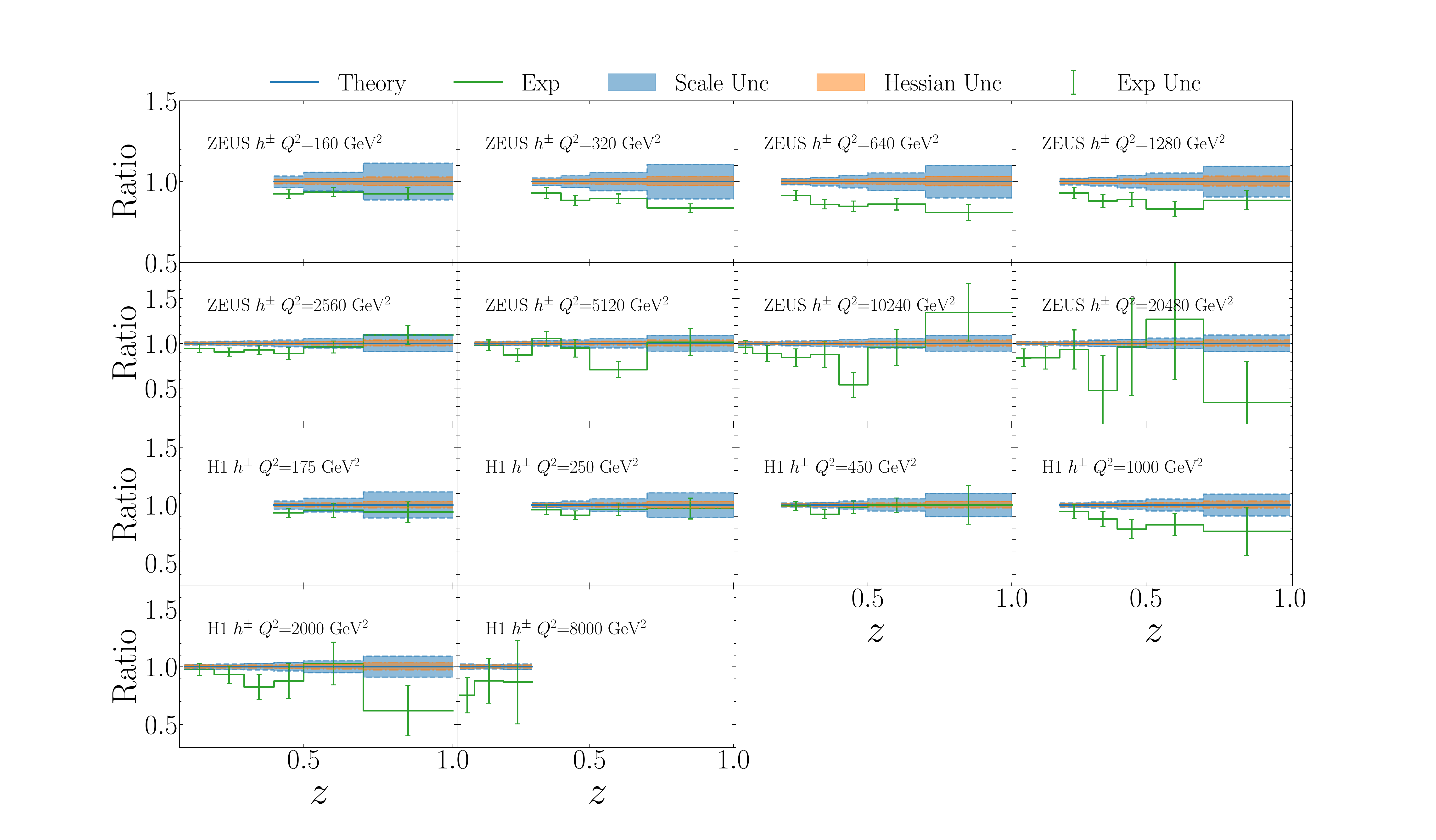}
	\caption{
 NLO predictions on cross sections of single inclusive hadron production in SIDIS comparing to various measurements from H1 and ZEUS, for kinematic region with $E_h>2$ GeV.
	}
  \label{Fig:sidissmallz}
\end{figure}
\begin{figure}[htbp]
  \centering
  \includegraphics[width=0.47\textwidth]{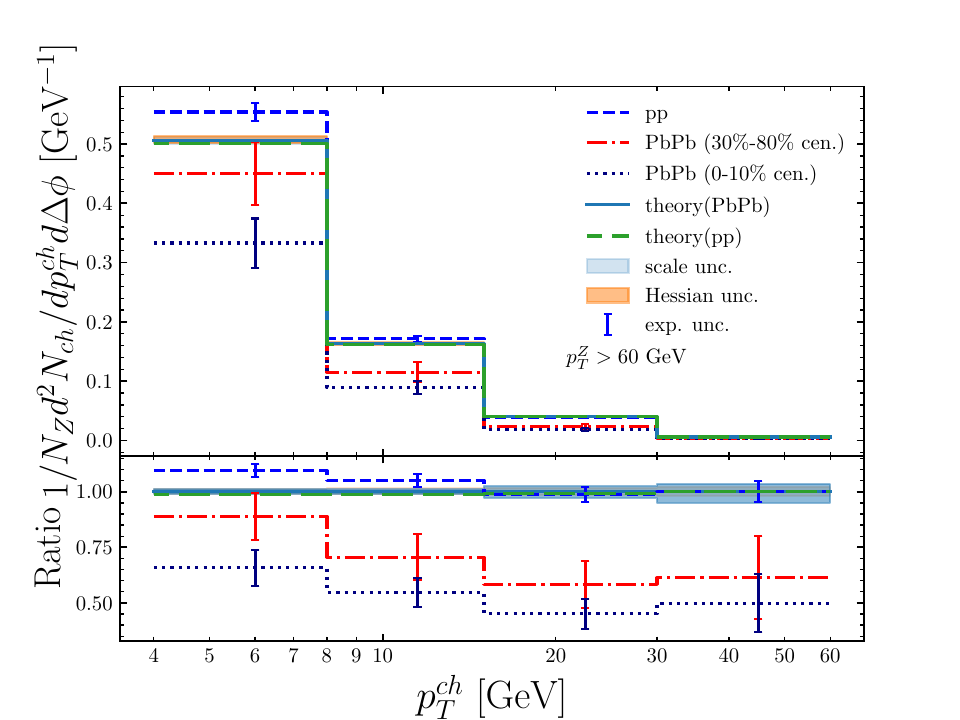}
  \includegraphics[width=0.47\textwidth]{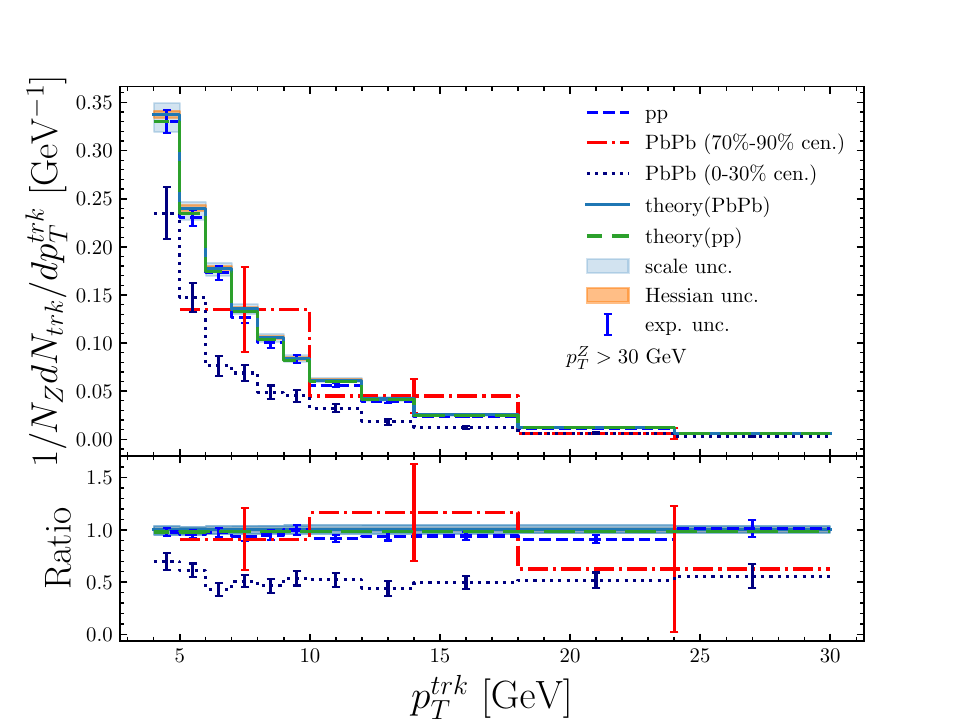}
	\caption{
 Comparison of theoretical predictions for $pp$ and $PbPb$ collisions with experimental data on $pp$ collisions and $PbPb$ collisions at more central and more peripheral regions. 
 Experimental uncertainties are represented by error bars. 
 The scale uncertainties and Hessian uncertainties are represented by shaded bands obtained similarly to those in Fig. \ref{Fig:atlas57}.
	}
  \label{Fig:pbpb}
\end{figure}

Apart from the datasets included in our global analysis, we also provide predictions and compare them to additional data which are either not directly used or excluded by our kinematic selections.
In Fig.~\ref{Fig:alice_2t} we show a comparison of our predictions to the inclusive charged pion production cross sections measured at ALICE 2.76 TeV and STAR.
Note that we only include ratios of identified charged hadron cross sections in our global analysis rather than the cross sections themselves. 
In the comparison we have normalized our predictions by an overall factor to align with data of the highest $p_T$ due to the unknown normalization appearing in denominator of the measured cross sections.
Our predictions undershoot the data at low $p_T\sim 4$ GeV by almost 50\% which is a common feature observed in predictions from all existing FFs as shown here for predictions from DSS~\cite{deFlorian:2007ekg} and NNFFs~\cite{Bertone:2017tyb}.
This gap could be filled by contributions from production mechanisms other than the hard scattering processes described by FFs.
In Fig.~\ref{Fig:siasmallz} we show comparison of our predictions to those measurements from SIA at $Z$-pole that are excluded by our kinematic selections, with $z\sim 0.05-0.088$.
We find very good agreement for all charged hadrons down to $z=0.05$ even though the data are not included in our fit.
That indicates the QCD factorization and perturbative calculations are still valid for SIA data with $z\sim 0.05$.
Similarly in Fig.~\ref{Fig:sidissmallz} we summarize comparison of our predictions to measurements from H1 and ZEUS that are excluded by our kinematic selections, with $E_h\sim 2-4$ GeV. 
Again we find consistency when extending into smaller $z$ values for SIDIS as can be seen from the few data points
on the left of each plot. 
Besides, we find the $Z$-tagged jet fragmentation data from LHCb at 8 TeV~\cite{LHCb:2019qoc} not consistent with the LHCb 13 TeV measurement, which thus are not included in our analysis.
Also the OPAL and TPC measurements on fragmentation to protons are excluded due to tension with other measurements from SIA. 
Measurements on jet fragmentation have been widely used as a probe of final state medium effects in heavy-ion collisions.
The probability and pattern of parton fragmentation in jet are supposed to be modified via medium induced radiations and medium response, see e.g. \cite{Chen:2020tbl, Zhang:2022bhq}.
Jet transport coefficients can be extracted by comparing measured hadron production cross sections in heavy-ion collisions to reference cross sections, namely those would be without final state medium effects, see e.g.~\cite{Xie:2022fak}.
We show predictions on the reference cross sections at NLO in QCD for $Z$-tagged jet fragmentation in $PbPb$ collisions at 5.02 TeV, and compare to the CMS and ATLAS measurements~\cite{CMS:2021otx,ATLAS:2020wmg} in Fig.~\ref{Fig:pbpb}.
We use nCTEQ15 PDFs~\cite{Kovarik:2015cma} for PDF inputs of $Pb$ nucleus.
Experimental setup of the measurements are the same as those of $pp$ collisions included in our global analysis.
Each plot show measurements from different centrality classes in $PbPb$ collisions, and measurements from $pp$ collisions as we used previously.
Firstly we can see the $PbPb$ reference cross sections are almost the same as predictions on $pp$ cross sections since the changes in PDFs only lead to small corrections on flavor composition of final state jets. 
For central $PbPb$ collisions the measured jet fragmentation cross sections are clearly suppressed as compared to the reference cross sections, by almost 50\%.
For peripheral collisions shown in the CMS plot, the measurements are in consistent with the reference cross sections within uncertainties. 

\section{SIDIS calculation in FMNLO}
\label{sec:fmnlo}
\subsection{Program}

In this study we have developed \texttt{v2.0} of the \texttt{FMNLO}
program.
Instructions on installation and usage of  \texttt{FMNLO} can be found in appendix A of  Ref.~\cite{Liu:2023fsq}. 
Here we highlight only the usage of the SIDIS component, which has been available since \texttt{v2.0}. 
We take the module used for the calculation of muon on proton target at the COMPASS experiment
as an example of the usage of the SIDIS component. 
This module, named \texttt{A4001}, is one of the default examples available in the \texttt{FMNLOv2.0} package.
The parameter card for this module corresponds to the file \texttt{FMNLO/mgen/A4001/proc.run}, and reads
\begin{verbatim}
sidis A4001
# subgrids with name tags
grid compass16x6y4
pdfname	'CT14nlo'
etag 'e+'
htag 'p'
obs 1
zdef 2
cut 0.02
q2d 10.0
q2u 100000.0
xbjd 0.14
xbju 0.18
yid 0.3
yiu 0.5
pdfmember 0
sqrtS 17.334
Rscale 1.0
Fscale 1.0
ncores 30
maxeval	1000000
iseed 13
end
\end{verbatim}
where
\begin{itemize}
\item \texttt{sidis} specifies the name of the directory that contains the module to be loaded.
\item \texttt{grid} is a string indicating the name of the running job.
\item \texttt{pdfname} and \texttt{pdfmember} specifies the proton PDFs to be used, which should be available in LHAPDF.
\item \texttt{etag} is the flavor of the lepton, can be either `e$-$' or `e$+$'. Note we always include contributions from the exchange of the $Z$ boson. 
\item \texttt{htag} specifies the target, which can be `p' for the proton,  `iss' for isoscalar nuclei. In the case of isoscalar, the nuclear PDFs are obtained from proton PDFs by assuming free nucleons with isospin symmetry.
\item \texttt{obs} specifies different distributions to be calculated: 
currently only \texttt{1} is available for distribution in hadron scaled momentum.
\item \texttt{zdef} indicates exact definition of the scaled momentum, \texttt{1} for using $x_p$ and \texttt{2} for using $z$.
\item \texttt{cut} gives the slicing parameter $\lambda$ and a value of $0.02$ is recommended.
\item \texttt{q2d} and \texttt{q2u}: lower and upper limit of $Q^2$, in unit of $\rm{GeV}^2$.
\item \texttt{xbjd}, \texttt{xbju}, \texttt{yid} and \texttt{yiu}: lower and upper limits of the DIS variables $x$ and $y$.
\item \texttt{sqrtS}: the c.m. energy $\sqrt{s}$ in GeV.
\item \texttt{Rscale}, \texttt{Fscale}: ratios of the renormalization and factorization scale w.r.t. our nominal choices.
\item \texttt{ncores, maxeval, iseed} indicate technical parameters of numerical calculations, including the number of CPU cores used, the maximum number of integrand evaluations, and seed for pseudo-random-number generation.
\end{itemize}

\subsection{Comparison to analytic results}

\begin{figure}[htbp]
\centering
  \includegraphics[width=0.6\textwidth]{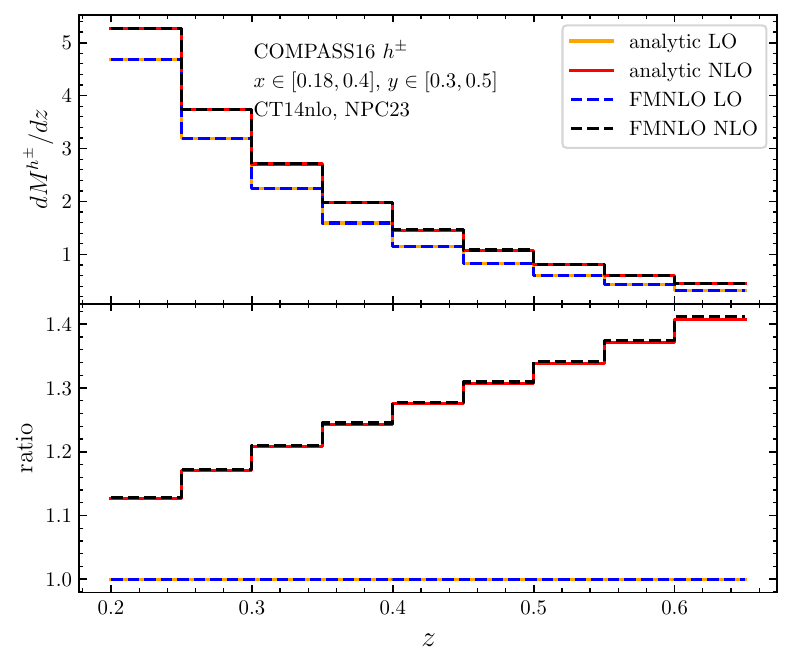}
	\caption{
Comparison of FMNLO results to the analytic predictions of the charged hadron multiplicities.
	}
  \label{Fig:comp-fmnlo-anal}
\end{figure}
In this section, we show the FMNLO predictions for SIDIS measurements.
In Fig.~\ref{Fig:comp-fmnlo-anal}, we present comparison of the FMNLO results on the charged hadron multiplicities to those obtained using analytic predictions,
corresponding to the settings of the COMPASS
measurements \cite{COMPASS:2016xvm,COMPASS:2016crr,Pierre:2019nry}. 
The multiplicity differential in scaled momentum $z$ for a hadron of type $h$ is defined in Eq.(\ref{Eq:hmul}).
As an example, we consider the COMPASS data bin with highest inelasticity, constrained by the kinematic cuts
$x\in[0.18, 0.4], y\in[0.3,0.5]$, at $s=300~\rm{GeV}^2$.
We use the proton PDF from CT14NLO~\cite{Dulat:2015mca} and the unidentified charged hadron FFs from NPC23.
The SIDIS cross section neglecting $Z$ boson contributions is given by
\begin{eqnarray}
     {\frac{d^3 \sigma^{\rm h}}{d x d y d z}} &
  = & \frac{4 \pi \alpha_e^2}{Q^2} \left\{ \frac{1 + (1 - y)^2}{2 y}
  \mathcal{F}_T^h (x, z, Q^2) + \frac{1 - y}{y} \mathcal{F}_L^h (x, z, Q^2)
  \right\},
\end{eqnarray}
where $ \mathcal{F}_{T/L}^h (x, z, Q^2)$  is related to the standard structure functions by 
$ \mathcal{F}_{T}=2F_1$,
$ \mathcal{F}_{L}=F_L/x$.
The latter can be factorized into PDFs $f$, FFs $D_i^h$, and partonic coefficient functions as 
\begin{align}
\mathcal{F}_i^h\left(x, z, Q^2\right)= & 
\sum_{f, f^{\prime}} \int_x^1 \frac{d \hat{x}}{\hat{x}} 
\int_z^1 \frac{d \hat{z}}{\hat{z}} 
f\left(\frac{x}{\hat{x}}, \mu_F\right) 
D_{f^{\prime}}^h\left(\frac{z}{\hat{z}}, \mu_D\right) \\
& \times \mathcal{C}_{f^{\prime} f}^i\left(\hat{x}, \hat{z}, \frac{Q^2}{\mu_R^2}, \frac{Q^2}{\mu_F^2}, \frac{Q^2}{\mu_D^2}, \alpha_s\left(\mu_R\right)\right).
\end{align}
The partonic coefficient functions have been calculated at NLO in QCD~\cite{Altarelli:1979kv,Nason:1993xx,Furmanski:1981cw,Graudenz:1994dq,deFlorian:1997zj,deFlorian:2012wk}, and at NNLO in QCD recently~\cite{Goyal:2023xfi, Bonino:2024qbh}.
We find very good agreement between the FMNLO results and the analytic predictions as shown in Fig.~\ref{Fig:comp-fmnlo-anal}, 
and the NLO corrections are around 10-40\%.

\begin{figure}[htbp]
\centering
  \includegraphics[width=0.6\textwidth]{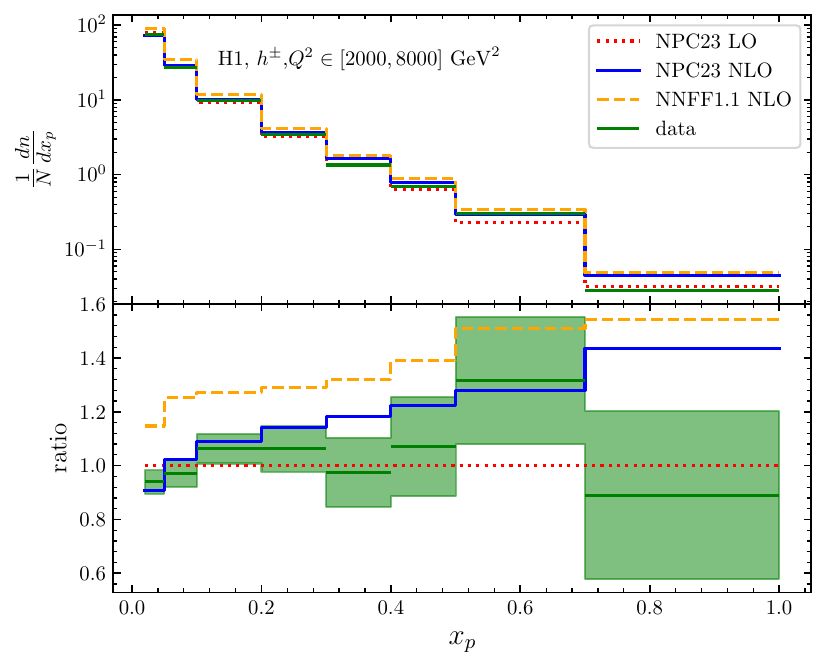}
	\caption{
    LO and NLO predictions for the hadron multiplicity differential in scaled momentum $x_p$ for SIDIS at $\sqrt{s}=318~{\rm GeV}$
    , in the photon virtuality $2000<Q^2<8000~{\rm GeV}^2$. 
    We show the predictions of the central FF from our fit and NNFF1.1.
	}
  \label{Fig:h1-2000-fmnlo}
\end{figure}

In Fig.~\ref{Fig:h1-2000-fmnlo}, we further present the LO and NLO predictions from FMNLO for the hadron multiplicity differential in scaled momentum $x_p$ for SIDIS at $\sqrt{s}=318~{\rm GeV}$, with the photon virtuality limited to $2000<Q^2<8000~{\rm GeV}^2$,
corresponding to the one of the high $Q^2$ bin of the H1 measurement~\cite{H1:2007ghd}.
Analytical results on hadron multiplicity differential in $x_p$ are not available.
We again use the unidentified charged hadron FFs from NPC23 while also include results using FFs from NNFF1.1 as well as the experimental data for comparisons.  
The NLO corrections are about 40\% in the large $x_p$ region.
The NLO predictions using of NNFF1.1 overshoot the data by 20\% or more in general. 

\section{LHAPDF6 grid}
\label{sec:lha6}
In this section, we provide a concise overview on the data base of our fragmentation functions. 
Interpolation tables of NPC23 FFs are formatted in LHAGRID1 format \cite{Buckley_2015}, the same format employed in PDF grid files.
Access to the FFs is facilitated through the unified interface of \texttt{LHAPDF6}, accessible via Fortran, CPP, and Python code.
To utilize these sets, one can extract the FF sets into the LHAPDF data directory, accessible through the \texttt{lhapdf --datadir} command or direct download from the website~\footnote{\url{https://lhapdf.hepforge.org/pdfsets.html}}.
The FFs we have fitted correspond to various partons fragmenting into $\pi^+, K^+, p$. 
They are named \texttt{NPC23\_PIp\_nlo, NPC23\_KAp\_nlo, and NPC23\_PRp\_nlo}, respectively.
The FFs for negative charge hadrons are obtained through charge conjugation, such as $D_u^{\pi^+} = D_{\bar u}^{\pi^-}$. 
They are released as \texttt{NPC23\_PIm\_nlo, NPC23\_KAm\_nlo, NPC23\_PRm\_nlo}. 
Additionally, FFs for $\pi^{\pm}, K^\pm, p(\bar p)$ combined are available as \texttt{NPC23\_PIsum\_nlo, NPC23\_KAsum\_nlo, NPC23\_PRsum\_nlo}, obtained by summing the fragmentation to positively and negatively charged hadrons.
Finally, FFs to positively, negatively charged hadron and unidentified charged hadrons are released as \texttt{NPC23\_CHHAp\_nlo}, \texttt{NPC23\_CHHAm\_nlo}, \texttt{NPC23\_CHHAsum\_nlo}.
Each of above fragmentation functions consist of 127 subsets, with one central set and 126 Hessian error sets corresponding to the positive and negative directions of 63 orthogonal directions.
For estimation of FFs uncertainties of any observable $X$, the following formula for asymmetric errors is employed~\cite{Nadolsky:2001yg}:
\begin{equation}\label{Eq:error}
\begin{aligned}
\delta^{+} X & =\sqrt{\sum_{i=1}^{N_d}\left[\max \left(X_{2 i-1}-X_0, X_{2 i}-X_0, 0\right)\right]^2}, \\
\delta^{-} X & =\sqrt{\sum_{i=1}^{N_d}\left[\max \left(X_0-X_{2 i-1}, X_0-X_{2 i}, 0\right)\right]^2}.
\end{aligned}
\end{equation}
Here, $X_0$ represents prediction obtained with the central set of FFs, and $X_{2i-1}(X_{2i})$ represents prediction obtained with the error set for the $i$-th eigenvector in positive (negative) direction.
Note when the observable involves FFs of different hadrons, one should take into account their correlations. 
For instance, for ratios of pion and kaon FFs, one should first calculate the ratios using consistent error sets of pions and kaons, and then apply Eq.~(\ref{Eq:error}) for estimation of the uncertainties. 
The interpolation tables of FFs utilize nodes at different momentum fraction $z$ and different fragmentation scale $Q$.
$z$ ranges from 0.003 to 1 with 99 nodes, while $Q$ spans from 4 to 4000 GeV with 32 nodes in total.
The grids also provide numerical values of $\alpha_S$ with $\alpha_S(M_Z)=0.118$ and two-loop running.
The number of active quark flavors is always fixed to 5.
The interpolator is set to use the default logcubic method, and the FFs are frozen when going out of ranges of $z$ or $Q$ specified above. 
\bibliography{npc232}

\end{document}